\documentclass[traditabstract]{aa}
\usepackage{graphicx}
\usepackage{txfonts}
\usepackage{array}
\usepackage{amssymb}
\usepackage{lscape}
\usepackage{rotating}
\usepackage{longtable}
\newcommand{\ci}{Cyg\,OB2\,\#5}

\newcommand{\xmm}{{\sc{XMM}}\emph{-Newton}}

\newcommand{\sw}{{\em Swift}}
\begin{document}

{\title{Quest for the tertiary component in \ci\thanks{Based on spectra obtained with the TIGRE telescope, located at La Luz observatory, Mexico (TIGRE is a collaboration of the Hamburger Sternwarte, the Universities of Hamburg, Guanajuato, and Li\`ege), as well as data collected at the Observatoire de Haute Provence, with the Neil Gehrels \sw\ Observatory, and with \xmm , an ESA Science Mission with instruments and contributions directly funded by ESA Member States and the USA (NASA).}\thanks{Table\,1 is only available in electronic form at the CDS via anonymous ftp to cdsarc.u-strasbg.fr (130.79.128.5) or via http://cdsweb.u-strasbg.fr/cgi-bin/qcat?J/A+A/.}}

\author{Gregor~Rauw\inst{1} \and Ya\"el~Naz\'e\inst{1}\thanks{F.R.S.-FNRS Research Associate.}
\and Fran~Campos\inst{2}
}

\institute{Groupe d'Astrophysique des Hautes Energies, STAR, Universit\'e de Li\`ege, Quartier Agora (B5c, Institut d'Astrophysique et de G\'eophysique), All\'ee du 6 Ao\^ut 19c, B-4000 Sart Tilman, Li\`ege, Belgium\\
  \email{g.rauw@uliege.be}
  \and Observatori Puig d’Agulles, Passatge Bosc 1, 08759 Vallirana, Barcelona, Spain}

\authorrunning{Rauw et al.}
\titlerunning{Monitoring of \ci }
%\date{}
  \abstract
   {}
   {The \ci\ system is thought to consist of a short-period (6.6\,d) eclipsing massive binary orbited by an OB-star orbiting with a period of $\sim$6.7\,yr; these stars in turn are orbited by a distant early B-star with a period of thousands of years. However, while the inner binary has been studied many times, information is missing on the other stars, in particular the third star whose presence was indirectly postulated from recurrent modulations in the radio domain. Besides, to this date, the X-ray light curve could not be fully interpreted, for example in the framework of colliding-wind emission linked to one of the systems.}
   {We obtained new optical and X-ray observations of \ci, which we combined to archival data. We performed a thorough and homogeneous investigation of all available data, notably revisiting the times of primary minimum in photometry. }
   {In the X-ray domain, \xmm\ provides scattered exposures over $\sim$5000\,d whilst \sw\ provides a nearly continuous monitoring for the last couple of years. Although the X-ray light curve reveals clear variability, no significant period can be found hence the high-energy emission cannot be explained solely in terms of colliding winds varying along either the short or intermediate orbits. The optical data reveal for the first time clear signs of reflex motion. The photometry indicates the presence of a 2366\,d (i.e.\ 6.5\,yr) period while the associated radial velocity changes are detected at the $3\sigma$ level in the systemic velocity of the He\,{\sc ii} $\lambda$\,4686 emission line. With the revised period, the radio light curve is interpreted consistently in terms of a wind interaction between the inner binary and the tertiary star. From these optical and radio data, we derive constraints on the physical properties of the tertiary star and its orbit.}
   {}

\maketitle

\section{Introduction}
Located in the highly reddened Cyg\,OB2 association, \ci\ (V729 Cyg, BD+40$^{\circ}$4220) was detected long ago to be an eclipsing binary with a period of 6.6\,d \citep{mic53}. Subsequent photometric and spectroscopic investigations of the system found the two components to be of similar brightness but with a primary roughly three times more massive than the secondary \citep{Boh76,Mas77,rau99}. Both stars (hereafter called A and B) are massive, and the primary and secondary have spectral types O\,6.5-7 and Ofpe/WN9, respectively \citep{rau99}. The second {\it Gaia} data release \citep[DR2;][]{Brown} quotes a parallax of 0.64$\pm$0.06\,mas, in good agreement with the radio parallax \citep[0.61$\pm$0.22\,mas,][]{dzi13}, and corresponding to a distance of $1501^{+142}_{-119}$\,pc \citep{bai18}.

At radio frequencies, the \ci\ binary is a clear emitter but it is not alone. It has a close (at 0.8'' north-east), faint companion of elongated shape \citep{mir94,con97}. Because of its shape, its location between the binary and an astrometric companion, and its non-thermal nature, \citet{con97} proposed that this additional radio emission arises in a collision between the winds of the binary and the astrometric companion, which is located at 0.98'' from the eclipsing binary. This star, hereafter called component D\footnote{Throughout this paper, we adopt the naming convention of \citet{ken10} which is different from that used in the Washington Double Star Catalog \citep{mas01}.}, is probably of spectral type B0--2\,V, and its orbital period would be about 8000\,yr at the {\it Gaia} distance. The situation became more complex when \citet{ken10} reported that the radio emission associated with the binary varies with a period of 6.7$\pm$0.3\,yr between a thermal nature and a mixed (thermal plus non-thermal) nature\footnote{This actually confirmed an early periodicity detection in the system from \citet{mir94}.}. The varying non-thermal emission was interpreted as also being due to colliding winds, this time between the binary and another, previously unknown, massive star that we shall designate as component C. The disappearance of this non-thermal component in Very Large Array ($VLA$) data of November 1987, April 1994, and mid-2000 is explained by the periastron passage, as the stars would then be so close that the collision takes place inside the radio photosphere. \citet{ort11} further imaged this non-thermal radio emission, revealing a bow-shock shape typical of colliding winds. They also noted the absence of flux variations on short timescales (hours), again in agreement with expectations for colliding winds. Using $VLBA$ data, \citet{dzi13} confirmed the previous results, including the late-O/early-B type for component C. These authors also reported a disappearance of the radio emission in mid-2012, which is earlier than expected from the favoured model of \citet{ken10}. 

Wind interactions are also likely to operate between the components of the eclipsing binary (stars A and B). Indeed, the H$\alpha$ and He\,{\sc ii} $\lambda$\,4686 emission lines display phase-locked line profile variations on the 6.6\,d cycle \citep{Vreux,rau99}. Doppler maps built from the H$\alpha$ line revealed that this emission arises in material moving from the primary towards the secondary and in the same orbital direction as the secondary, as would be expected from a wind-wind collision between the binary components \citep{lin09}. \ci\ thus appears as a complex system containing four stars and no less than three wind-wind collisions.

A direct signature of component C has not yet been reported. However, an indirect detection could be possible as that star and the close binary orbit around their common centre-of-mass. Reflex motion with a period of $\sim 6.7$\,yr should thus be detected for the binary and it can be used to estimate the properties of the companion. \citet{ken10} reported tentative hints of such a reflex motion in the radial velocities (RVs) of absorption lines reported by \citet{rau99}. However, the RVs of the absorption lines of the eclipsing binary are affected by large uncertainties and the small number of data further limited the significance of the results. \citet{caz14} reported no difference between the RV curve of the He\,{\sc ii} $\lambda$\,4686 emission as measured on old data and a new set of data covering a single 6.6\,d orbit in mid-2013, which is close to the expected periastron time of the tertiary in the favoured model of \citet{ken10}. Since the ephemeris of \citet{ken10} is preliminary, as revealed by the difference in periastron passage measured by \citet{dzi13}, this question is not settled yet and awaits a full monitoring of the $\sim 6.7$\,yr period. On the other hand, as the inner binary in \ci\ is eclipsing, observing cyclic time delays of the observed primary minimum with respect to the best ephemeris also provides a means to uncover reflex motion. Regarding the ephemeris, period changes linked to mass-exchange and mass-loss implying the need to use a quadratic ephemeris were investigated by several authors \citep{lin09,yas14,lau15}, but a cyclic ephemeris was only envisaged by \citet{qia07} and \citet{caz14}. Unfortunately, the former authors do not provide details on their conclusion that cyclic variations exist and the latter paper provided only hints for their presence (as there was notably one discrepant point). A new solution, accounting for both quadratic and cyclic changes, is thus required.

Another indirect way to obtain information on Star C is through the analysis of the X-ray emission of the system. \ci\ was amongst the first massive stars detected in X-rays \citep{har79}. Further X-ray observations were reported by different authors \citep{lin09,yos11,caz14}. \ci\ was found to be brighter than usual ($\log{(L_{\rm X}/L_{\rm BOL})}=-6.4$, \citealt{lin09}) as well as harder (presence of a plasma at $kT=1-2$\,keV; \citealt{caz14}). Variations were also detected \citep{lin09,yos11,caz14}. These three characteristics are typical of X-rays arising in colliding wind interactions. However, no fully coherent folding with the short 6.6\,d period of the binary could be achieved in these three papers. The folding with the 6.7\,yr period was more promising \citep{caz14} but its coverage was very patchy and requires confirmation.

Several questions on \ci\ therefore remain without clear answers: What is exactly the orbit of Star C, what are its properties, and where does the X-ray emission come from? In this paper, we tackle these problems by performing a long-term optical and X-ray monitoring (presented in Sect.\,\ref{obs}) and by combining it with archival data. The results of our analyses for the X-ray and optical domains are presented in Sects.\,\ref{Xrays} and \ref{opt}, respectively. Section\,\ref{radiolc} rediscusses the radio light curve in view of our results, while Sect.\,\ref{summary} summarizes our findings.

\section{Observations and data reduction\label{obs}} 
\subsection{Optical spectroscopy}
Optical spectra of \ci\ in the blue domain were collected with the Aur\'elie spectrograph \citep{Gillet} at the 1.52\,m telescope of the Observatoire de Haute Provence (OHP, France). The data were taken during six observing campaigns of six nights each between June 2013 and August 2018. Aur\'elie was equipped with a $2048 \times 1024$\,CCD with a pixel size of 13.5\,$\mu$m squared. For all campaigns, except that of June 2015, we used a 600\,l\,mm$^{-1}$ grating providing a reciprocal dispersion of 16\,\AA\,mm$^{-1}$. The resolving power, measured on the Thorium-Argon calibration exposures, was 7000 over the wavelength range from 4440\,\AA\ to 4890\,\AA. In June 2015, we instead used a 1200\,l\,mm$^{-1}$ grating providing resolving power that is twice as high over the smaller wavelength domain from 4580\,\AA\ to 4770\,\AA. Typical integration times were 1 -- 2\,hours. The data were reduced using the {\sc midas} software (version 17FEBpl\,1.2).

Between 2015 and 2019, we also collected a number of high-resolution echelle spectra of \ci\ with the fully robotic 1.2\,m Telescopio Internacional de Guanajuato Rob\'otico Espectros\'opico \citep[TIGRE,][]{Schmitt} installed at La Luz Observatory near Guanajuato (Mexico). The TIGRE telescope features the refurbished HEROS echelle spectrograph covering the wavelength domain from 3800 to 8800\,\AA, with a small 100\,\AA\ wide gap near 5800\,\AA. The resolving power is about 20\,000. The data reduction was performed with the dedicated TIGRE/HEROS reduction pipeline \citep{Mittag}. 

\subsection{Optical photometry}
Dedicated differential photometry of \ci\ was obtained in the $V$ filter at the private observatory of one of us (F.C.). It is situated in Vallirana (near Barcelona, Spain) and it is equipped with a Newton telescope of 20\,cm diameter (with f/4.7 and a German equatorial mount). The camera is a CCD SBIG ST-8XME (KAF 1603ME) providing a field of view of 50\arcmin\ $\times$ 33\arcmin. The exposures were typically of 180\,s duration. The images were corrected for bias, dark current, and flat-field in the usual way using the data reduction software Maxim Dl v5. The photometry was extracted with the FotoDif v3.93 software, using as comparison star SAO~49783 (TYC~3157-195-1) which has a $V$-band magnitude very similar to our target. No colour transformation was applied. Four stars (TYC~3157-1310-1, TYC~3157-603-1, TYC~3161-1269-1, and TYC~3157-463-1) were further measured for checking the stability of the photometric reduction. The new photometric data of \ci\ are made available as Table\,1 at CDS.
\addtocounter{table}{1}

\begin{figure*}[thb]
\begin{minipage}{8.5cm}
  \includegraphics[width=8.5cm]{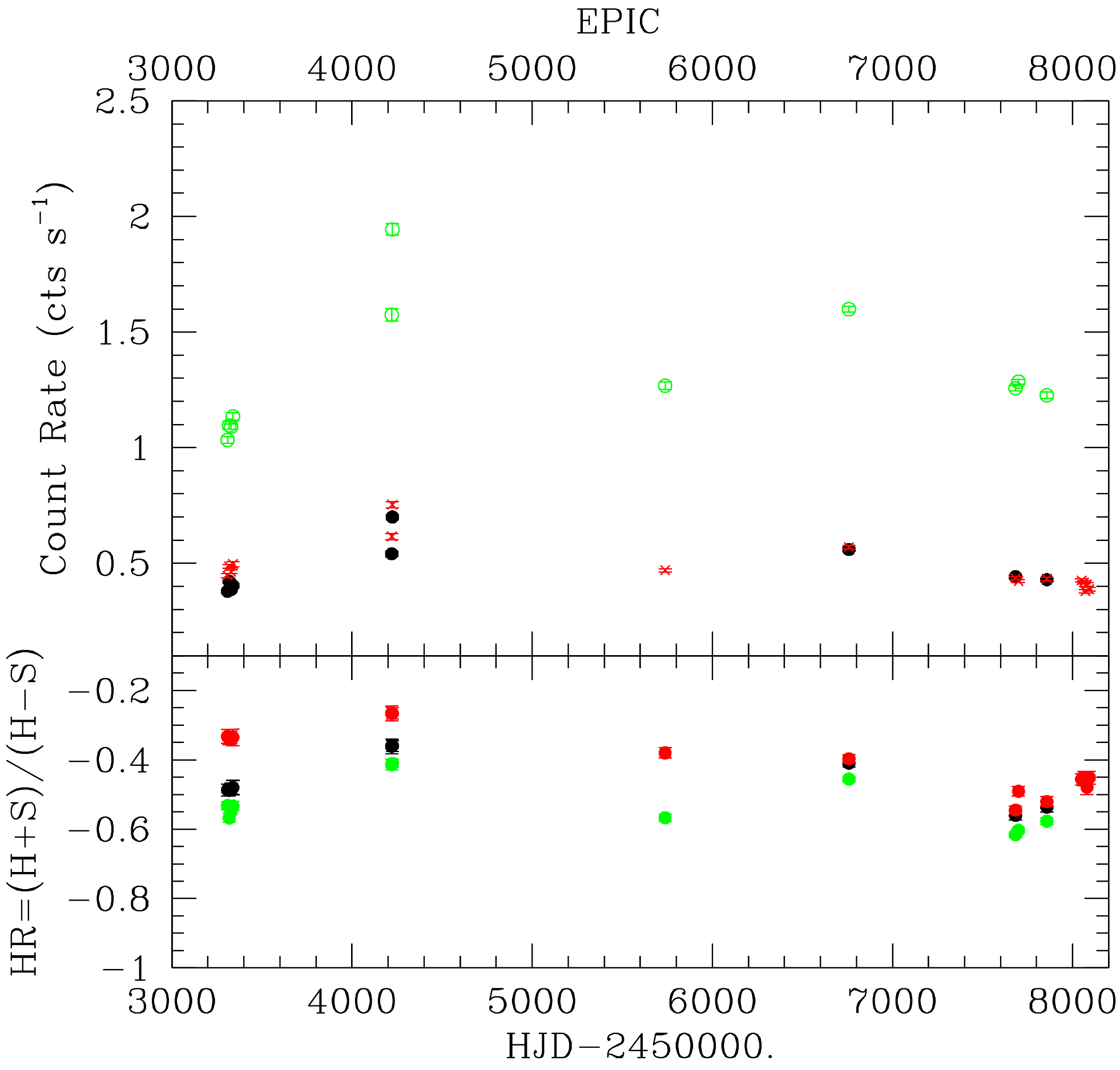}
\end{minipage}
\hfill
\begin{minipage}{8.5cm}
  \includegraphics[width=8.5cm]{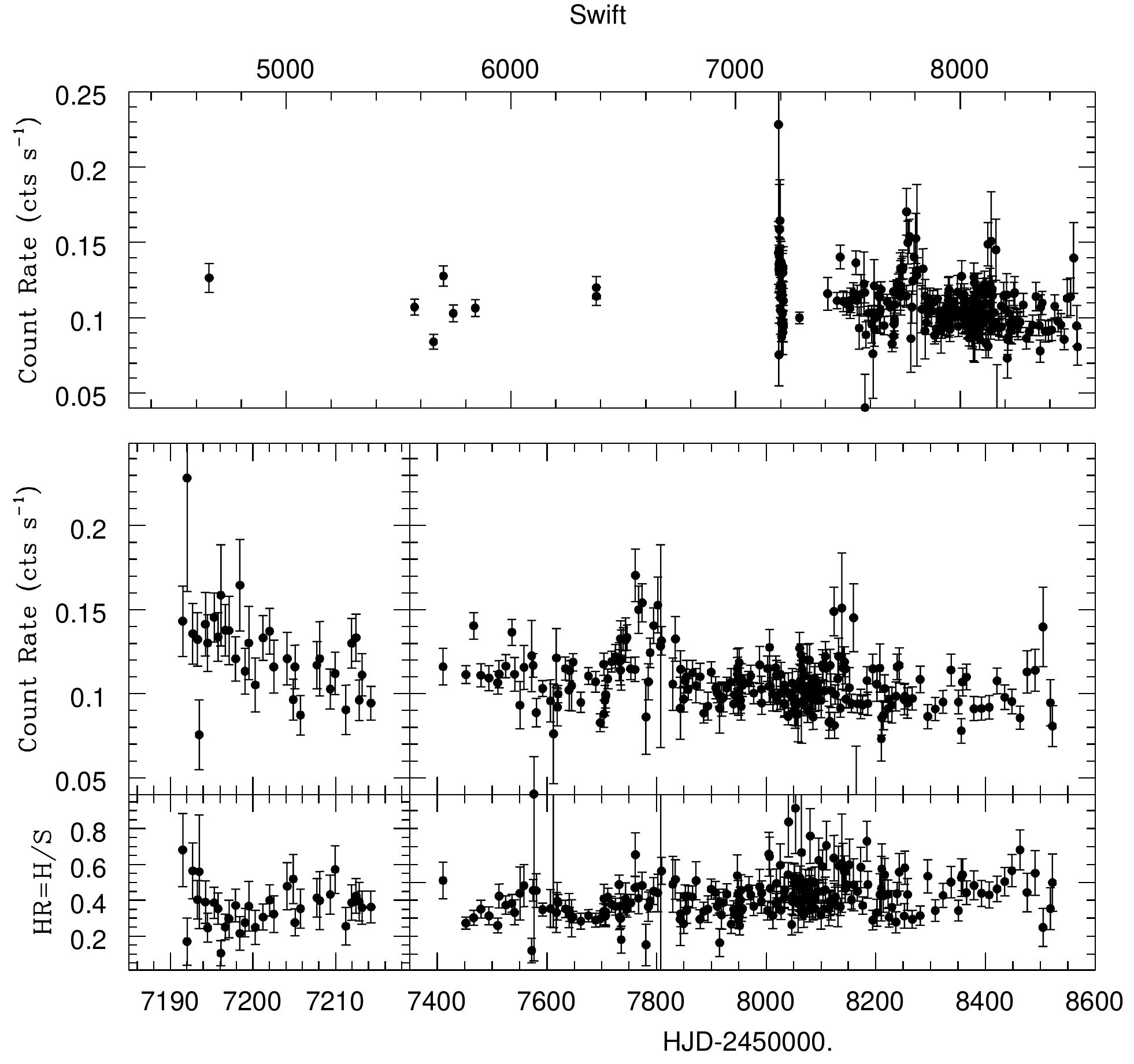}
  \end{minipage}
\hfill
\begin{minipage}{8.5cm}
  \includegraphics[width=8.5cm]{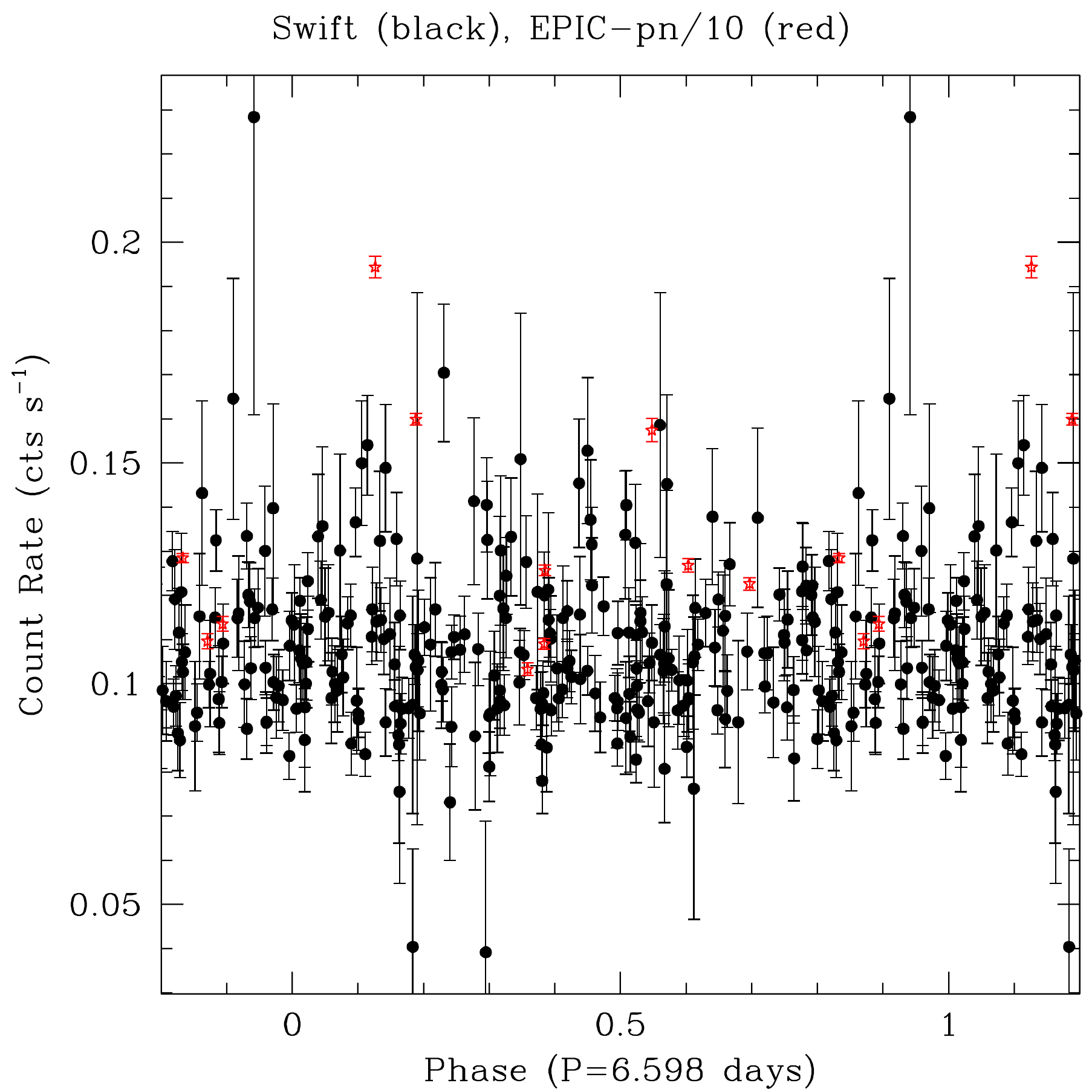}
  \end{minipage}
\hfill
\begin{minipage}{8.5cm}
  \includegraphics[width=8.5cm]{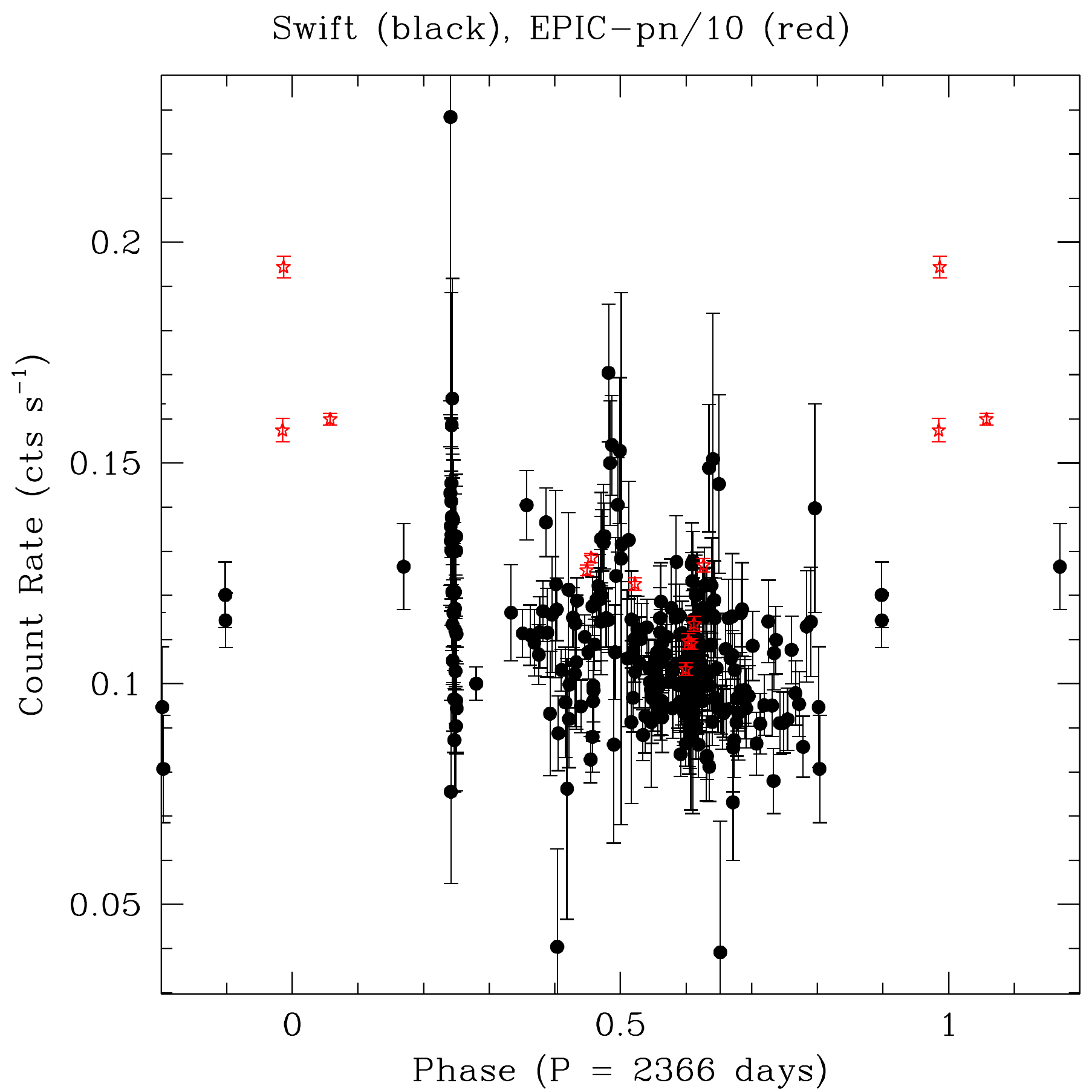}
\end{minipage}
\caption{{\it Top panels:} Light curves of \ci\ and evolution of $HR$ for \xmm\ (left, pn in green, MOS1 in black, and MOS2 in red) and \sw\ data (right). {\it Bottom panels:} Light curves folded with the two orbital periods of \ci. \xmm-pn data were divided by 10 and are shown as red stars, while \sw\ data are shown as black dots. On the left, quadratic ephemeris from Eq.\,\ref{ephem2} were used to derive the phase, while the ephemeris of the long orbit from Sect.\,\ref{outerorbit} were used for the right panel.}
\label{lc}
\end{figure*}

\subsection{X-ray domain}
\subsubsection{\xmm }
Since the launch of \xmm, we have obtained ten observations of the Cyg\,OB2 region \citep[see][for presentation of the older data]{deb06,rau11,naz12,caz14}. The association also appears off-axis in observations centred on PSR~J2032+4127. All datasets were reduced with SAS v16.0.0 using calibration files available in Fall 2017 and following the recommendations of the \xmm\ team\footnote{SAS threads, see \\ http://xmm.esac.esa.int/sas/current/documentation/threads/ }. 

The EPIC observations were taken in the full-frame mode and with the medium filter (to reject optical and UV light), except for ObsID 0677980601 for which the large window mode was used for the MOS cameras to avoid pile-up in Cyg\,OB2\,\#9 (which was then at its maximum). After pipeline processing, the data were filtered for keeping only best-quality data ({\sc{pattern}} of 0--12 for MOS and 0--4 for pn). Background flares were detected in several observations (Revs 0896, 0911, 1353, 1355, 2114, 3097). Only times with a count rate for photons of energy above 10.\,keV lower than 0.2--0.3\,cts\,s$^{-1}$ (for MOS) or 0.4\,cts\,s$^{-1}$ (for pn) were kept. A source detection was performed on each EPIC dataset using the task {\it edetect\_chain} on the 0.3--2.0 (soft) and 2.0--10.0 (hard) energy bands and for a log-likelihood of 10. This task searches for sources using a sliding box and determines the final source parameters from point spread function fitting; the final count rates correspond to equivalent on-axis, full point spread function count rates (Table~\ref{journalxmm}).

\subsubsection{\sw}
We also obtained some dedicated observations of the Cyg\,OB2 region with the Neil Gehrels \sw\ Observatory. Additional \sw\ X-ray Telescope (XRT) data of \ci\ exist in the archives as another campaign on Cyg\,OB2 was performed in June-July 2015 and observations dedicated to PSR J2032+4127, 3EGJ2033+4118, or WR\,144 serendipitously encompass \ci. All these XRT data taken in PC mode were retrieved from the HEASARC archive centre and were further processed locally using the XRT pipeline of HEASOFT v6.22.1 with calibrations v20170501. In addition, corrected count rates in the same energy bands as \xmm\ were obtained for each observation from the UK on-line tool\footnote{http://www.swift.ac.uk/user\_objects/} (Table~\ref{journalswift}). No optical loading is expected for XRT data because of the severe extinction towards Cyg\,OB2.

\begin{figure}
\includegraphics[width=8.5cm]{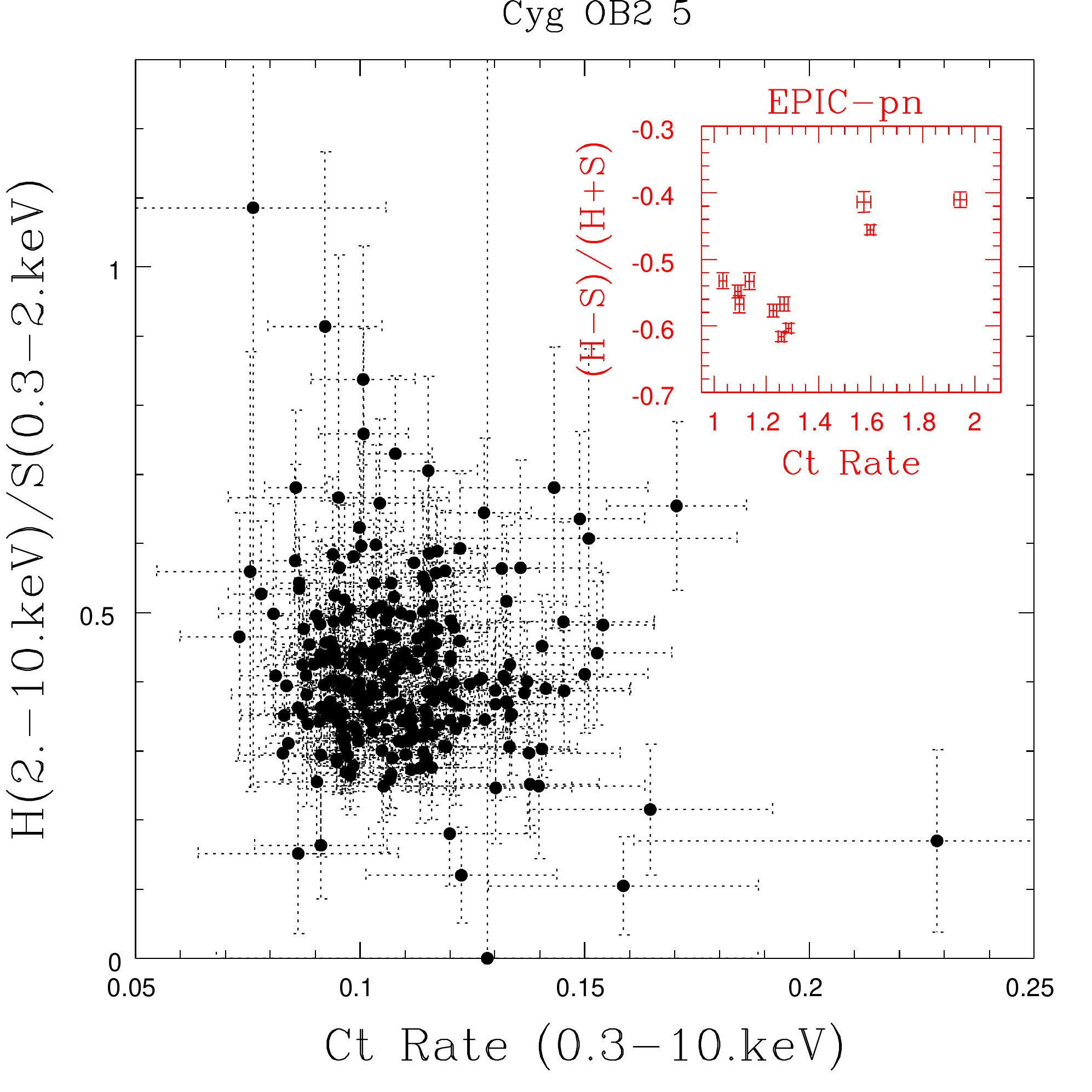}
\caption{Count rates compared to hardness ratios for \sw\ and \xmm\ (in red, inset). For the \xmm\ data, the vertical axis represents $HR = (H+S)/(H-S)$ rather than $H/S$.}
  \label{cthr}
\end{figure}

\section{Long-term X-ray monitoring \label{Xrays}}

The top panels of Fig.\,\ref{lc} present the \xmm\ and \sw\ light curves. The \xmm\ data are few in number and scattered over 12\,yr, but \sw\ has regularly monitored \ci\ every week (on average) since 2016. It is immediately clear from the light curves that the X-ray emission of \ci\ significantly varies. The hardness ratio changes too, but these differences are only truly significant in \xmm\ data. The correlation between flux and hardness (Fig. \ref{cthr}) amounts to 0.77 for \xmm-pn data, implying that the spectrum hardens as it brightens. In the spectral fits of \citet{caz14}, this hardening appears linked to an increase in absorption. Yet, we note that the \sw\ data reveal no clear correlation between count rates and brightness ratio (most probably because of their much larger noise). 

Some remarkable events can be spotted in the \sw\ light curve: a monotonic decrease during a month-long campaign in summer 2015 (around $HJD$=2\,457\,200) and a complex rise/decrease event in winter 2015--2016 (around $HJD$=2\,457\,750). In the past, the nearly doubling of the X-ray flux between the first \xmm\ data of November 2004 and the second \xmm\ set of April/May 2007 appeared surprising \citep{lin09,caz14}, but this change possibly reflects another event of the kind revealed by the long-term \sw\ observations. However, while changes of large amplitude exist, it is difficult to see a clear periodicity in the behaviour of \ci.

To assess this in a quantitative way, we applied a set of period search algorithms: (1) the Fourier algorithm adapted to sparse/uneven datasets \citep[a method rediscovered recently by \citealt{zec09}\footnote{These papers also note that the method of \citet{sca82}, while popular, is not fully correct, statistically.}]{hmm,gos01}, (2) three binned analyses of variances (\citealt{whi44}, \citealt{jur71}, which is identical with no bin overlap to the ``pdm'' method of \citealt{ste78}, and \citealt{cuy87}, which is identical to the ``AOV'' method of \citealt{sch89}), and (3) conditional entropy \citep[see also \citealt{gra13}]{cin99,cin99b}. We calculated periodograms for different datasets: the full \sw\ dataset, the \sw\ data before 2016, the \sw\ data from 2016, the \sw\ data from 2017, the \sw\ data from 2018, and the \xmm\ and \sw\ data combined, using pn count rates divided by a factor of 10. Results are shown in Figure \ref{fourier}. No coherent and significant period emerges from the analysis. In particular, the two main periods do not show up in a significant way.

\begin{figure*}[thb]
\includegraphics[width=6.0cm]{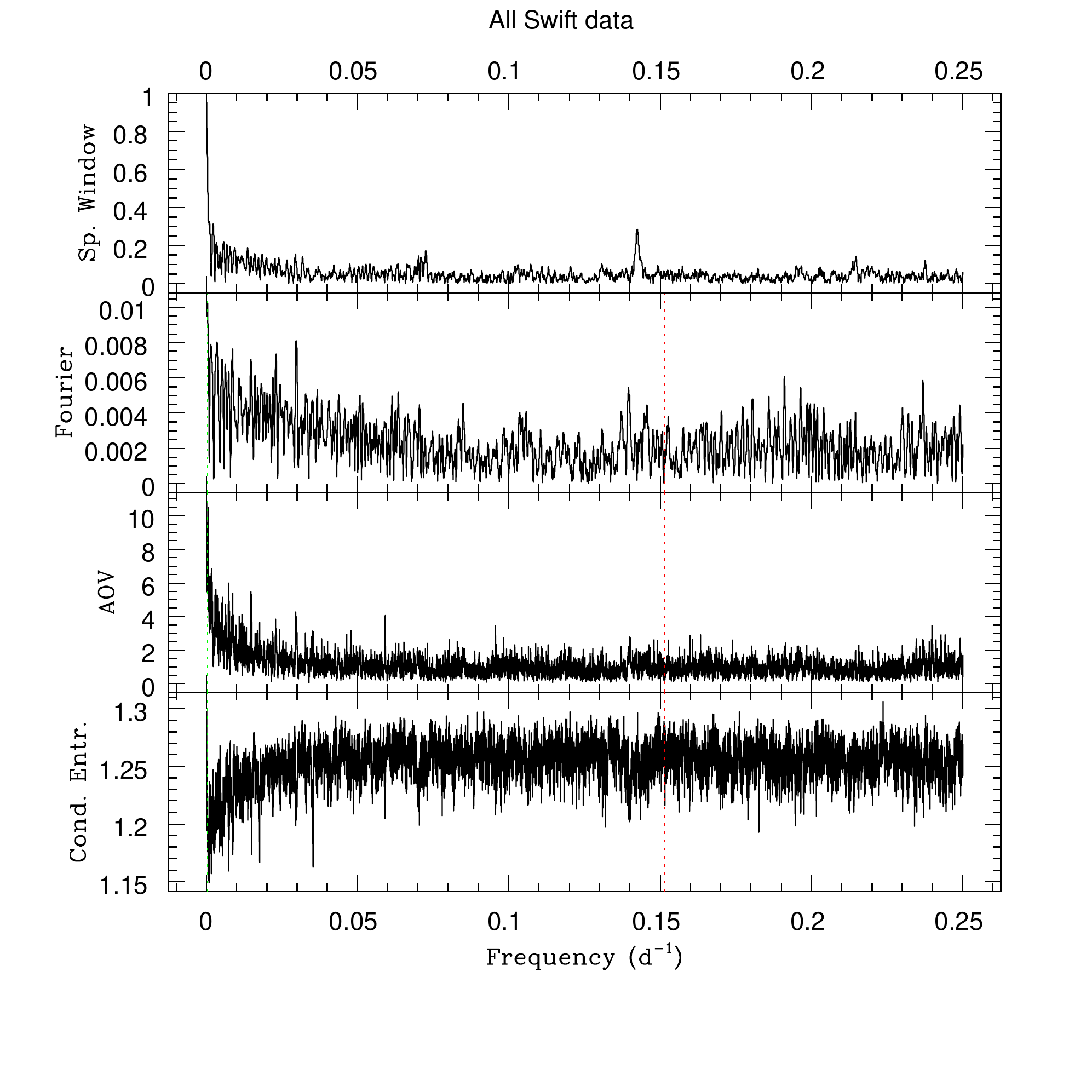}
\includegraphics[width=6.0cm]{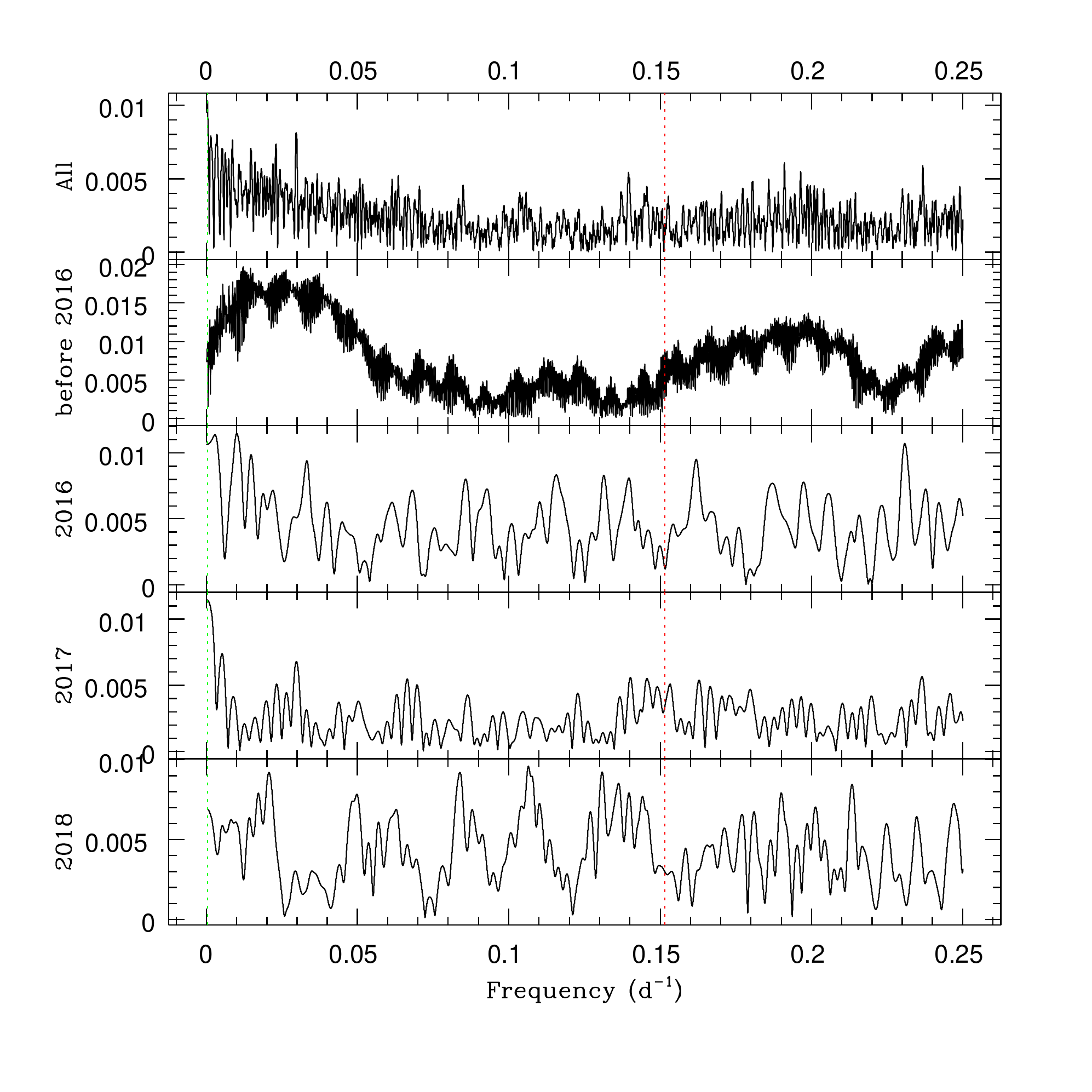}
\includegraphics[width=6.0cm]{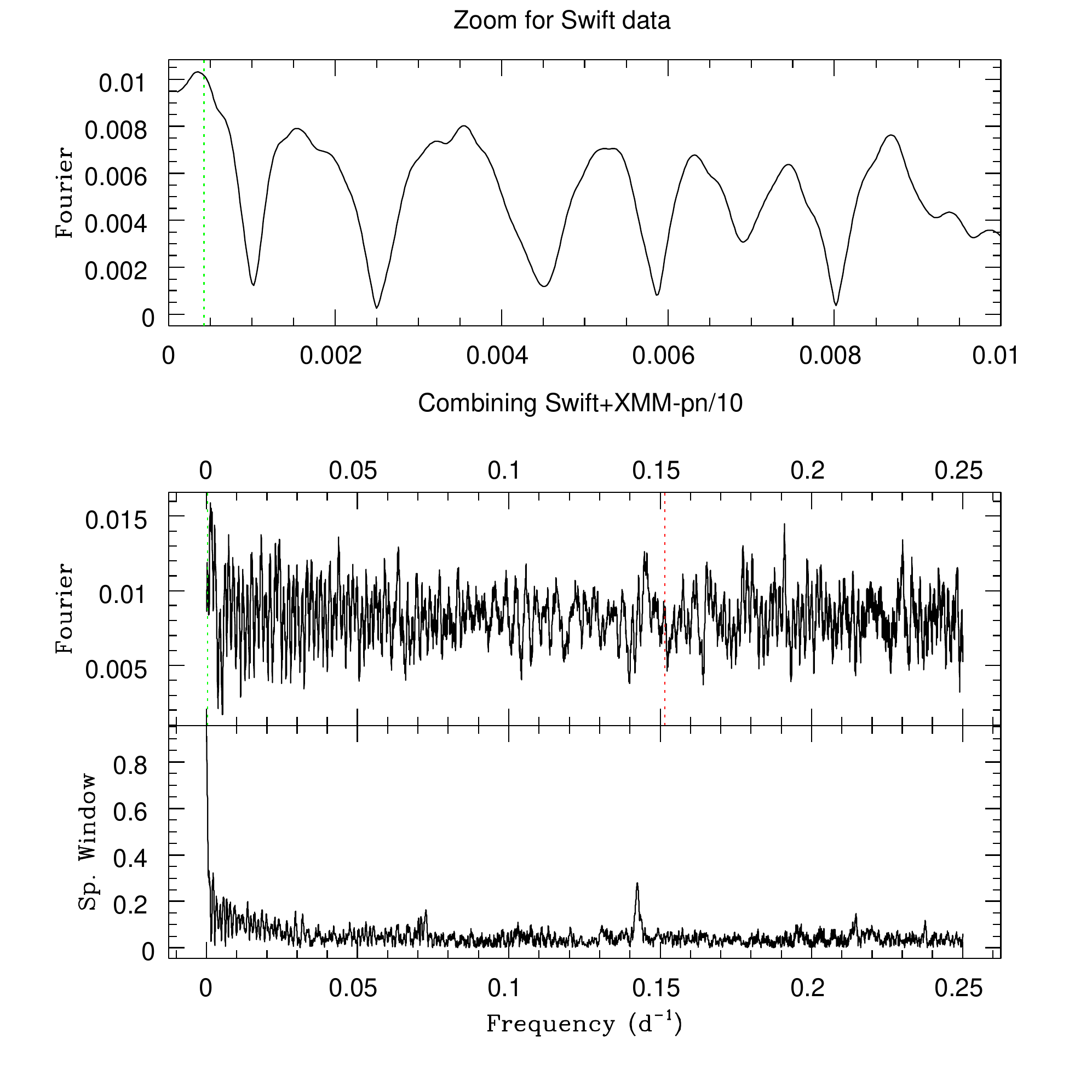}
\caption{{\it Left:} Modified Fourier (with its spectral window on top), AOV, and conditional entropy periodograms considering all \sw\ data. {\it Middle:} Modified Fourier periodograms for different sets of \sw\ data. {\it Right, top:} Zoom on the low-frequency content of the modified Fourier periodogram calculated on all \sw\ data. {\it Right, bottom:} Modified Fourier periodogram, with its spectral window, for \sw\ and \xmm-pn data combined. The pn data were arbitrarily scaled by a factor 0.1. The two known periods (see also Sect.\,\ref{opt}) are shown by dotted lines: green for 2366\,d and red for 6.6\,d.}
\label{fourier}
\end{figure*}

We further assess the compatibility of the X-ray data with the proposed periods (6.6\,d and 6.5\,yr, see Sect.\,\ref{outerorbit}) by folding the light curves with them (bottom panels of Fig. \ref{lc}). As is obvious from these figures, there seems to be no smooth and coherent behaviour with phase, whatever the period considered. In particular, the densely sampled \sw\ light curve appears scattered and does not exhibit the variations with orientation \citep[genuine eclipses or modulations of wind absorption; e.g.][]{Willis,Rau14,Lom15,Gos16} or orbital separation \citep[minimum near apastron passage for long-period systems; e.g.][]{naz12,Pan14} typical of colliding wind systems \citep[for a review on the X-ray emission of such systems, see][]{rau16}. In addition, \citet{dzi13} mentioned a radio minimum in mid-2012. Whilst no X-ray data are available for that year, observations exist $\sim$6.5\,yr later in 2018-2019 and no remarkable event is detected. There is a lack of \sw\ data around periastron of the 6.5\,yr cycle, however. This gap is filled by some scarce \xmm\ data. These data seem to indicate an increased flux, as could be expected for a wind-wind collision, but the behaviour is not smooth as two points taken a few days apart have count rates differing by nearly 25\%. Therefore, we cannot draw a firm conclusion as to the exact origin of the emission and its variations.

Since our study reveals that the X-ray emission of \ci\ is not simply phase-locked with either the short or long orbital periods, we might wonder whether the X-rays could come from another source? Confusion with an unresolved foreground or background source is unlikely. Indeed, though Cyg\,OB2 hosts a population of low-mass pre-main sequence (PMS) stars which can be relatively X-ray bright, the X-ray emission of \ci\ is much brighter than that of flaring PMS objects in Cyg\,OB2 \citep[see Fig.\,1 in][]{rau11}. Moreover, PMS sources are most of the time in quiescence, and thus cannot account for a persistent overluminous X-ray emission as seen for \ci. Confusion with a background AGN or X-ray binary is also very unlikely as such sources have X-ray spectra that are very different from the spectrum of \ci. Hence the X-ray overluminosity of \ci\ most-likely arises within the \ci\ system, its stars and its multiple wind interaction regions.

Yet, we found that there is considerable variability on timescales of weeks or months, i.e.\ unrelated to the orbital periods. One possibility to explain the variability of the X-ray emission could be variations of the wind outflow of the components of the inner eclipsing binary. Such variations, if they exist, should also affect the equivalent width (EW) of the H$\alpha$ emission. We have thus measured the EWs on our TIGRE spectra after correcting for telluric absorptions by means of the {\sc iraf} software and using the \citet{hin00} template of telluric lines. We integrated the flux of the normalized spectra between 6519 and 6600\,\AA. For each date of observation, we also computed the orbital phase of the eclipsing binary using our quadratic ephemerides (Eq.\,\ref{ephem2}, Sect.\,\ref{opt}). Typical errors on the EWs are $\leq 1$\,\AA. The raw EWs display two maxima near phases 0.0 and 0.5, i.e.\ at the times of photometric minimum. This situation indicates that a significant fraction of the line emission arises from an extended region that is not directly concerned by the eclipses. We thus corrected the EWs for the variations of the continuum level using our empirical $V$-band light curve built in Sect.\,\ref{secphotom}. At a given epoch, most of the variations on timescales of the 6.6\,d orbital cycle are removed in this way (Fig.\,\ref{Halpha}). However, there is considerable scatter in this curve: EWs measured at different epochs, but at similar orbital phases differ by 2 - 4\,\AA, i.e.\ 12 - 25\% of the mean EW of H$\alpha$. The bottom panel of Fig.\,\ref{Halpha} illustrates the variations of the corrected EWs with epoch. Whilst this result does not prove that the X-ray variability is indeed due to variations of the mass-loss rate with time, it tells us that such variations of the mass-loss rate are very likely present in \ci. 
\begin{figure}[htb]
  \begin{center}
    \resizebox{8cm}{!}{\includegraphics{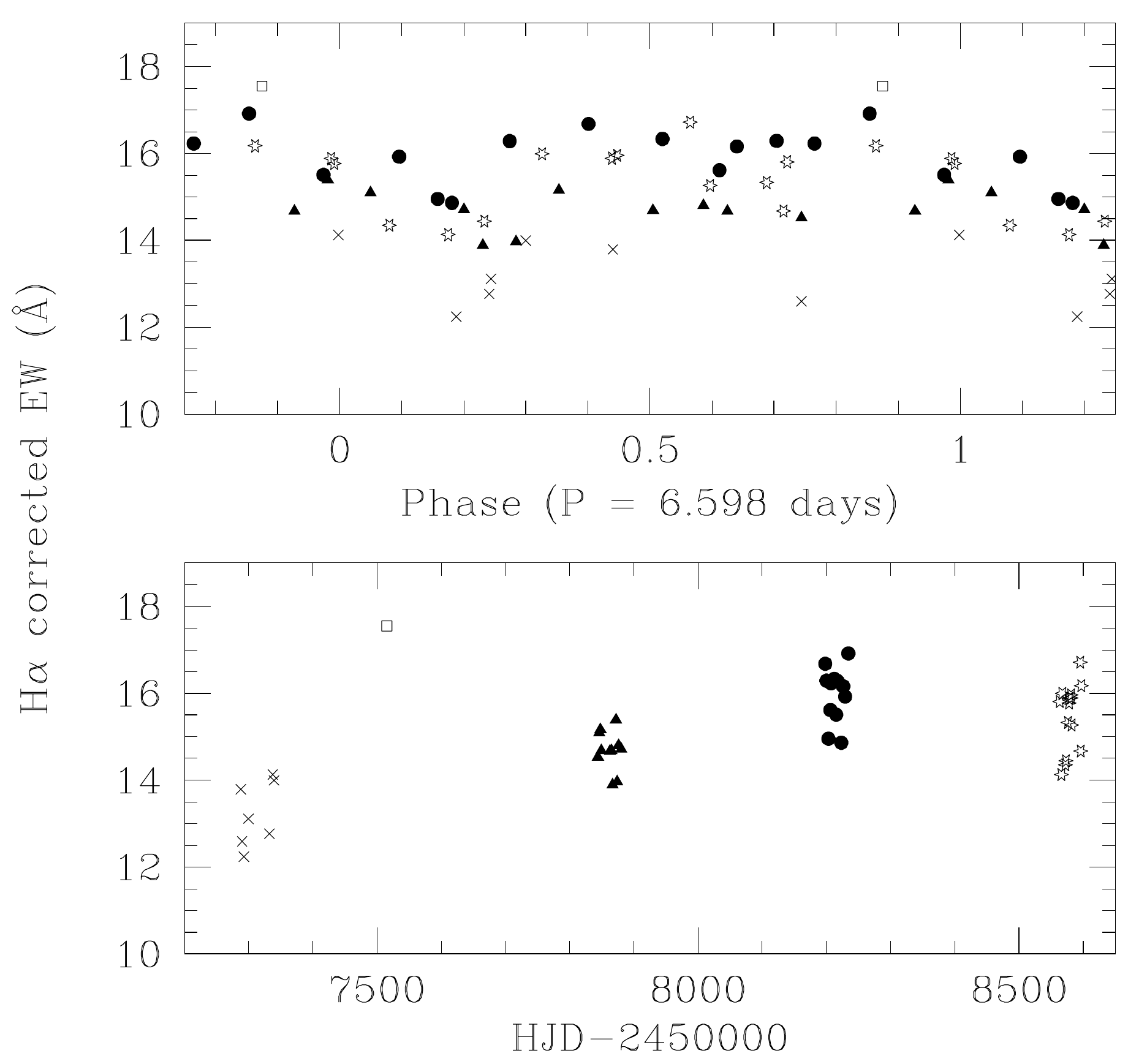}}
\end{center}  
  \caption{Equivalent width of the H$\alpha$ emission line measured on spectra collected with the HEROS spectrograph and corrected for the phase-locked variations of the continuum level. The data are shown as a function of orbital phase of the eclipsing binary (top panel, computed with the quadratic ephemerides, Eq.\,\ref{ephem2}) and as a function of date (bottom panel). Different symbols stand for different observing campaigns: crosses, open squares, filled triangles, filled circles and stars indicate data collected in 2015, 2016, 2017, 2018 and 2019, respectively. \label{Halpha}}
\end{figure}

\section{Long-term optical monitoring \label{opt}}
\subsection{Photometry \label{secphotom}}
The eclipses of the inner binary of \ci\ were first reported by \citet{mic53}. The primary eclipse (eclipse of the primary star by the secondary) is $\sim$0.3\,mag deep, which is slightly more than the secondary eclipse which has a depth of $\sim$0.25\,mag. The eclipses have been repeatedly observed since then and three types of information are available in the literature: (1) full datasets \citep[see also the Hipparcos, INTEGRAL Optical Monitoring Camera (OMC), and ASAS-SN databases\footnote{Available from Vizier catalogue I/239/hip\_va\_1, http://sdc.cab.inta-csic.es/omc/ and https://asas-sn.osu.edu/, respectively; the few data from the Northern Survey for Variable Stars  do not well cover the light curve.}]{mic53,hal74,lin09,yas14,kum17,lau17}; (2) times of primary or secondary eclipses \citep{saz61,hau64,rom69,kur85,hub07,zas17}; and (3) light-curve modelling, often considering quadratic ephemeris \citep{hal74,ls78,lin09,yas14,lau15,ant16}. 

\begin{table*}
\centering
\footnotesize
\caption{Revised times of primary photometric minimum resulting from the light-curve fitting}
\label{photom}
\begin{tabular}{lccccr}
\hline\hline
Source & N & HJD interval & $t_0$ & $E$ & $(O-C)_2$\\
\hline
\citet{mic53}   & 36 & 34216-34355 & 34218.443$\pm$0.020 &    0 &  $0.0229$ \\
\citet{hal74}-1 & 23 & 39751-39807 & 39753.959$\pm$0.036 &  839 & $-0.0840$ \\
\citet{hal74}-2 & 67 & 40376-40557 & 40380.833$\pm$0.010 &  934 & $-0.0116$ \\
\citet{hal74}-3 & 49 & 40711-40864 & 40717.364$\pm$0.048 &  985 &  $0.0262$ \\
\citet{kum17}-1 & 89 & 45591-45975 & 45593.251$\pm$0.049 & 1724 &  $0.0496$ \\
\citet{kum17}-2 & 63 & 46233-46325 & 46233.289$\pm$0.034 & 1821 &  $0.0877$ \\
\citet{kum17}-3 & 50 & 52082-52143 & 52085.631$\pm$0.059 & 2708 &  $0.0410$ \\
Hipparcos-1     & 62 & 47860-48443 & 47862.891$\pm$0.044 & 2068 & $-0.0017$ \\
Hipparcos-2     & 58 & 48500-49044 & 48502.907$\pm$0.027 & 2165 &  $0.0119$ \\
\citet{lin09}   & 320& 51047-51060 & 51049.720$\pm$0.017 & 2551 &  $0.0127$ \\
OMC-1           & 464& 53189-53216 & 53194.101$\pm$0.019 & 2876 &  $0.0515$ \\
OMC-2           & 90 & 53313-53331 & 53319.402$\pm$0.047 & 2895 & $-0.0092$ \\
OMC-3           & 214& 53685-54017 & 53688.860$\pm$0.039 & 2951 & $-0.0393$ \\
OMC-4           & 221& 54439-54609 & 54441.025$\pm$0.033 & 3065 & $-0.0450$ \\
OMC-5           & 147& 54771-54792 & 54777.588$\pm$0.048 & 3116 &  $0.0205$ \\
OMC-6           & 113& 54937-54974 & 54942.509$\pm$0.034 & 3141 & $-0.0084$ \\
OMC-7           & 289& 55501-55543 & 55503.370$\pm$0.026 & 3226 &  $0.0230$ \\
OMC-8           & 59 & 57193-57362 & 57199.058$\pm$0.033 & 3483 &  $0.0255$ \\
\citet{yas14}-1 & 165& 55440-55498 & 55443.987$\pm$0.036 & 3217 &  $0.0230$ \\
\citet{yas14}-2 & 180& 55722-55759 & 55727.730$\pm$0.082 & 3260 &  $0.0521$ \\
\citet{yas14}-3 & 77 & 55792-55819 & 55793.672$\pm$0.023 & 3270 &  $0.0131$ \\
\citet{lau17}-1 & 130& 55630-55775 & 55635.335$\pm$0.016 & 3246 &  $0.0281$ \\
\citet{lau17}-2 & 92 & 55780-55927 & 55780.484$\pm$0.036 & 3268 &  $0.0221$ \\
\citet{lau17}-3 & 151& 55990-56151 & 55991.595$\pm$0.010 & 3300 & $-0.0028$ \\
\citet{lau17}-4 & 157& 56177-56297 & 56182.925$\pm$0.031 & 3329 & $-0.0159$ \\
\citet{lau17}-5 & 149& 56375-56487 & 56380.860$\pm$0.018 & 3359 & $-0.0209$ \\
\citet{lau17}-6 & 156& 56503-56626 & 56506.209$\pm$0.007 & 3378 & $-0.0340$ \\
ASAS-SN-1       & 151& 57102-57353 & 57106.717$\pm$0.039 & 3469 &  $0.0556$ \\
ASAS-SN-2       & 159& 57463-57722 & 57469.585$\pm$0.014 & 3524 &  $0.0342$ \\
ASAS-SN-3       & 166& 57831-58080 & 57832.454$\pm$0.069 & 3579 &  $0.0126$ \\
ASAS-SN-4       & 130& 58207-58431 & 58208.532$\pm$0.029 & 3636 &  $0.0029$ \\
this paper-1    & 566& 58262-58333 & 58267.927$\pm$0.010 & 3645 &  $0.0168$ \\
this paper-2    & 207& 58358-58426 & 58360.290$\pm$0.043 & 3659 &  $0.0076$ \\
\hline
\end{tabular}
\tablefoot{The second column provides the number of points in the dataset. The HJDs are given as HJD-2\,400\,000. Column four quotes the times of primary minimum $t_0$ along with their estimated uncertainties. Columns five and six yield the epoch (i.e.\ the binary cycle) since the first $t_0$ and the $(O-C)_2$ residual evaluated with respect to the quadratic ephemerides.}
\end{table*}

The presence of Star C, a gravitationally bound companion to the inner, eclipsing binary should lead to light-time effects on the times of its photometric minima. These effects can be studied by comparing the observed times of primary minimum to those computed from the ephemeris of the eclipsing binary. To this aim, precise eclipse timing is needed. Unfortunately, discrepancies exist among the reported values. For example, \citet{mic53} provided a primary eclipse time in Julian day of 2\,434\,265.73 whereas \citet{hal74} quoted Miczaika's time as 2\,434\,218.463; \citet{kur85} asserted that the eclipse times given by \citet{mic53} and \citet{saz61} are wrong while those of \citet{rom69} deviate a lot from the ephemeris, suggesting they are all inaccurate. \citet{yas14} also excluded some of the \citet{rom69} times, although without justifying their decision to do so.

\begin{figure*}
  \begin{minipage}{8cm}
    \includegraphics[width=8cm]{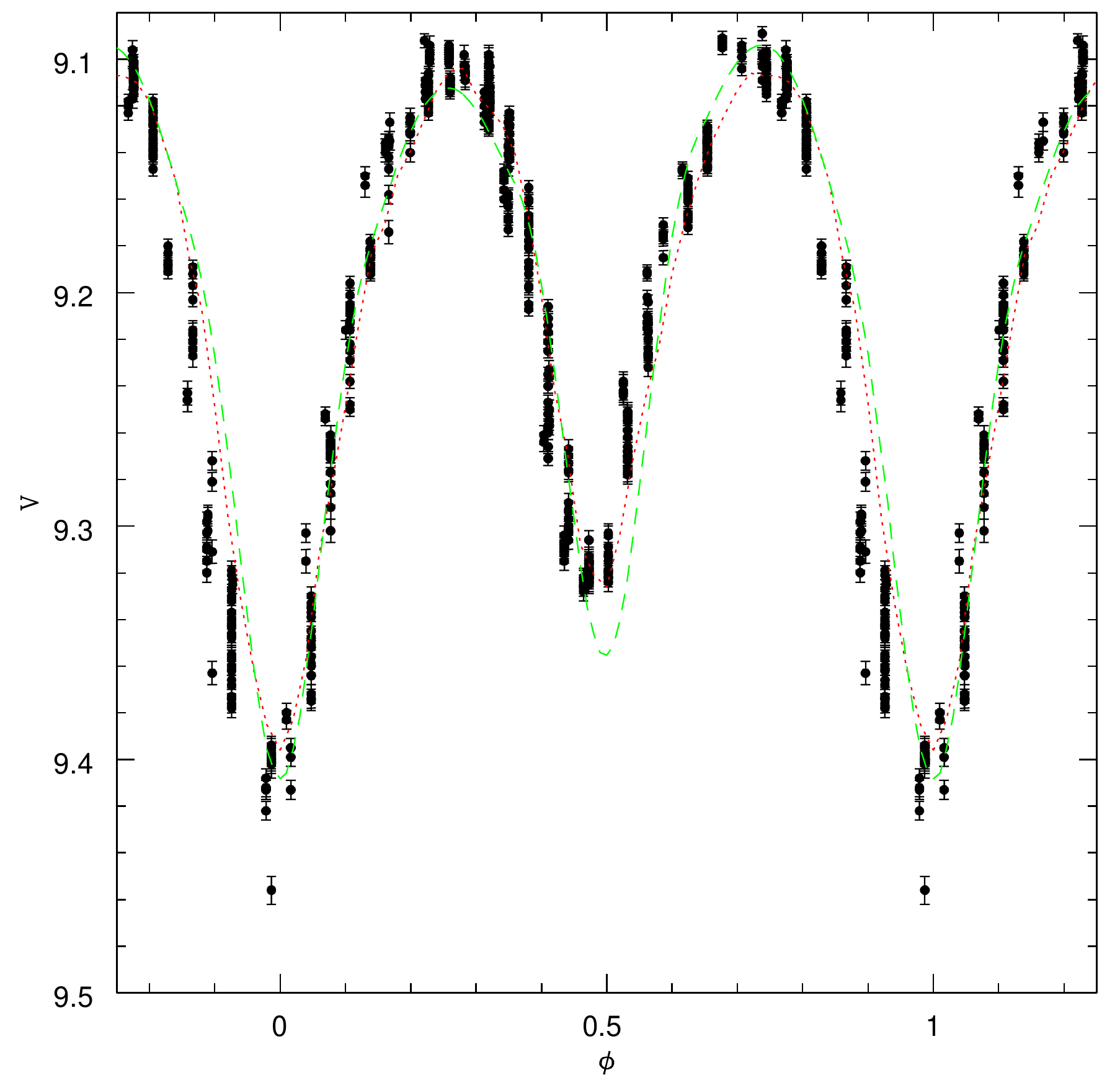}
  \end{minipage}
  \hfill
  \begin{minipage}{8cm}    
    \includegraphics[width=8cm]{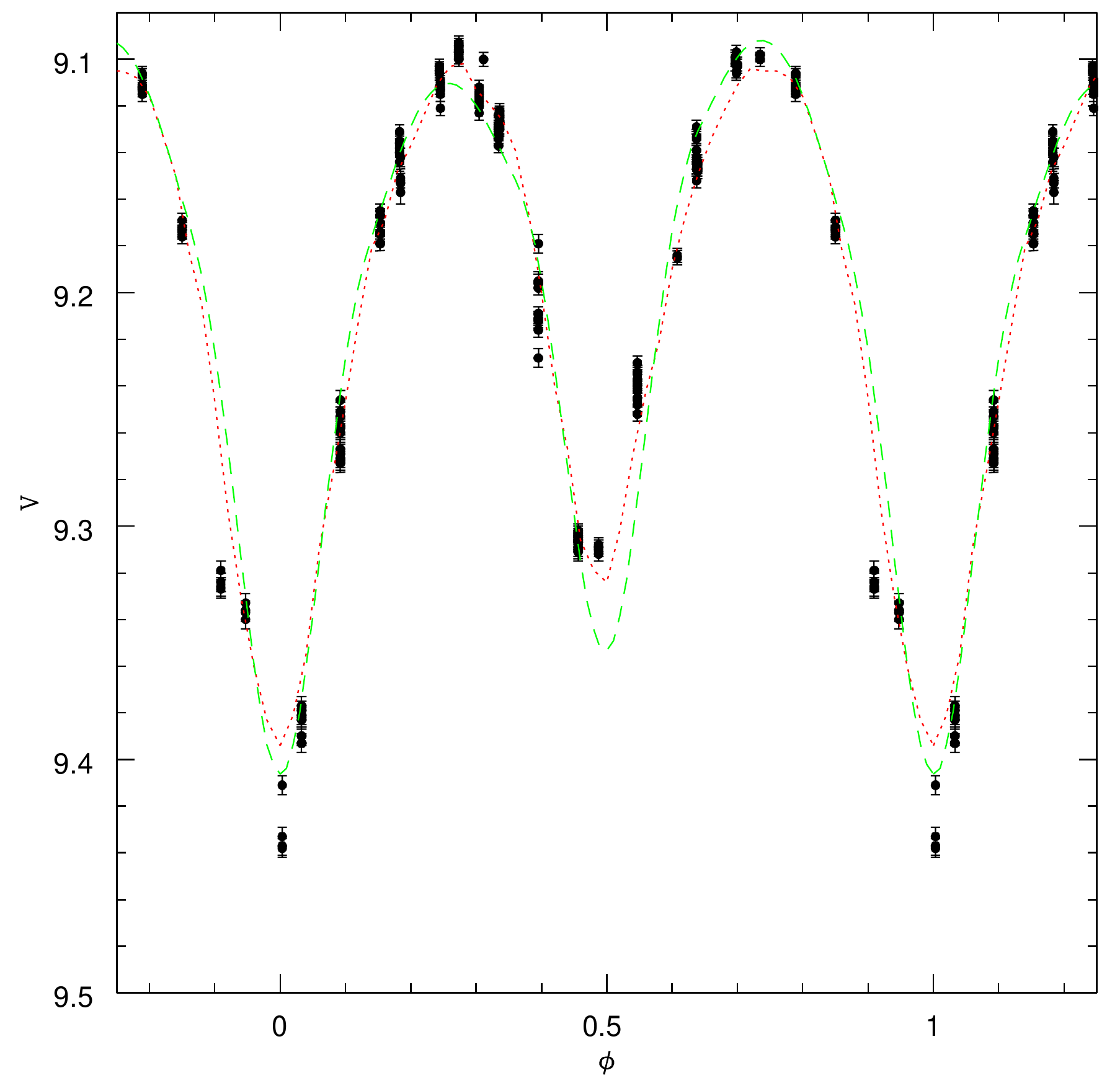}
  \end{minipage}
\caption{Our new photometry folded with $P=6.59787$\,d and the best-fit eclipse times of Table \ref{photom}. The green dashed curve corresponds to the theoretical light curve of \citet{lin09}, whereas the red dotted curve corresponds to the empirical median light curve. The left and right panels correspond to the epochs (HJD-2\,400\,000) 58\,262 -- 58\,333 and 58\,358 -- 58\,426, respectively. }
\label{folded}
\end{figure*}

Given these problems, we rederived the primary eclipse times $t_0$ and their uncertainties in a homogeneous way for all available data\footnote{Unfortunately many publications only provide the value of $t_0$ and do not quote the original photometric data on which the analysis was based. For these references it is impossible to establish an estimate of the uncertainty on $t_0$.}. Long datasets were cut into smaller chunks, checking that the light-curve shape was indeed well sampled by individual subsets. Table \ref{photom} provides the considered time intervals and number of data points. No cleaning of the data was performed, except for the OMC datasets where some clear outliers exist\footnote{Points that were kept have $V<9.45$ for all phases and $V<9.35$ in $\phi=0.1-0.2$, $V<9.2$ in $\phi=0.2-0.4$, $V<9.35$ in $\phi=0.4-0.65$, $V<9.2$ in $\phi=0.65-0.7$, $V<9.15$ in $\phi=0.7-0.85$, $V<9.2$ in $\phi=0.85-0.9$.}.

Using a $\chi^2$ criterion, we then searched for the primary eclipse times and magnitude offsets minimizing the deviations between the (folded) individual data points and the theoretical light curve of \citet{lin09}. Our method is similar to the semi-automatic fitting procedure (AFP) used by \citet{zas14}. The steps in time and offsets were $2\times10^{-4}$\,d and 0.001\,mag, respectively. Fitting offsets is necessary because some sources provide the magnitude of the star, whilst others provide its relative magnitude with respect to comparison stars. Moreover, the datasets use different filters that are not always perfectly calibrated to the standard $V$ filter. Owing to the strong reddening, the observed spectral energy distribution of \ci\ is quite steep and the calibration of the magnitude outside eclipse is thus quite sensitive to the passband of the filter that was used. Yet, whilst this affects the zero point of the light curve, it does not impact its shape. \citet{lin09} indeed showed that light curves collected with narrow-band filters with central wavelengths ranging between 4686\,\AA\ and 6051\,\AA\ could all be fitted with the same synthetic light curve (see their Figs.\,1 and 2). Hence, whilst different $V$-band filters have different zero points, the morphology of the light curve is identical, allowing us to perform the correlation with a single template. We note that individual errors on data points are available in all but two cases: from the standard deviations of points taken the same night, we found that typical errors are $\sim$0.010\,mag and 0.020\,mag for data from \citet{hal74} and \citet{kum17}, respectively, and we used these values for the $\chi^2$ calculation.

Since the orbital period of the inner binary may vary with time, we performed this $\chi^2$ minimization considering periods ranging between 6.59777 and 6.59820\,d, which bracket literature values, by steps of $10^{-5}$\,d. We found that the minimum $\chi^2$ values do not depend on the period. In fact, the individual datasets are usually too short for such minute variation of the period to become detectable. However, this procedure allows us to assess how sensitive the best-fit parameters are and the resulting range of $t_0$ values provides a first estimate of the uncertainty on $t_0$.

As a second step, we derived an empirical photometric light curve consisting of normal points obtained by phase-folding all datasets using the best-fit individual $t_0$ times and offsets and by performing a median in 50 phase bins. We then repeated the $\chi^2$ minimization procedure using this light curve as comparison. As could be expected, $\chi^2$ values were smaller than in previous cases and the best-fit $t_0$/offset values were slightly different. Shifting the theoretical or empirical light curves also revealed in some cases the presence of several local minima in eclipse times. As an example of the adjustment of the phase-folded light curves we show the fit to the two epochs of new photometric measurements in Fig.\,\ref{folded}.

To quantify the uncertainty on $t_0$ we performed Monte Carlo simulations to account for the intrinsic variability. To do so, we proceeded in the following way. For each dataset, we first computed the residuals of the observed light curve over the \citet{lin09} model taken as a template. We then computed the Fourier periodogram of these residuals using the method of \citet{hmm} and \citet{gos01}. For the highest quality datasets this periodogram revealed power that decreased as a function of frequency, although there were no dominant peaks. This situation is reminiscent of the red noise that was found in various massive stars, including the massive binary HD~149\,404 \citep{Rauw19}. For lower quality data, the periodogram was typically dominated by white noise.
We then used the characteristics of the Fourier periodogram as input to our Monte Carlo code \citep{Rauw19} to simulate 5000 synthetic light curves with the same temporal sampling as the actual data. These artificial light curves were again fitted with the \citet{lin09} model template and the resulting $t_0$ and photometric offsets were used to build the distributions of these parameters. In most cases, we find a nearly normal distribution for $t_0$ resulting in $\sigma$ values of about 0.01 to 0.02\,d. In several cases however the distributions are asymmetric (e.g.\ strongly peaked with a wing extending mostly to one side) or display several peaks (due to local minima of the $\chi^2$). In those cases, the errors were enlarged so that the $\pm1\sigma$ interval encompasses the two neighbouring minima; the resulting $\sigma$ values are significantly larger and in the worst case up to 0.08\,d. 

All these tests were done to get a realistic value of the errors. Indeed, the light curves present some scatter around the eclipse variations that exceed the photometric errors on individual data points. This comes from the presence of intrinsic variability on top of the light curve of the eclipsing system \citep{lin09}, which is a major contributor to the uncertainties on the determination of the times of primary minimum. Hence the $\chi^2$ for the best-fit $t_0$ and offset values are larger than expected. Using the usual $\chi^2+1$ method to derive $1\sigma$ errors therefore leads to underestimates. Comparing the scatter in eclipse times derived by shifting the two comparison curves for a set of periods as well as from the Monte Carlo simulations provides a more secure estimate, albeit with larger error values. To be as conservative as possible, we decided to adopt as the best estimate of the error on $t_0$ the maximum of the uncertainty estimates between half of the full range of $t_0$ values derived by the shifting method and the $\sigma$ obtained in the Monte Carlo simulations. Table \ref{photom} yields the final, average values of the eclipse times, along with their errors.

\begin{figure}[htb]
  \begin{center}
  \resizebox{8cm}{!}{\includegraphics{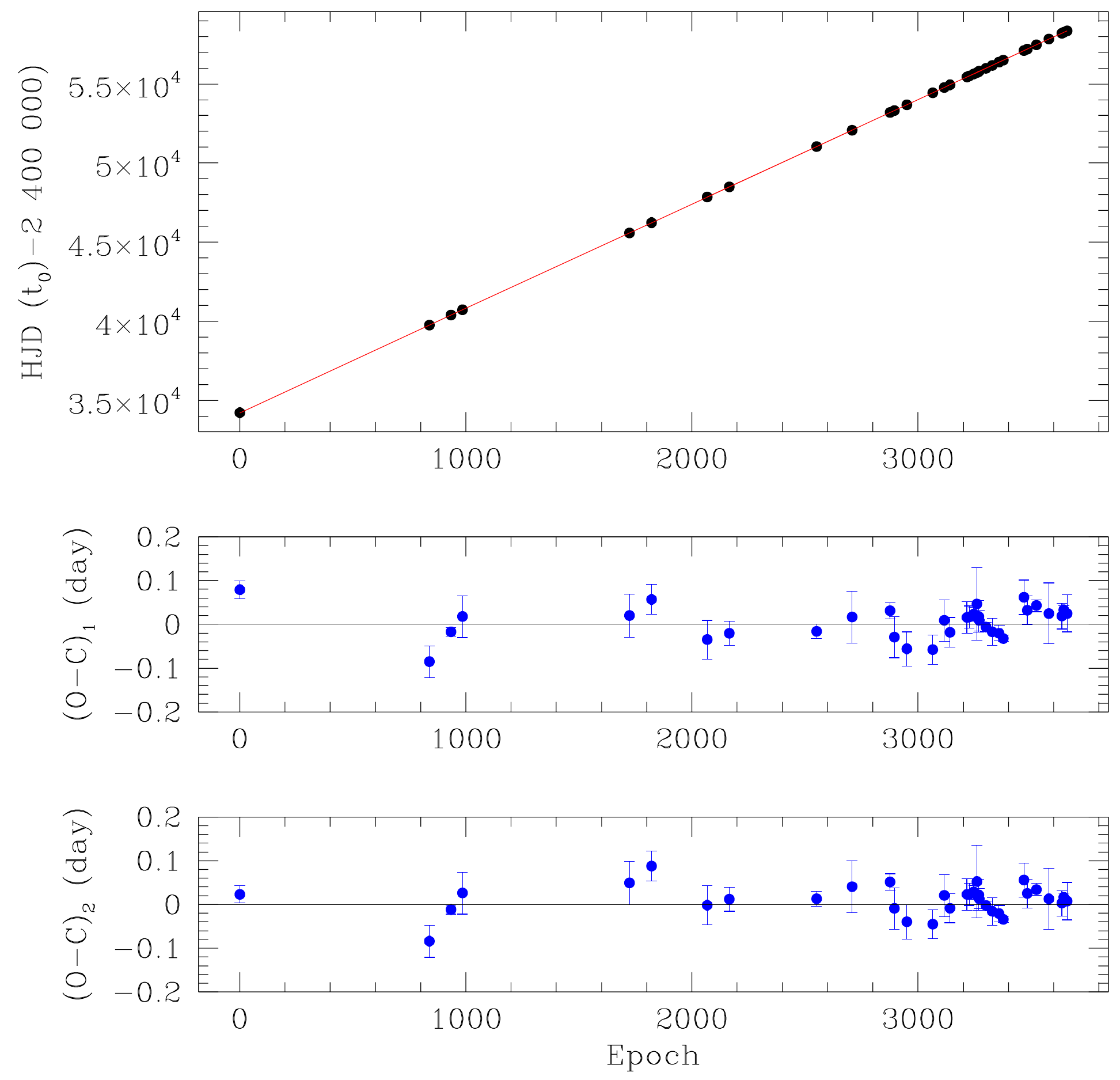}}
\end{center}  
\caption{{\it Top panel}: observed times of primary minimum as a function of epoch, along with the best-fit quadratic ephemerides (red line). {\it Middle and bottom panels}: $O-C$ of the times of primary minimum as a function of epoch $E$ for the linear ephemerides (middle panel) and the quadratic ephemerides (bottom panel). \label{OCt0}}
\end{figure}
With these values at hand, we then computed the best-fit ephemerides.
The best-fit linear ephemerides are written as
\begin{eqnarray}
{\rm HJD}(t_0) & = & {\rm HJD}_0 + P_0\,E \nonumber \\ 
& = & (2\,434\,218.3638 \pm 0.0011) \nonumber \\
& + & (6.597866 \pm 0.000001)\,E
\label{ephem1}
.\end{eqnarray}
We computed the residuals, labelled $(O-C)_1$ in Fig.\,\ref{OCt0}, of the $t_0$ values. The Fourier periodogram of the $(O-C)_1$ residuals indicates significant power at low frequencies. The highest peak corresponds to a timescale of about 33\,yr which is half the total duration spanned by the data. We concluded that there exist long-term trends on top of the linear ephemerides, most probably associated to the mass loss of the inner binary system \citep{SinghChaubey}. To account for these long-term trends we thus considered quadratic ephemerides.

\begin{figure}[htb]
  \begin{center}
  \resizebox{8cm}{!}{\includegraphics{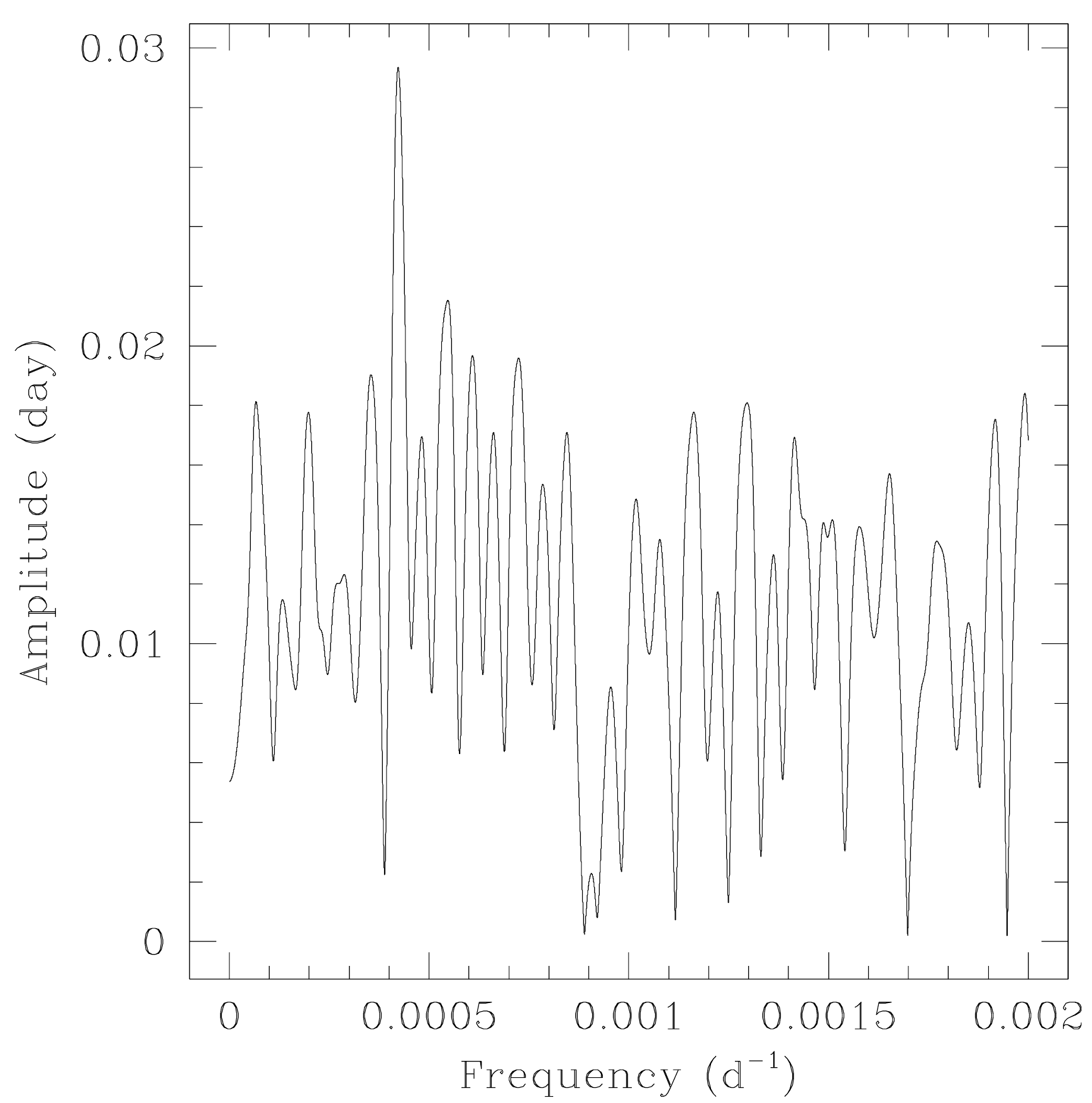}}
\end{center}  
\caption{Fourier periodogram of the $(O-C)_2$ values.\label{spOC2}}
\end{figure}

The best-fit quadratic ephemerides are written as
\begin{eqnarray}
{\rm HJD}(t_0) & = & {\rm HJD}_0 + P_0\,E + \frac{1}{2}\,\dot{P}\,P_0\,E^2 \nonumber \\ 
& = & (2\,434\,218.4196 \pm 0.0032)\nonumber \\
& + & (6.597866 \pm 0.000009)\,E \nonumber \\
& + & (2.03 \pm 0.18) \times 10^{-8}\,E^2
\label{ephem2}
.\end{eqnarray}
The quadratic ephemerides lead to a reduction of the $\chi^2$ of the fit from 89.2 (31 degrees of freedom) for Eq.\,\ref{ephem1} to 69.6 (30 degrees of freedom) for Eq.\,\ref{ephem2}.

Compared to the quadratic ephemerides of \citet{lin09} and \citet{lau15}, we find a value of $\dot{P} = (0.615 \pm 0.055)\,10^{-8}$\,s\,s$^{-1} = (0.19 \pm 0.02)$\,s\,yr$^{-1}$, which is about one-third of the old value. The same reduction by a factor $\sim 1/3$ applies to the mass-loss rate \citep[computed according to case IV of][]{SinghChaubey} which would now be $(7.7 \pm 2.1)\,10^{-6}$\,M$_{\odot}$\,yr$^{-1}$. The lower value of $\dot{P}$ is due to the much longer time series considered in this work, hence leading to a much better constrained value of $\dot{P}$, and to the fact that we rederived the values of $t_0$ in a self-consistent way rather than relying on published values. We note that this value of $\dot{M}$ is lower than that derived by \citet{ken10} from the thermal part of the radio emission, which was obtained assuming a larger distance than found by {\it Gaia}. The scaling of the result of \citet{ken10} is written as
\begin{equation}
  \dot{M}_{\rm radio} = \frac{2.8\,10^{-5}\,{\rm M_{\odot}\,yr^{-1}}}{\sqrt{F}}\left(\frac{v_{\infty}}{1500\,{\rm km\,s^{-1}}}\right)\,\left(\frac{d}{1.5\,{\rm kpc}}\right)^{3/2}
,\end{equation}
where $F \leq 1$ stands for the volume filling factor. For reasonable values of $F$ and $v_{\infty}$, the radio mass-loss rate appears thus larger than the optical value. However, in this comparison, we need to keep in mind that the primary radio component of \ci\ contains two wind interaction zones (between the winds of stars A and B and between the combined wind of stars A+B and the wind of Star C). Aside from the non-thermal radio emission that is seen to vary, these wind interaction zones can also contribute an extra thermal radio emission that could bias the determination of $\dot{M}$ from the radio flux towards higher values \citep{Pittard2010}.

Figure\,\ref{OCt0} shows the residuals, labelled $(O-C)_2$, of the $t_0$ values over the quadratic ephemerides; we then computed their Fourier periodogram (Fig.\,\ref{spOC2}). For frequencies below 0.002\,d$^{-1}$, the highest peak is now present at a frequency\footnote{Inspecting the periodogram at higher frequencies, we noted several peaks of similar strength. However, they correspond to timescales of less than 500\,d, which are actually not well sampled by our $(O-C)_2$ data.} of $(0.0004227 \pm 0.0000041)$\,d$^{-1}$, which corresponds to a period of $(2366 \pm 23)$\,d or 6.48\,yr. We note that this frequency is close to that of the second highest peak (at a frequency near 0.00038\,d$^{-1}$) in the periodogram of the $(O-C)_1$ data. This periodicity is interpreted in terms of reflex motion in Sect.\,\ref{outerorbit}. Finally, for completeness, Appendix\,\ref{AppendixD} presents a detailed analysis of the combined light curve of the eclipsing binary. 

\begin{figure*}[htb]
  \begin{center}
    \resizebox{15cm}{!}{\includegraphics{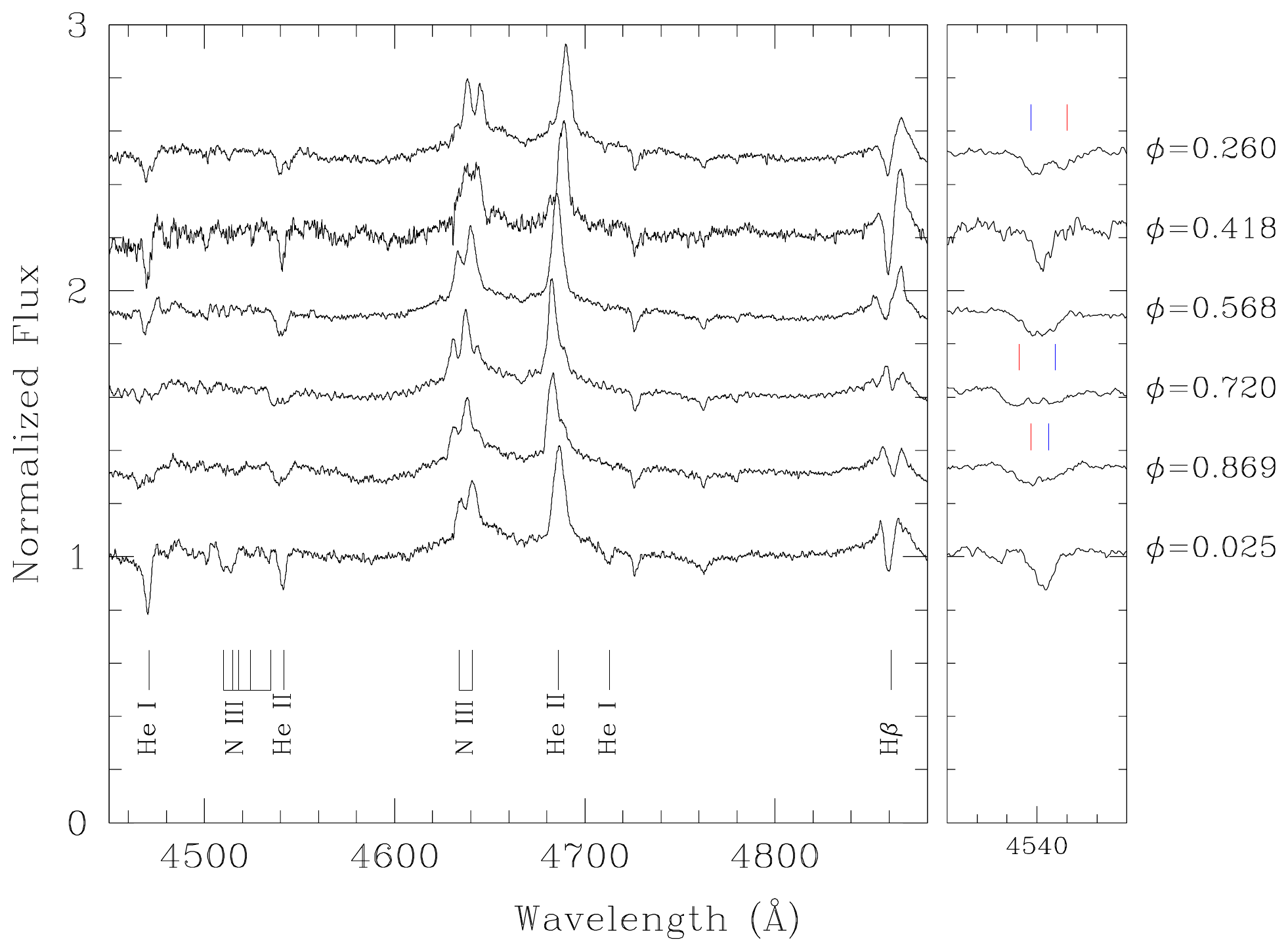}}
\end{center}  
  \caption{Blue spectrum of \ci\ as a function of orbital phase as observed with the Aur\'elie spectrograph during our June 2014 OHP campaign. For clarity the consecutive spectra are shifted vertically by 0.3 continuum units. The right panel provides a zoom on the He\,{\sc ii} $\lambda$\,4542 line. The blue and red tick marks above the spectra indicate the position of the primary and secondary absorptions, respectively, for those spectra where the lines were resolved. \label{montage}}
\end{figure*}
\subsection{Spectroscopy}
Figure\,\ref{montage} illustrates the phase dependence of the blue spectrum of \ci\ during our OHP campaign of 2014. Although the spectra clearly reveal the signature of the orbital motion of the eclipsing binary, assessing the nature and the properties of the components of this eclipsing binary is notoriously difficult. Indeed, as already noted by previous studies \citep[e.g.][]{rau99}, the visibility and nature of the absorption lines in the spectrum of \ci\ change considerably with orbital phase. For instance, the H$\gamma$ and He\,{\sc i} $\lambda$\,4471 lines display a P-Cygni type profile at phases around 0.35 -- 0.55 and near phase 0.0. Whilst the orbital motion of the primary can clearly be traced by the RVs of the He\,{\sc ii} $\lambda\lambda$\,4542, 5412 and O\,{\sc iii} $\lambda$\,5592 lines, measuring the RVs of the secondary is more difficult because of the appearance of P-Cygni absorption troughs at some orbital phases. 

The spectrum also features a number of emission lines. Some of these emissions (H$\alpha$, He\,{\sc i} $\lambda$\,5876) arise mostly from the wind-wind collision zone \citep{rau99}. Others, such as He\,{\sc ii} $\lambda$\,4686, N\,{\sc iii} $\lambda\lambda$\,4634-40, Si\,{\sc iv} $\lambda\lambda$\,4089, 4116, 6668, 6701 and S\,{\sc iv} $\lambda\lambda$\,4486, 4504 closely follow the orbital motion of the secondary, although with a phase shift probably due to the contribution of a wind interaction region \citep{rau99}. On the contrary, the C\,{\sc iii} $\lambda$\,5696 emission follows the orbital motion of the primary.

Table\,\ref{journaloptical} lists our RV measurements of various absorption lines and of the peak of the He\,{\sc ii} $\lambda$\,4686 emission. In Appendix\,\ref{AppendixC}, we use these new RVs of the absorption lines to revise the orbital solution of the eclipsing binary. Unfortunately though, the RVs of the absorption lines are clearly not appropriate to detect a possible reflex motion of the eclipsing binary. Instead, we measured the RVs of the peak of the He\,{\sc ii} $\lambda$\,4686 emission line. These RVs describe a sine wave that is slightly shifted in phase with respect to the RV curve of the secondary \citep[see][]{rau99}. This feature is most likely related to the wind-wind interaction in the eclipsing binary and turned out to be remarkably stable over more than two decades.

The red dots in Fig.\,\ref{RVsHeII4686} show the measured heliocentric velocities from \citet{rau99} along with our new data folded with the quadratic ephemerides. The data were taken over 14 observing seasons (see Table\,\ref{obsseason}). For each season, we adjusted an S-wave relation 
\begin{equation}
  v(\phi) = -v_x\,\cos{(2\,\pi\,\phi)}+v_y\,\sin{(2\,\pi\,\phi)}+v_z
,\end{equation}
where the amplitude $\sqrt{v_x^2+v_y^2}$ was requested to be the same at all epochs. In this way, we thus obtained the values of the systemic velocity $v_z$ as a function of epoch (Table\,\ref{obsseason}). We tested various values of the sine-wave amplitudes between 229 and 239\,km\,s$^{-1}$ and found stable values of the epoch-dependent $v_z$ velocities. The black dots in Fig.\,\ref{RVsHeII4686} indicate the RVs of the peak of He\,{\sc ii} $\lambda$\,4686 after correcting for the variations of $v_z$ with epoch. Whilst the raw RVs have $|O-C| = 16.9$\,km\,s$^{-1}$ about the best-fit curve, the RVs corrected for the epoch dependence of $v_z$ have $|O-C| = 13.6$\,km\,s$^{-1}$. This may at first sight seem a small difference, but we show in next section its significance for the study of reflex motion.

\begin{figure}[htb]
  \begin{center}
    \resizebox{8cm}{!}{\includegraphics{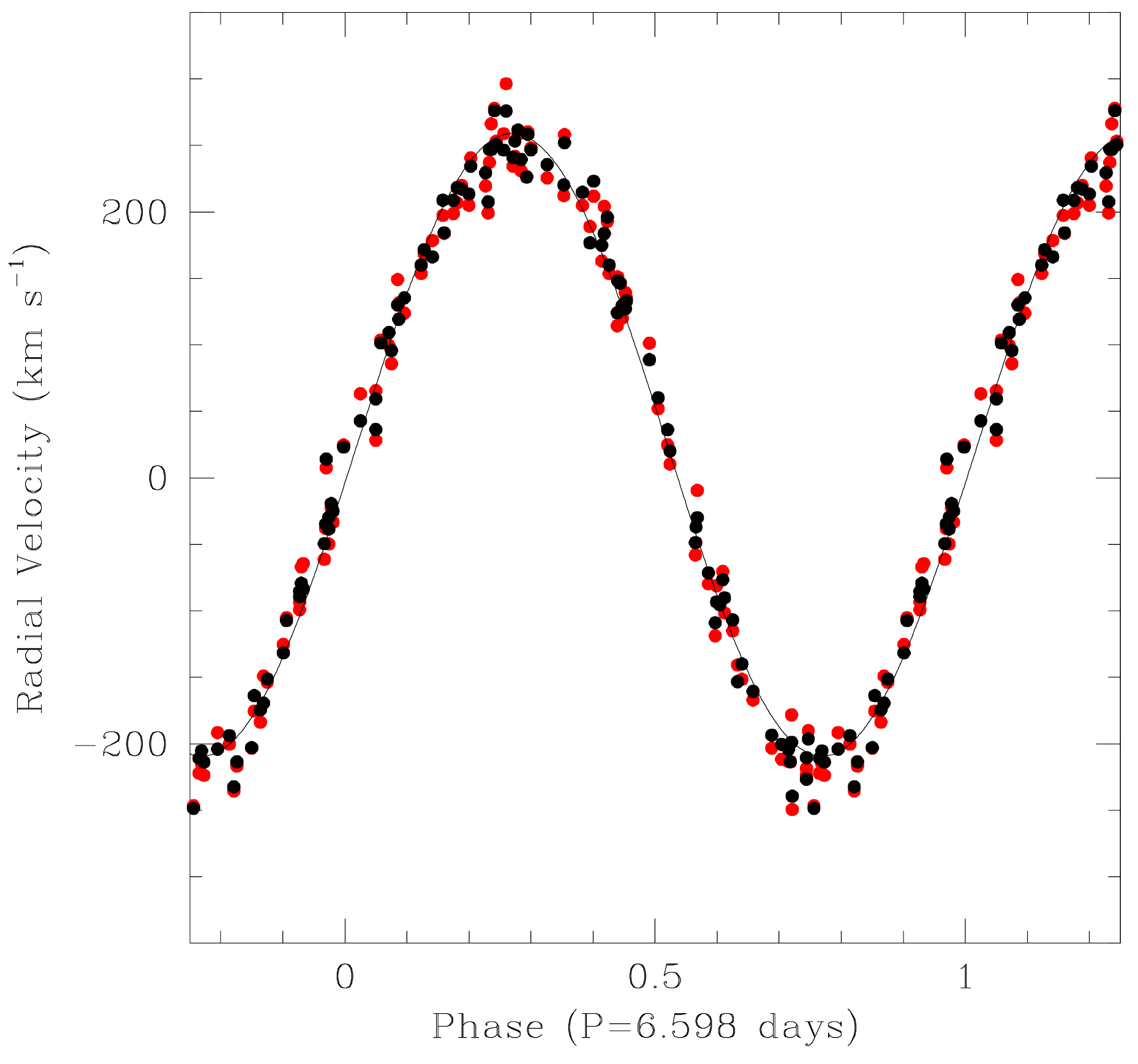}}
\end{center}  
  \caption{Radial velocities of the peak of the He\,{\sc ii} $\lambda$\,4686 emission line as a function of phase of the eclipsing binary. The orbital phases are computed with the quadratic ephemerides. The red dots yield the raw heliocentric RVs, whereas the black dots were corrected for the epoch-dependence of the systemic velocity. The best-fit S-wave relation (corresponding to $v_x = 28.8$\,km\,s$^{-1}$, $v_y = 232.3$\,km\,s$^{-1}$ and $v_z = 24.8$\,km\,s$^{-1}$) is shown overplotted on the data.\label{RVsHeII4686}}
\end{figure}

\begin{table}
  \begin{center}
  \caption{Epoch-dependence of the systemic velocity of He\,{\sc ii} $\lambda$\,4686 \label{obsseason}}
  \begin{tabular}{c r c}
    \hline
    Date & \multicolumn{1}{c}{$v_z$} & N \\
    (HJD-2\,400\,000) & \multicolumn{1}{c}{km\,s$^{-1}$} & \\
    \hline
49552.5273 & $33.4  \pm 10.0$ &  3 \\
49915.3715 & $34.3  \pm  5.0$ & 10 \\
50310.0288 & $ 8.5  \pm  7.5$ &  5 \\
50640.9228 & $23.1  \pm  7.5$ &  6 \\
56457.9372 & $30.2  \pm  7.5$ &  6 \\ 
56813.9965 & $54.4  \pm  7.5$ &  6 \\
57179.0896 & $40.9  \pm  7.5$ &  6 \\
57311.3448 & $29.1  \pm  7.5$ &  7 \\
57545.2619 & $28.5  \pm  5.0$ &  9 \\
57862.1242 & $ 9.3  \pm  5.0$ & 11 \\
58004.6100 & $16.0  \pm  5.0$ &  8 \\
58214.1248 & $18.0  \pm  5.0$ & 12 \\
58355.7685 & $16.3  \pm 10.0$ &  3 \\
58579.6787 &  $3.6  \pm  5.0$ & 11 \\
\hline
  \end{tabular}
  \tablefoot{The last column indicates the number of available RV data points per epoch. The first four dates correspond to data presented in \citet{rau99}, while the remaining entries are found from the analysis of new data presented in Table\,\ref{journaloptical}.}
  \end{center}
\end{table}

\subsection{Orbital signatures of Star C \label{outerorbit}}
Two observational pieces of evidence point towards the presence of reflex motion in the A+B system in \ci\ due to the presence of Star C: regularly changing eclipse times and changes in systemic velocities. Figure\,\ref{combinedfit} shows the $(O-C)_2$ data folded with the period of 2366\,d. The modulation displays a peak-to-peak amplitude around 0.1\,d. Provided we are dealing with a hierarchical triple system we can express the light-time effect as
\begin{equation}
  (O-C)_2 = \frac{a_{\rm AB}\,\sin{i_{\rm AB+C}}\,(1 - e_{\rm AB+C}^2)}{c\,(1 + e_{\rm AB+C}\,\cos{\phi_{\rm AB+C}(t)})}\,\sin{(\phi_{\rm AB+C}(t) + \omega_{\rm AB})}
  \label{OC2}
,\end{equation}
where $a_{\rm AB}$, $e_{\rm AB+C}$, $i_{\rm AB+C}$, $\phi_{\rm AB+C}(t)$ and $\omega_{\rm AB}$ indicate, respectively, the semi-major axis of the orbit of the A+B inner binary around the centre of mass of the A+B+C triple system, the eccentricity of this outer orbit, its inclination, the true anomalie at time $t$ of the inner binary on the outer orbit, and the argument of periastron measured from the ascending node of the inner binary on the outer orbit.   

We then folded the $v_z$ values with the 2366\,d period (Fig.\,\ref{combinedfit}). Assuming they reflect an SB1 orbital motion, we can express them via 
\begin{eqnarray}
  v_z(t) & = & v_{z,0} + \frac{a_{\rm AB}\,\sin{i_{\rm AB+C}}}{\sqrt{1 - e_{\rm AB+C}^2}}\,\left(\frac{2\,\pi}{P_{\rm AB+C}}\right)\,\left[\cos{(\phi_{\rm AB+C}(t) + \omega_{\rm AB})}\right. \nonumber \\
   &   & \left. + e_{\rm AB+C}\,\cos{\omega_{\rm AB}}\right]
  \label{vz}
.\end{eqnarray}
Combining Eqs.\,\ref{OC2} and \ref{vz}, the orbital motion of the inner binary around centre of mass of the triple system can be described by five parameters: $a_{\rm AB}\,\sin{i_{\rm AB+C}}$, $e_{\rm AB+C}$, $\omega_{\rm AB}$, $v_{z,0}$ and $t_{\rm AB+C,0}$. The last of these parameters stands for the time of periastron passage of the outer orbit which is used along with $P_{\rm AB+C} = 2366$\,d in Kepler's equation to compute the true anomaly $\phi_{\rm AB+C}(t)$.
We searched for the combination of these five parameters that provides the best simultaneous fit to Eqs.\,\ref{OC2} and \ref{vz}. The best fit is illustrated in Fig.\,\ref{combinedfit}, and the $1\,\sigma$ and 90\% confidence contours projected onto five parameter planes are shown in Fig.\,\ref{chiplanes}.  
\begin{figure}[htb]
  \begin{center}
    \resizebox{8cm}{!}{\includegraphics{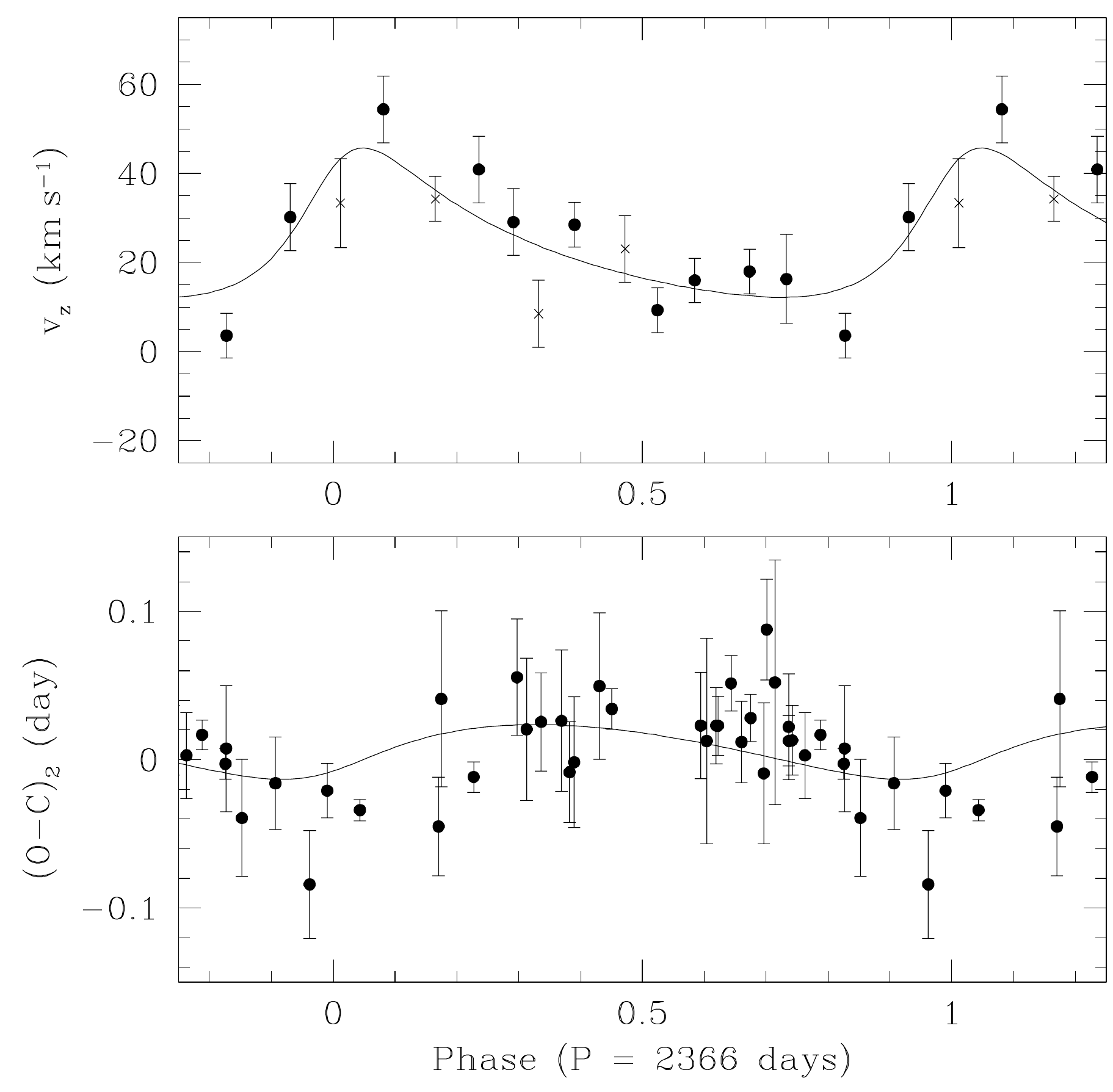}}
\end{center}  
  \caption{{\it Top}: $v_z$ velocity of the peak of the He\,{\sc ii} $\lambda$\,4686 emission line as a function of epoch folded with the 2366\,d period. {\it Bottom}: $(O-C)_2$ of the times of primary minimum folded with the 2366\,d period.\label{combinedfit}}
\end{figure}

\begin{figure*}[htb]
  \begin{center}
    \resizebox{6cm}{!}{\includegraphics{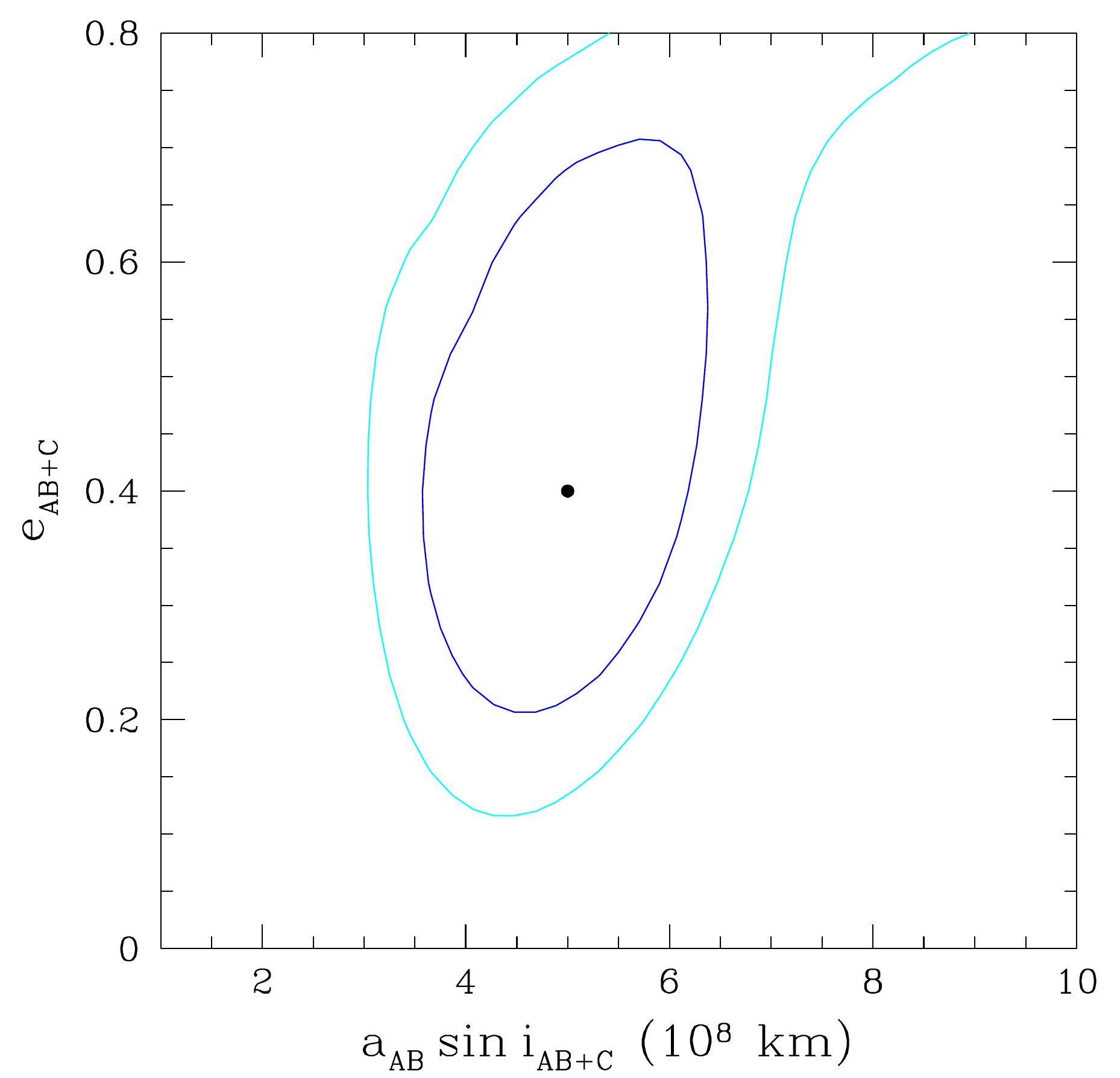}}
    \resizebox{6cm}{!}{\includegraphics{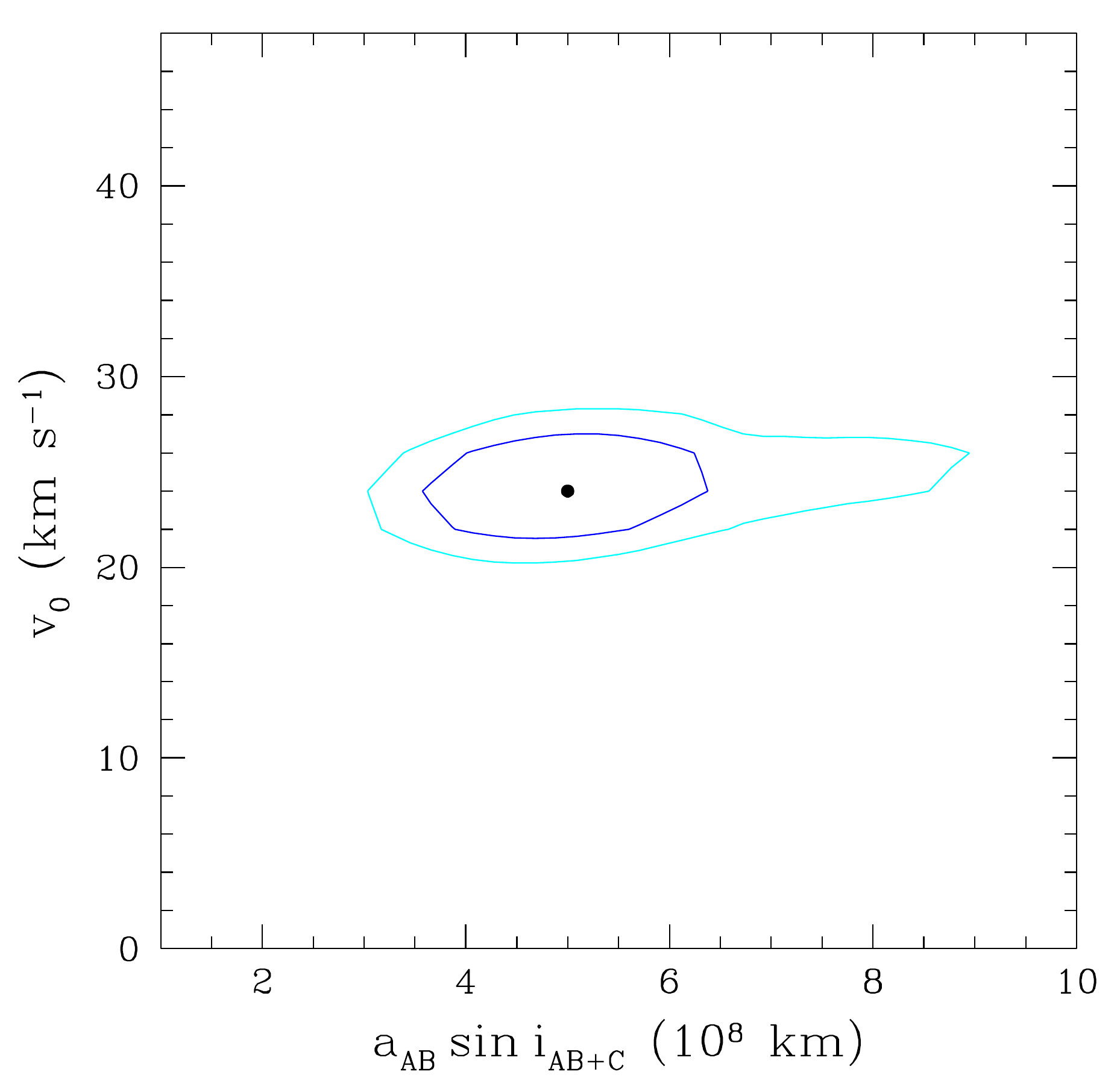}}
    \resizebox{6cm}{!}{\includegraphics{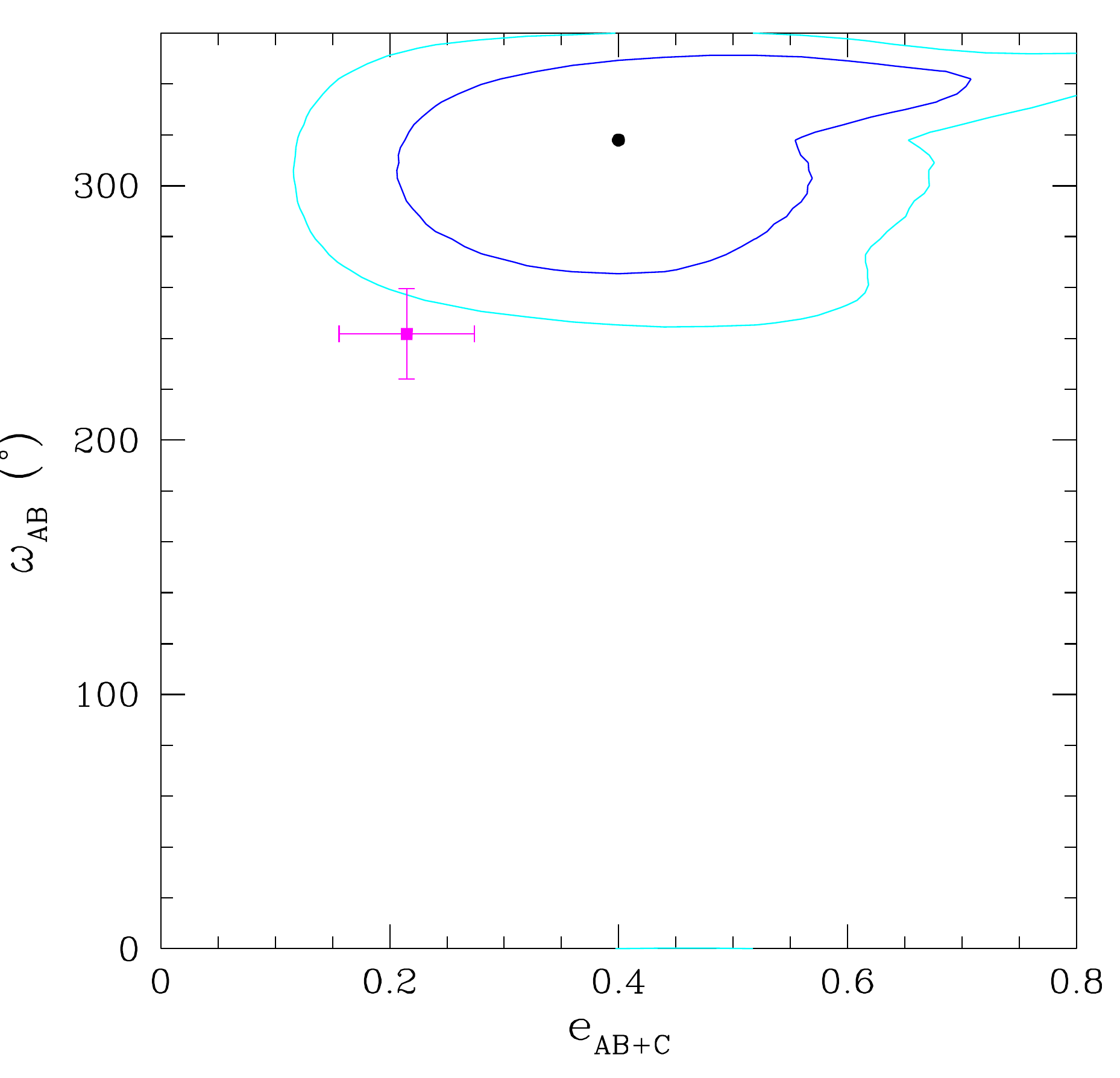}}
    \resizebox{6cm}{!}{\includegraphics{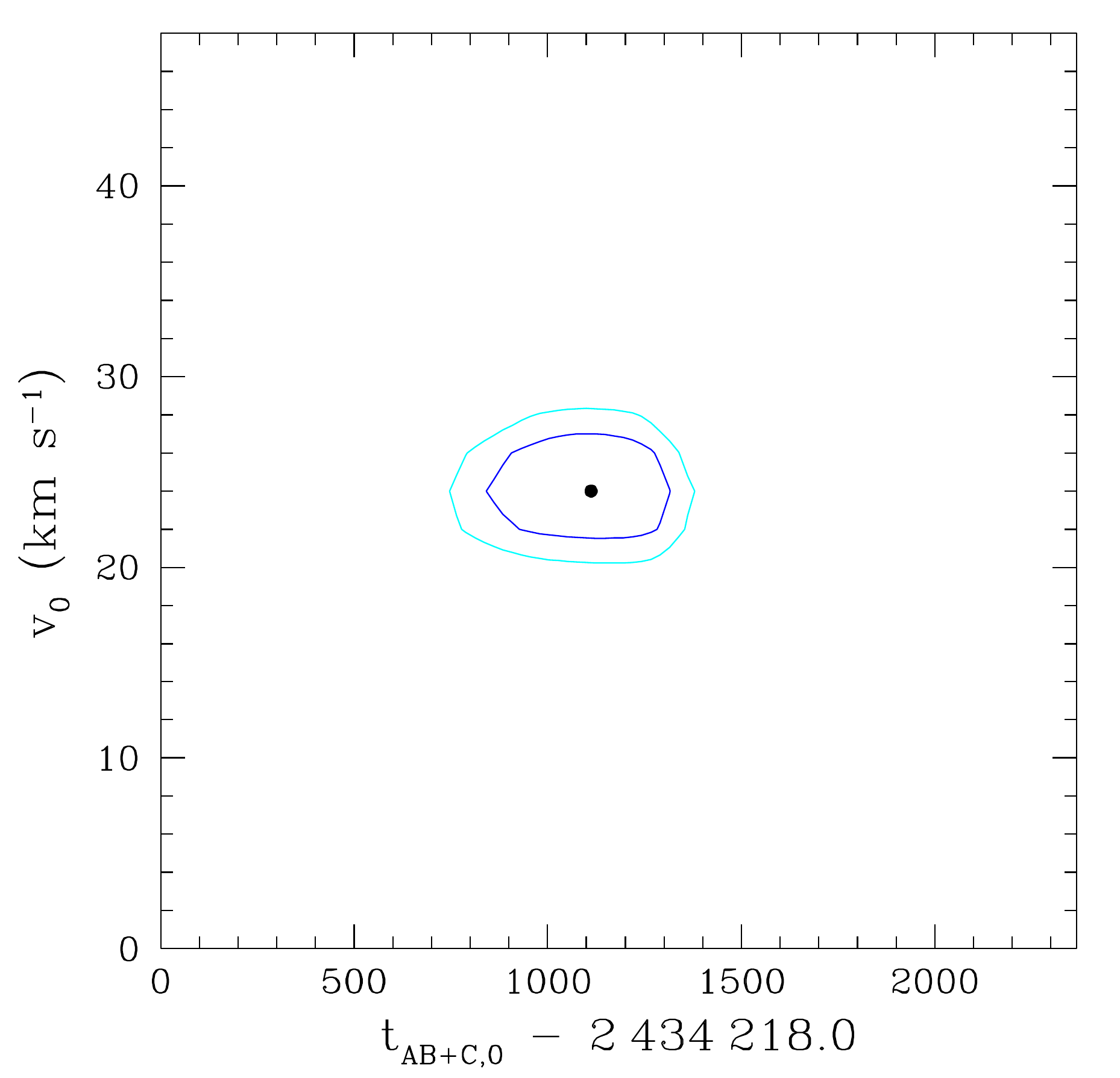}}
    \resizebox{6cm}{!}{\includegraphics{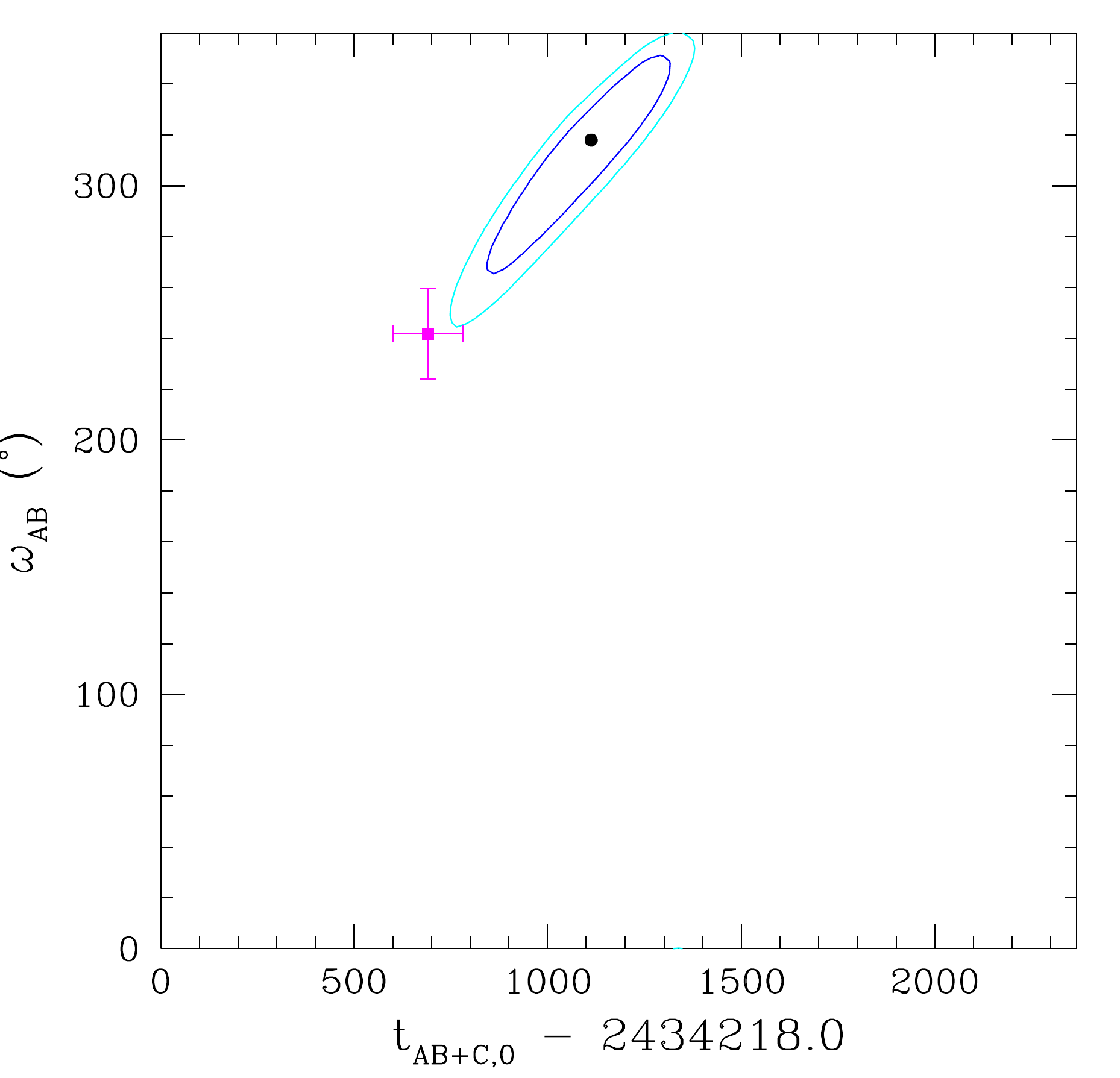}}
\end{center}  
  \caption{$\chi^2$ contours of the combined fit of the $6.5$\,yr orbit corresponding to uncertainties of $1\sigma$ (blue contour) and the 90\% confidence range (cyan contour) projected on planes consisting of various pairs of parameters. The best-fit solution is shown by the black dots. The magenta square (with error bars) in the third and fifth figure yields the parameters and their $1\sigma$ errors as determined from the fit of the radio light curve for $s=2$ (see Sect.\,\ref{radiolc}). \label{chiplanes}}
\end{figure*}

At the $1\sigma$ level, the best-fit parameters are $v_{z,0} = 24^{+3}_{-3}$\,km\,s$^{-1}$, $a_{\rm AB}\,\sin{i_{AB+C}} = (5.0^{+1.3}_{-1.4})\,10^8$\,km, $e_{\rm AB+C} = 0.40^{+0.30}_{-0.19}$, $t_{\rm AB+C,0} = 2\,435\,112^{+202}_{-265}$, and $\omega_{\rm AB} = 318^{+44}_{-53}$. Some parameters span a wide range of possible values. This is especially the case for $e_{\rm AB+C}$. Other parameters exhibit correlations (i.e., $t_{\rm AB+C,0}$ and $\omega_{\rm AB}$). From these results, we can estimate the mass function of the outer orbit as
\begin{eqnarray}
  f(m_{\rm C}) & = & \frac{m_{\rm C}^3\,\sin^3{i_{\rm AB+C}}}{(m_{\rm A} + m_{\rm B} + m_{\rm C})^2}  = \frac{4\,\pi^2\,a^3_{\rm AB}\,\sin^3{i_{\rm AB+C}}}{G\,P_{\rm AB+C}^2}
  \nonumber \\
  & = & 0.89 \pm 0.87\,{\rm M}_{\odot}
.\end{eqnarray}

Considering the masses of the A and B components (see Appendix\,\ref{AppendixC}) and the probable inclination of the orbit (see Sect.\,\ref{radiolc}), we derive $m_{\rm C} = 14.0^{+4.5}_{-10.6}$\,M$_{\odot}$, placing this star in the B-type range if it is a supergiant as the A and B components. Our value of $m_{\rm C}$ is a factor 1.6 lower than the estimate of \citet{ken10}, although both estimates overlap within their errors. Our best values of the orbital parameters are summarized in Table\,\ref{orbitalparam} along with those of the inner orbit (derived in Appendix\,\ref{AppendixC}).

From the third light value found in Appendix\,\ref{AppendixD} and accounting for the contributions from Star D, we can estimate a magnitude difference between Star C and the eclipsing binary of about 2.1. Whilst spectral lines associated to Star C could easily remain hidden in the complex A+B spectrum, a direct detection of Star C is certainly within the reach of modern long baseline interferometry. At a distance of 1.5\,kpc, the semi-major axis of the tertiary orbit (13.2\,AU) corresponds to an angular separation of 0.009\arcsec. These parameters along with the near-infrared magnitude of \ci\ make it a promising target for the refurbished Center of High Angular Resolution Astronomy (CHARA) interferometric array. The most favourable time for detecting Star C via this technique would probably be around the next apastron passage, which is predicted for mid-August 2023.
\begin{table}
  \begin{center}
    \caption{Orbital parameters of the triple system in \ci \label{orbitalparam}}
  \begin{tabular}{l c c c}
    \hline\hline
    & \multicolumn{2}{c}{Eclipsing binary} & Tertiary orbit \\
    \hline
    & Star A & Star B & A+B  \\
    \hline
    Period\,(d) & \multicolumn{2}{c}{see Eq.\,\ref{ephem2}} & $2366 \pm 23$ \\
    \vspace*{-3mm}\\
    $t_0$ (HJD$-$2\,400\,000) & \multicolumn{2}{c}{see Eq.\,\ref{ephem2}} & $35\,330^{+205}_{-265}$ \\
    \vspace*{-3mm}\\
    $e$ & \multicolumn{2}{c}{0.0} & $0.40^{+0.30}_{-0.19}$ \\
    \vspace*{-3mm}\\
    $\omega$\,($^{\circ}$) &   &   & $318^{+44}_{-53}$ \\
    \vspace*{-3mm}\\
    $K$\,(km\,s$^{-1}$) & $87.7 \pm 6.1$ & $280 \pm 31$ & $16.8 \pm 5.3$ \\
    \vspace*{-3mm}\\
    $v_0$\,(km\,s$^{-1}$) & $7.0 \pm 4.5$ & $-100 \pm 20$ & $24^{+3}_{-3}$ \\
    \vspace*{-3mm}\\
    $a\,\sin{i}$\,(R$_{\odot}$) & $11.4$ & $36.4$ & $720^{+185}_{-200}$ \\
    \vspace*{-3mm}\\
    $m\,\sin^3{i}$\,(M$_{\odot}$) & $25.8 \pm 8.6$ & $8.1 \pm 1.9$ & \\
    $f(m)$\,(M$_{\odot}$) & & & $0.89 \pm 0.87$ \\
      \hline
  \end{tabular}
  \end{center}
\end{table}
      
\section{Revisiting the radio light curve \label{radiolc}}
The first evidence for the presence of Star C in \ci\ came from radio observations. In this section, we revisit the published radio data in view of our optical results. We first folded the 4.8\,GHz and 8.4\,GHz {\it VLA} data from \citet{ken10} and the 8.4\,GHz {\it VLBA} data from \citet{dzi13} with our ephemerides. The {\it VLBA} data indicate fluxes that are significantly lower than those of the {\it VLA} observations. \citet{dzi13} estimated that about one-fourth of the non-thermal flux was resolved out by the {\it VLBA}. In addition, most of the more extended thermal flux is also resolved out. To match the level of the two sets of data, we have thus multiplied the {\it VLBA} fluxes by a factor four-thirds and added 5\,mJy. This scaling is somewhat arbitrary, and we include the {\it VLBA} data only for illustrative purpose. The results are shown in Fig.\,\ref{radio}. Except for one strongly deviating {\it VLBA} point, our best-fit period of 2366\,d yields a smooth radio lightcure and allows us to reconcile the \citet{ken10} and \citet{dzi13} data. We also note that the radio minima coincide well with the time of periastron.

We then implemented the model of \citet{ken10}, originally proposed by \citet{wil90}, to describe the variations of the radio emission of \ci. In this model, the observed radio flux (in mJy) is given by the sum of the constant free-free thermal emission of the wind and a phase-dependent non-thermal emission associated with the shock between the winds of A+B and C which undergoes a phase-dependent free-free absorption by the wind of A+B, i.e.
\begin{equation}
  S_{\nu}(t) = 2.5\,\left(\frac{\nu}{4.8\,{\rm GHz}}\right)^{0.6} + S_{4.8}(t)\,\left(\frac{\nu}{4.8\,{\rm GHz}}\right)^{\alpha}\,\exp{[-\tau_{\nu}(t)]}
  \label{Williams}
.\end{equation}
In this equation, $\nu$, $S_{4.8}$, $\alpha$, and $\tau_{\nu}(t)$ are the frequency (in GHz), the level of the non-thermal emission at 4.8\,GHz, its spectral index, and the optical depth of the wind, respectively \citep{wil90,ken10}. We assumed that $S_{4.8}(t)$ scales with the orbital separation as
\begin{equation}
  S_{4.8}(t) = S\,\left(\frac{r}{a_{\rm AB+C}}\right)^{-s}
,\end{equation}
and that the optical depth varies with frequency as
\begin{equation}
  \tau_{\nu}(t) = \tau_{4.8}(t)\,\left(\frac{\nu}{4.8\,{\rm GHz}}\right)^{-2.1}
.\end{equation}
The optical depth $\tau_{4.8}(t)$ is expressed as a function of time and of the orbital parameters \citep[see][and Appendix\,\ref{tau}]{wil90}.

We compared the radio fluxes of \citet{ken10} folded with our period to grids of synthetic radio light curves computed with Eq.\,\ref{Williams}. We computed four grids, corresponding to the values of $s$ considered by \citet{ken10}, i.e.\ 0.0, 0.5, 1.0, and 2.0. The parameter space spans seven dimensions as our grid samples $e_{\rm AB+C}$ between 0.0 and 0.8 (21 steps), $t_0$ with 100 steps of 0.01\,$P_{\rm AB+C}$, $\omega_{\rm C}$ between 0 and $2\,\pi$ (120 steps), $\tau_0$ between 0.4 and 4.0 (10 steps), $\alpha$ between $-1.1$ and $-0.2$ (10 steps), $S$ between 6.0\,mJy and 13.0\,mJy (15 steps), and $i_{\rm AB+C}$ between $45^{\circ}$ and $90^{\circ}$ (4 steps).

\citet{ken10} found inclinations very close to $90^{\circ}$, although with rather large uncertainties. Our calculations confirmed this situation. Overall, the best-fit quality is obtained for $s=2.0$. Considering all models that are acceptable at the $1\sigma$ level, we find the following parameters from the radio light curve: $e_{\rm AB+C} = 0.215 \pm 0.059$, $\omega_{\rm C} = (61.7 \pm 17.8)^{\circ}$, $t_{\rm AB+C,0} = 34909 \pm 90$, $\tau_0 = 3.11 \pm 0.68$, $\alpha = -0.80 \pm 0.14$, $S = (10.50 \pm 1.65)$\,mJy, and $i_{\rm AB+C} = (86.0 \pm 7.5)^{\circ}$. The synthetic curves corresponding to these parameters are shown on top of the observations in Fig.\,\ref{radio} and some parameters and their errors are shown in Fig.\,\ref{chiplanes}. We find a fair agreement between the two totally independent determination of those parameters ($e_{\rm AB+C}$, $\omega_{AB+C}$, $t_{\rm AB+C,0}$) that are in common with the orbit determination in Sect.\,\ref{outerorbit}. This further supports our determination of the properties of Star C.

\begin{figure}[htb]
  \begin{center}
    \resizebox{8cm}{!}{\includegraphics{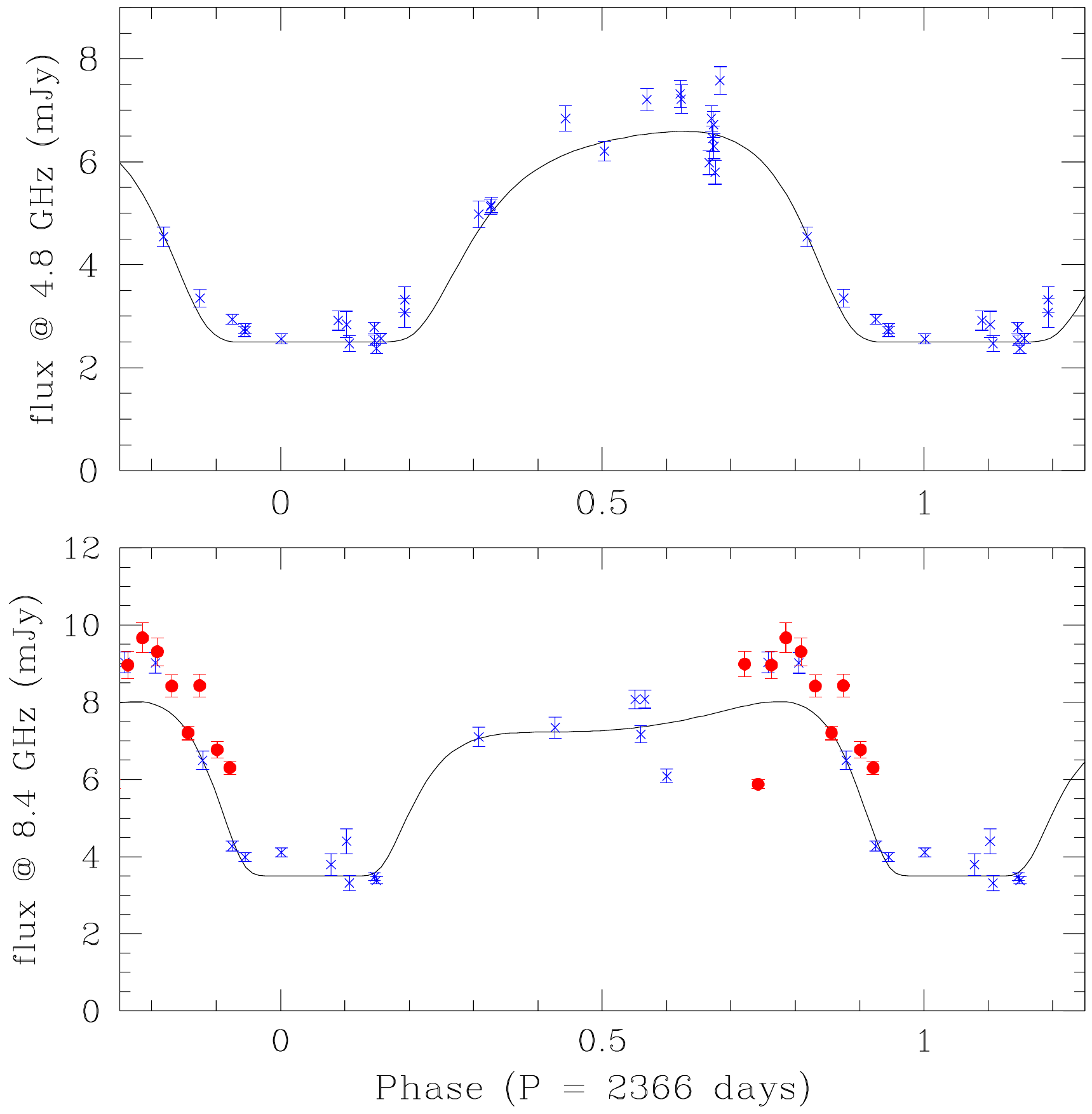}}
\end{center}  
  \caption{Radio fluxes of \ci\ as a function of phase in the 2366\,d cycle (with $t_{\rm AB+C,0} = 34909$) along with the best-fit model with $s=2$ (see text). The crosses stand for the {\it VLA} data from \citet{ken10}, whereas the filled dots indicate the {\it VLBA} fluxes from \citet{dzi13} multiplied by a factor 4/3 and with a 5\,mJy offset added (see text). The {\it VLBA} data were not used in the fit but they nevertheless agree well with VLA results.\label{radio}}
\end{figure}

\section{Discussion and conclusions \label{summary}}
\ci\ is a highly interesting, but extremely challenging massive multiple system. In particular, previous radio studies of the system hinted at the possible presence of a star orbiting the inner binary, which remained to be confirmed. In this context, a number of open issues exist, notably on the exact nature of the radio and X-ray emission.

Using photometric data spanning more than six decades, we have shown for the first time that there exists a light-time effect in the times of primary eclipses of the innermost binary. This indicates a periodic motion with a period of 6.5\,yr. We further found the signature of this reflex motion in the systemic velocity of the He\,{\sc ii} $\lambda$\,4686 emission line. Whilst the orbital parameters are still subject to relatively large uncertainties, we note that they favour a moderate eccentricity. Folding the radio data with our newly determined period, we found that they are best fitted with a model where the intrinsic non-thermal radio emission from the wind interaction zone falls off with orbital separation as $r^{-2}$.

Concerning the X-ray emission, we found considerable variability on timescales of months, but no clear indication of a phase-locked variability with either the period of the eclipsing binary or the 6.5\,yr period of the triple system. Though the X-ray overluminosity of \ci\ most likely stems from the existence of several wind interaction zones in this system, the variability cannot be directly connected to the orbital phases of the A+B or AB+C systems and thus seems to have different origins. We tentatively suggest that transient variations of the mass-loss rate of the eclipsing binary might be responsible for part of these changes of the X-ray emission.

Future work, notably using a proper treatment of atmospheric eclipses, may further shed light on the exact nature of Star B and the remaining discrepancy in distance estimate (see Appendix\,\ref{AppendixD}). 

\begin{acknowledgements}
We acknowledge support from the Fonds National de la Recherche Scientifique (Belgium), the Communaut\'e Fran\c caise de Belgique (including notably support for the observing runs at OHP), the European Space Agency (ESA) and the Belgian Federal Science Policy Office (BELSPO) in the framework of the PRODEX Programme (contract XMaS). GR acknowledges the continuous efforts of the administration of Li\`ege University, especially the finance department, to make his life more complicated. We thank the \sw\ team for their kind assistance as well as Drs Yakut, Laur, and Paschke for their help in locating data. Our thanks also go to Dietmar Bannuscher and Klaus H\"au\ss ler for providing us with a copy of the Harthaer Beobachtungszirkulare and to Dr P.M.\ Williams for advices on the modelling of the radio light curve. We thank Dr Zasche for his constructive referee report. ADS and CDS were used for this research. 
\end{acknowledgements}

\appendix
\section{Journal of X-ray and optical spectroscopy observations}
In this appendix, we provide the detailed list of the \xmm\ and \sw\ observations of \ci\ (Tables\,\ref{journalxmm} and \ref{journalswift}), as well as the list of the new optical spectra and the corresponding RVs (Table\,\ref{journaloptical}).
\begin{table*}[h!]
\centering
\footnotesize
\caption{Journal of the \xmm\ observations. }
\label{journalxmm}
\begin{tabular}{lcccc}
\hline\hline
Rev/ObsID & HJD-2\,450\,000 & \multicolumn{3}{c}{Count Rates (cts\,s$^{-1}$) \& $HR$}\\
& & MOS1 & MOS2 & pn \\
\hline
0896/0200450201 & 3308.581 &0.379$\pm$0.007 (--0.487$\pm$0.017) &0.446$\pm$0.009 (--0.333$\pm$0.020) &1.033$\pm$0.014 (--0.532$\pm$0.012) \\
0901/0200450301 & 3318.559 &0.422$\pm$0.008 (--0.488$\pm$0.016) &0.487$\pm$0.009 (--0.339$\pm$0.018) &1.097$\pm$0.017 (--0.567$\pm$0.013) \\
0906/0200450401 & 3328.543 &0.386$\pm$0.007 (--0.483$\pm$0.015) &0.463$\pm$0.008 (--0.335$\pm$0.018) &1.090$\pm$0.012 (--0.548$\pm$0.010) \\
0911/0200450501 & 3338.505 &0.403$\pm$0.009 (--0.4800$\pm$0.020) &0.496$\pm$0.011 (--0.335$\pm$0.023) &1.135$\pm$0.016 (--0.533$\pm$0.013) \\
1353/0505110301 & 4220.354 &0.541$\pm$0.012 (--0.361$\pm$0.021) &0.616$\pm$0.013 (--0.266$\pm$0.022) &1.574$\pm$0.026 (--0.414$\pm$0.016) \\
1355/0505110401 & 4224.169 &0.700$\pm$0.011 (--0.360$\pm$0.015) &0.753$\pm$0.013 (--0.266$\pm$0.017) &1.944$\pm$0.025 (--0.410$\pm$0.012) \\
2114/0677980601 & 5738.256 &0.000$\pm$0.000 (-0.0000$\pm$0.000) &0.470$\pm$0.007 (--0.380$\pm$0.015) &1.268$\pm$0.016 (--0.567$\pm$0.011) \\
2625/0740300101 & 6758.209 &0.559$\pm$0.007 (--0.410$\pm$0.011) &0.570$\pm$0.007 (--0.397$\pm$0.011) &1.599$\pm$0.013 (--0.456$\pm$0.007) \\
3089/0780040101 & 7683.229 &0.442$\pm$0.007 (--0.560$\pm$0.012) &0.436$\pm$0.006 (--0.545$\pm$0.012) &1.257$\pm$0.012 (--0.616$\pm$0.007) \\
3097/0793183001 & 7699.380 &0.000$\pm$0.000 (-0.0000$\pm$0.000) &0.423$\pm$0.006 (--0.491$\pm$0.014) &1.285$\pm$0.010 (--0.604$\pm$0.007) \\
3176/0800150101 & 7856.833 &0.429$\pm$0.007 (--0.537$\pm$0.014) &0.433$\pm$0.007 (--0.521$\pm$0.015) &1.226$\pm$0.014 (--0.577$\pm$0.010) \\
3273/0801910201 & 8050.426 &                                   & 0.426$\pm$0.007 (--0.456$\pm$0.017) &                                      \\
3280/0801910301 & 8063.415 &                                   & 0.417$\pm$0.008 (--0.454$\pm$0.021) &                                      \\
3284/0801910401 & 8071.063 &                                   & 0.378$\pm$0.008 (--0.456$\pm$0.019) &                                      \\
3288/0801910501 & 8079.696 &                                   & 0.408$\pm$0.008 (--0.479$\pm$0.020) &                                      \\
3294/0801910601 & 8092.351 &                                   & 0.388$\pm$0.007 (--0.452$\pm$0.018) &                                      \\
\hline
\end{tabular}
\tablefoot{HJD correspond to dates at mid-exposure. Columns 3 to 5 provide the count rates in the total band followed by the hardness ratio between brackets ($HR=(H-S)/(H+S)$) for the MOS1, MOS2, and pn cameras. The energy bands are: 0.3--10.0\,keV (total), 0.3--2.0\,keV (S, soft), and 2.0--10.0\,keV (H, hard).}
\end{table*}

\begin{table*}
\footnotesize
\caption{Journal of the \sw\ observations. }
\label{journalswift}
\begin{minipage}{8.5cm}
  \centering
\begin{tabular}{lcc}
\hline\hline
ObsID & HJD-2\,450\,000 & Count Rates (cts\,s$^{-1}$) \& $HR$\\
\hline
00037920001 & 4657.335 & 0.127$\pm$0.010 ( 0.40$\pm$0.07 ) \\ 
00031904001 & 5571.617 & 0.107$\pm$0.005 ( 0.39$\pm$0.04 ) \\ 
00031904002 & 5655.834 & 0.084$\pm$0.005 ( 0.31$\pm$0.04 ) \\ 
00031904003 & 5700.082 & 0.128$\pm$0.007 ( 0.35$\pm$0.04 ) \\ 
00031904004 & 5743.841 & 0.103$\pm$0.005 ( 0.33$\pm$0.04 ) \\ 
00031904005 & 5842.171 & 0.106$\pm$0.006 ( 0.50$\pm$0.06 ) \\ 
00032767001 & 6380.437 & 0.120$\pm$0.007 ( 0.44$\pm$0.06 ) \\ 
00032767002 & 6380.874 & 0.114$\pm$0.006 ( 0.55$\pm$0.06 ) \\ 
00033818001 & 7191.528 & 0.143$\pm$0.021 ( 0.68$\pm$0.20 ) \\ 
00033818002 & 7192.049 & 0.228$\pm$0.067 ( 0.17$\pm$0.13 ) \\ 
00033818003 & 7192.738 & 0.136$\pm$0.018 ( 0.56$\pm$0.16 ) \\ 
00033818005 & 7193.317 & 0.132$\pm$0.016 ( 0.40$\pm$0.11 ) \\ 
00033818004 & 7193.513 & 0.076$\pm$0.021 ( 0.56$\pm$0.32 ) \\ 
00033818006 & 7194.261 & 0.141$\pm$0.019 ( 0.39$\pm$0.12 ) \\ 
00033818007 & 7194.532 & 0.130$\pm$0.017 ( 0.25$\pm$0.08 ) \\ 
00033818008 & 7195.319 & 0.145$\pm$0.015 ( 0.39$\pm$0.09 ) \\ 
00033818009 & 7195.784 & 0.134$\pm$0.015 ( 0.35$\pm$0.09 ) \\ 
00033818010 & 7196.132 & 0.159$\pm$0.030 ( 0.10$\pm$0.07 ) \\ 
00033818011 & 7196.661 & 0.138$\pm$0.015 ( 0.25$\pm$0.07 ) \\ 
00033818012 & 7197.114 & 0.138$\pm$0.020 ( 0.30$\pm$0.10 ) \\ 
00033818013 & 7197.917 & 0.121$\pm$0.013 ( 0.37$\pm$0.09 ) \\ 
00033818014 & 7198.438 & 0.165$\pm$0.027 ( 0.22$\pm$0.09 ) \\ 
00033818016 & 7199.052 & 0.113$\pm$0.014 ( 0.28$\pm$0.08 ) \\ 
00033818017 & 7199.512 & 0.130$\pm$0.022 ( 0.37$\pm$0.14 ) \\ 
00033818018 & 7200.299 & 0.105$\pm$0.016 ( 0.25$\pm$0.10 ) \\ 
00033818020 & 7201.233 & 0.133$\pm$0.013 ( 0.31$\pm$0.07 ) \\ 
00033818022 & 7202.031 & 0.137$\pm$0.014 ( 0.40$\pm$0.09 ) \\ 
00033818023 & 7202.535 & 0.116$\pm$0.016 ( 0.32$\pm$0.10 ) \\ 
00033818025 & 7204.158 & 0.121$\pm$0.016 ( 0.48$\pm$0.13 ) \\ 
00033818026 & 7204.891 & 0.097$\pm$0.012 ( 0.52$\pm$0.14 ) \\ 
00033818027 & 7205.091 & 0.116$\pm$0.013 ( 0.28$\pm$0.08 ) \\ 
00033818028 & 7205.757 & 0.087$\pm$0.012 ( 0.35$\pm$0.11 ) \\ 
00033818032 & 7207.752 & 0.117$\pm$0.014 ( 0.41$\pm$0.11 ) \\ 
00033818033 & 7208.098 & 0.121$\pm$0.022 ( 0.40$\pm$0.16 ) \\ 
00033818035 & 7209.362 & 0.103$\pm$0.012 ( 0.43$\pm$0.11 ) \\ 
00033818036 & 7209.959 & 0.112$\pm$0.013 ( 0.57$\pm$0.13 ) \\ 
00033818038 & 7211.251 & 0.090$\pm$0.015 ( 0.26$\pm$0.10 ) \\ 
00033818039 & 7211.956 & 0.130$\pm$0.015 ( 0.39$\pm$0.10 ) \\ 
00033818040 & 7212.489 & 0.133$\pm$0.014 ( 0.42$\pm$0.10 ) \\ 
00033818041 & 7212.876 & 0.096$\pm$0.012 ( 0.39$\pm$0.11 ) \\ 
00033818042 & 7213.210 & 0.111$\pm$0.013 ( 0.36$\pm$0.09 ) \\ 
00033818044 & 7214.291 & 0.094$\pm$0.010 ( 0.36$\pm$0.09 ) \\ 
00032767003 & 7284.945 & 0.100$\pm$0.004 ( 0.38$\pm$0.03 ) \\ 
00034282001 & 7410.533 & 0.116$\pm$0.011 ( 0.51$\pm$0.10 ) \\ 
00034282002 & 7452.349 & 0.111$\pm$0.005 ( 0.27$\pm$0.03 ) \\ 
00034282003 & 7466.312 & 0.140$\pm$0.008 ( 0.30$\pm$0.04 ) \\ 
00034282004 & 7479.606 & 0.111$\pm$0.007 ( 0.35$\pm$0.05 ) \\ 
00034282005 & 7494.293 & 0.109$\pm$0.008 ( 0.31$\pm$0.05 ) \\ 
00034282006 & 7510.368 & 0.107$\pm$0.007 ( 0.26$\pm$0.04 ) \\ 
00034282007 & 7512.656 & 0.112$\pm$0.008 ( 0.42$\pm$0.07 ) \\ 
00034282008 & 7525.100 & 0.116$\pm$0.007 ( 0.37$\pm$0.05 ) \\ 
00034282009 & 7536.168 & 0.137$\pm$0.008 ( 0.38$\pm$0.05 ) \\ 
00034282010 & 7540.990 & 0.112$\pm$0.010 ( 0.33$\pm$0.07 ) \\ 
00034282011 & 7550.719 & 0.093$\pm$0.014 ( 0.44$\pm$0.15 ) \\ 
00034282012 & 7558.217 & 0.116$\pm$0.013 ( 0.48$\pm$0.12 ) \\ 
00034282013 & 7572.285 & 0.123$\pm$0.021 ( 0.12$\pm$0.07 ) \\ 
00034282014 & 7574.919 & 0.117$\pm$0.007 ( 0.46$\pm$0.06 ) \\ 
00034282015 & 7576.331 & 0.040$\pm$0.022 ( 1.44$\pm$1.38 ) \\ 
00034282016 & 7580.567 & 0.089$\pm$0.008 ( 0.45$\pm$0.09 ) \\ 
00034282018 & 7592.128 & 0.103$\pm$0.005 ( 0.35$\pm$0.04 ) \\ 
00034282019 & 7606.350 & 0.096$\pm$0.012 ( 0.35$\pm$0.10 ) \\ 
00034282020 & 7612.145 & 0.076$\pm$0.030 ( 1.09$\pm$0.84 ) \\ 
00034282021 & 7617.279 & 0.121$\pm$0.017 ( 0.33$\pm$0.11 ) \\ 
00034282022 & 7619.060 & 0.092$\pm$0.011 ( 0.40$\pm$0.10 ) \\ 
00034282023 & 7620.467 & 0.100$\pm$0.005 ( 0.39$\pm$0.05 ) \\ 
00034282024 & 7633.726 & 0.115$\pm$0.007 ( 0.35$\pm$0.05 ) \\ 
00034282025 & 7640.273 & 0.102$\pm$0.012 ( 0.34$\pm$0.10 ) \\ 
00081975001 & 7641.655 & 0.114$\pm$0.006 ( 0.32$\pm$0.04 ) \\ 
\hline
\end{tabular}
\end{minipage}
%\hfill
\begin{minipage}{8.5cm}
\centering
\begin{tabular}{lcc}
\hline\hline
ObsID & HJD-2\,450\,000 & Count Rates (cts\,s$^{-1}$) \& $HR$\\
\hline
00034282026 & 7645.126 & 0.105$\pm$0.015 ( 0.30$\pm$0.10 ) \\ 
00034282027 & 7647.775 & 0.119$\pm$0.005 ( 0.31$\pm$0.03 ) \\ 
00034282028 & 7661.928 & 0.095$\pm$0.006 ( 0.28$\pm$0.04 ) \\ 
00034282029 & 7675.716 & 0.111$\pm$0.005 ( 0.31$\pm$0.03 ) \\ 
00034282030 & 7688.885 & 0.107$\pm$0.006 ( 0.29$\pm$0.04 ) \\ 
00034282031 & 7697.339 & 0.083$\pm$0.005 ( 0.30$\pm$0.04 ) \\ 
00034282033 & 7703.612 & 0.118$\pm$0.007 ( 0.34$\pm$0.04 ) \\ 
00034282032 & 7703.878 & 0.088$\pm$0.008 ( 0.41$\pm$0.08 ) \\ 
00034282035 & 7705.812 & 0.096$\pm$0.009 ( 0.33$\pm$0.07 ) \\ 
00034282034 & 7706.941 & 0.100$\pm$0.008 ( 0.31$\pm$0.06 ) \\ 
00034282036 & 7707.534 & 0.098$\pm$0.009 ( 0.34$\pm$0.07 ) \\ 
00034282037 & 7711.159 & 0.109$\pm$0.006 ( 0.42$\pm$0.05 ) \\ 
00034282038 & 7717.962 & 0.119$\pm$0.006 ( 0.38$\pm$0.04 ) \\ 
00034282039 & 7725.504 & 0.122$\pm$0.007 ( 0.37$\pm$0.05 ) \\ 
00034282041 & 7731.771 & 0.120$\pm$0.006 ( 0.49$\pm$0.05 ) \\ 
00034282040 & 7732.292 & 0.119$\pm$0.011 ( 0.31$\pm$0.07 ) \\ 
00034282043 & 7734.319 & 0.114$\pm$0.009 ( 0.30$\pm$0.05 ) \\ 
00034282042 & 7734.519 & 0.133$\pm$0.011 ( 0.37$\pm$0.07 ) \\ 
00034282044 & 7735.556 & 0.120$\pm$0.018 ( 0.18$\pm$0.08 ) \\ 
00034282045 & 7739.302 & 0.132$\pm$0.007 ( 0.37$\pm$0.04 ) \\ 
00034282046 & 7743.517 & 0.132$\pm$0.013 ( 0.41$\pm$0.09 ) \\ 
00034282047 & 7746.210 & 0.133$\pm$0.007 ( 0.35$\pm$0.04 ) \\ 
00034282048 & 7752.887 & 0.115$\pm$0.007 ( 0.38$\pm$0.05 ) \\ 
00034282050 & 7760.755 & 0.114$\pm$0.007 ( 0.47$\pm$0.06 ) \\ 
00034282049 & 7761.388 & 0.170$\pm$0.016 ( 0.65$\pm$0.12 ) \\ 
00034282052 & 7767.162 & 0.150$\pm$0.014 ( 0.41$\pm$0.09 ) \\ 
00034282053 & 7773.816 & 0.154$\pm$0.011 ( 0.48$\pm$0.08 ) \\ 
00034282054 & 7780.730 & 0.086$\pm$0.022 ( 0.15$\pm$0.12 ) \\ 
00034282055 & 7785.178 & 0.107$\pm$0.011 ( 0.36$\pm$0.08 ) \\ 
00034282056 & 7788.407 & 0.124$\pm$0.009 ( 0.40$\pm$0.06 ) \\ 
00034282057 & 7794.808 & 0.141$\pm$0.011 ( 0.45$\pm$0.07 ) \\ 
00034282058 & 7802.420 & 0.153$\pm$0.017 ( 0.44$\pm$0.10 ) \\ 
00034282059 & 7807.305 & 0.128$\pm$0.060                   \\ 
00034282060 & 7809.061 & 0.132$\pm$0.008 ( 0.56$\pm$0.08 ) \\ 
00034282062 & 7829.627 & 0.106$\pm$0.016 ( 0.49$\pm$0.15 ) \\ 
00034282063 & 7834.403 & 0.133$\pm$0.013 ( 0.52$\pm$0.11 ) \\ 
00034282064 & 7843.522 & 0.091$\pm$0.018 ( 0.29$\pm$0.15 ) \\ 
00034282065 & 7844.019 & 0.115$\pm$0.011 ( 0.33$\pm$0.07 ) \\ 
00093148001 & 7849.623 & 0.097$\pm$0.008 ( 0.27$\pm$0.05 ) \\ 
00093146001 & 7852.308 & 0.108$\pm$0.007 ( 0.42$\pm$0.06 ) \\ 
00093148002 & 7854.833 & 0.110$\pm$0.010 ( 0.34$\pm$0.07 ) \\ 
00034282067 & 7857.733 & 0.103$\pm$0.006 ( 0.42$\pm$0.06 ) \\ 
00093146002 & 7865.591 & 0.112$\pm$0.007 ( 0.42$\pm$0.06 ) \\ 
00034282068 & 7872.172 & 0.105$\pm$0.010 ( 0.51$\pm$0.11 ) \\ 
00093146003 & 7879.549 & 0.110$\pm$0.008 ( 0.29$\pm$0.05 ) \\ 
00034282070 & 7886.292 & 0.088$\pm$0.006 ( 0.34$\pm$0.05 ) \\ 
00093146004 & 7893.802 & 0.093$\pm$0.008 ( 0.35$\pm$0.07 ) \\ 
00034282072 & 7899.750 & 0.113$\pm$0.006 ( 0.46$\pm$0.06 ) \\ 
00093146005 & 7907.787 & 0.103$\pm$0.007 ( 0.44$\pm$0.06 ) \\ 
00093148003 & 7910.909 & 0.100$\pm$0.006 ( 0.37$\pm$0.05 ) \\ 
00034282074 & 7914.332 & 0.099$\pm$0.006 ( 0.39$\pm$0.06 ) \\ 
00034282073 & 7915.252 & 0.091$\pm$0.015 ( 0.16$\pm$0.08 ) \\ 
00034282075 & 7918.609 & 0.097$\pm$0.010 ( 0.32$\pm$0.08 ) \\ 
00093146006 & 7921.606 & 0.098$\pm$0.008 ( 0.39$\pm$0.07 ) \\ 
00034282076 & 7928.276 & 0.103$\pm$0.006 ( 0.43$\pm$0.06 ) \\ 
00093146007 & 7936.157 & 0.107$\pm$0.008 ( 0.27$\pm$0.05 ) \\ 
00093148004 & 7940.613 & 0.094$\pm$0.006 ( 0.35$\pm$0.05 ) \\ 
00034282078 & 7942.762 & 0.099$\pm$0.006 ( 0.33$\pm$0.05 ) \\ 
00034282079 & 7947.330 & 0.115$\pm$0.012 ( 0.54$\pm$0.12 ) \\ 
00034282080 & 7947.992 & 0.112$\pm$0.010 ( 0.32$\pm$0.07 ) \\ 
00034282081 & 7948.855 & 0.108$\pm$0.009 ( 0.43$\pm$0.08 ) \\ 
00034282082 & 7949.580 & 0.095$\pm$0.007 ( 0.36$\pm$0.06 ) \\ 
00093146008 & 7950.012 & 0.095$\pm$0.006 ( 0.29$\pm$0.05 ) \\ 
00034282083 & 7950.782 & 0.119$\pm$0.009 ( 0.39$\pm$0.07 ) \\ 
00034282084 & 7951.704 & 0.107$\pm$0.009 ( 0.26$\pm$0.05 ) \\ 
00034282085 & 7953.295 & 0.096$\pm$0.008 ( 0.40$\pm$0.07 ) \\ 
00034282086 & 7954.301 & 0.092$\pm$0.008 ( 0.46$\pm$0.09 ) \\ 
00034282087 & 7954.823 & 0.109$\pm$0.009 ( 0.34$\pm$0.06 ) \\ 
\hline
\end{tabular}
\end{minipage}
\end{table*}
\addtocounter{table}{-1}
\begin{table*}
\footnotesize
\caption{Continued}
\begin{minipage}{8.5cm}
  \centering
\begin{tabular}{lcc}
\hline\hline
ObsID & HJD-2\,450\,000 & Count Rates (cts\,s$^{-1}$) \& $HR$\\
\hline
00034282089 & 7956.693 & 0.105$\pm$0.006 ( 0.35$\pm$0.05 ) \\ 
00093146010 & 7968.102 & 0.107$\pm$0.007 ( 0.47$\pm$0.06 ) \\ 
00093148005 & 7971.798 & 0.111$\pm$0.006 ( 0.43$\pm$0.05 ) \\ 
00093146011 & 7977.409 & 0.100$\pm$0.006 ( 0.37$\pm$0.05 ) \\ 
00034282090 & 7988.247 & 0.117$\pm$0.011 ( 0.48$\pm$0.10 ) \\ 
00034282091 & 7991.196 & 0.103$\pm$0.008 ( 0.39$\pm$0.06 ) \\ 
00093146012 & 7991.963 & 0.094$\pm$0.007 ( 0.44$\pm$0.07 ) \\ 
00093148006 & 8002.625 & 0.115$\pm$0.008 ( 0.34$\pm$0.05 ) \\ 
00034282092 & 8005.019 & 0.104$\pm$0.010 ( 0.66$\pm$0.12 ) \\ 
00093146013 & 8006.341 & 0.128$\pm$0.011 ( 0.64$\pm$0.11 ) \\ 
00093148007 & 8007.729 & 0.105$\pm$0.011 ( 0.47$\pm$0.11 ) \\ 
00034282093 & 8012.089 & 0.100$\pm$0.010 ( 0.31$\pm$0.07 ) \\ 
00093148008 & 8017.773 & 0.115$\pm$0.007 ( 0.39$\pm$0.05 ) \\ 
00034282094 & 8018.919 & 0.111$\pm$0.009 ( 0.49$\pm$0.08 ) \\ 
00093146014 & 8020.457 & 0.096$\pm$0.007 ( 0.39$\pm$0.06 ) \\ 
00034282095 & 8026.067 & 0.100$\pm$0.010 ( 0.60$\pm$0.12 ) \\ 
00093146015 & 8027.055 & 0.111$\pm$0.007 ( 0.35$\pm$0.05 ) \\ 
00034282096 & 8032.880 & 0.094$\pm$0.008 ( 0.45$\pm$0.09 ) \\ 
00093148009 & 8033.170 & 0.105$\pm$0.007 ( 0.41$\pm$0.06 ) \\ 
00093146016 & 8033.867 & 0.093$\pm$0.007 ( 0.46$\pm$0.07 ) \\ 
00034282097 & 8040.248 & 0.086$\pm$0.005 ( 0.54$\pm$0.07 ) \\ 
00093146017 & 8040.950 & 0.101$\pm$0.012 ( 0.84$\pm$0.19 ) \\ 
00034282098 & 8046.624 & 0.098$\pm$0.009 ( 0.26$\pm$0.06 ) \\ 
00093146018 & 8047.623 & 0.106$\pm$0.009 ( 0.39$\pm$0.07 ) \\ 
00093148011 & 8047.952 & 0.098$\pm$0.008 ( 0.42$\pm$0.08 ) \\ 
00088016001 & 8050.016 & 0.097$\pm$0.009 ( 0.40$\pm$0.08 ) \\ 
00034282099 & 8051.280 & 0.094$\pm$0.007 ( 0.53$\pm$0.08 ) \\ 
00093148012 & 8051.774 & 0.090$\pm$0.009 ( 0.50$\pm$0.11 ) \\ 
00034282100 & 8053.531 & 0.092$\pm$0.013 ( 0.91$\pm$0.25 ) \\ 
00093146019 & 8055.455 & 0.087$\pm$0.007 ( 0.48$\pm$0.08 ) \\ 
00034282101 & 8058.015 & 0.104$\pm$0.010 ( 0.38$\pm$0.08 ) \\ 
00034282102 & 8058.611 & 0.088$\pm$0.017 ( 0.38$\pm$0.16 ) \\ 
00034282103 & 8061.171 & 0.127$\pm$0.009 ( 0.41$\pm$0.07 ) \\ 
00093146020 & 8061.944 & 0.108$\pm$0.007 ( 0.52$\pm$0.07 ) \\ 
00093148013 & 8063.528 & 0.123$\pm$0.007 ( 0.34$\pm$0.04 ) \\ 
00034282104 & 8064.580 & 0.095$\pm$0.025 ( 0.67$\pm$0.35 ) \\ 
00034282106 & 8065.948 & 0.114$\pm$0.010 ( 0.36$\pm$0.07 ) \\ 
00034282107 & 8068.316 & 0.111$\pm$0.010 ( 0.32$\pm$0.07 ) \\ 
00093146021 & 8069.013 & 0.093$\pm$0.007 ( 0.40$\pm$0.06 ) \\ 
00093148014 & 8069.702 & 0.104$\pm$0.006 ( 0.45$\pm$0.06 ) \\ 
00034282108 & 8070.399 & 0.100$\pm$0.009 ( 0.40$\pm$0.08 ) \\ 
00034282109 & 8072.421 & 0.097$\pm$0.007 ( 0.49$\pm$0.08 ) \\ 
00034282110 & 8075.189 & 0.120$\pm$0.009 ( 0.43$\pm$0.07 ) \\ 
00093146022 & 8076.113 & 0.090$\pm$0.007 ( 0.43$\pm$0.07 ) \\ 
00034282111 & 8076.674 & 0.105$\pm$0.008 ( 0.44$\pm$0.07 ) \\ 
00093148015 & 8078.070 & 0.103$\pm$0.007 ( 0.50$\pm$0.07 ) \\ 
00088268001 & 8079.102 & 0.120$\pm$0.010 ( 0.34$\pm$0.06 ) \\ 
00034282112 & 8079.463 & 0.101$\pm$0.010 ( 0.44$\pm$0.10 ) \\ 
00034282113 & 8080.459 & 0.101$\pm$0.010 ( 0.76$\pm$0.15 ) \\ 
00034282114 & 8081.862 & 0.099$\pm$0.010 ( 0.44$\pm$0.10 ) \\ 
00093146023 & 8082.750 & 0.104$\pm$0.007 ( 0.51$\pm$0.07 ) \\ 
00034282115 & 8083.678 & 0.101$\pm$0.009 ( 0.38$\pm$0.07 ) \\ 
00093148016 & 8084.678 & 0.099$\pm$0.007 ( 0.43$\pm$0.07 ) \\ 
00034282116 & 8085.672 & 0.086$\pm$0.007 ( 0.36$\pm$0.07 ) \\ 
00034282117 & 8087.739 & 0.107$\pm$0.009 ( 0.41$\pm$0.07 ) \\ 
00034282118 & 8089.069 & 0.109$\pm$0.009 ( 0.50$\pm$0.09 ) \\ 
00093146024 & 8089.668 & 0.096$\pm$0.007 ( 0.35$\pm$0.06 ) \\ 
00093148017 & 8093.357 & 0.105$\pm$0.007 ( 0.45$\pm$0.06 ) \\ 
00034282119 & 8095.882 & 0.100$\pm$0.010 ( 0.62$\pm$0.12 ) \\ 
00010451001 & 8096.174 & 0.100$\pm$0.006 ( 0.45$\pm$0.06 ) \\ 
00093148018 & 8099.936 & 0.096$\pm$0.010 ( 0.32$\pm$0.08 ) \\ 
00034282120 & 8102.614 & 0.117$\pm$0.009 ( 0.59$\pm$0.09 ) \\ 
00093146025 & 8104.042 & 0.115$\pm$0.008 ( 0.45$\pm$0.07 ) \\ 
00010451002 & 8105.767 & 0.102$\pm$0.006 ( 0.35$\pm$0.05 ) \\ 
\hline
\end{tabular}
\end{minipage}
%\hfill
\begin{minipage}{8.5cm}
\centering
\begin{tabular}{lcc}
\hline\hline
ObsID & HJD-2\,450\,000 & Count Rates (cts\,s$^{-1}$) \& $HR$\\
\hline
00093148019 & 8108.129 & 0.122$\pm$0.009 ( 0.46$\pm$0.07 ) \\ 
00034282121 & 8109.887 & 0.115$\pm$0.011 ( 0.71$\pm$0.14 ) \\ 
00093148020 & 8114.604 & 0.083$\pm$0.010 ( 0.35$\pm$0.09 ) \\ 
00010451003 & 8116.128 & 0.084$\pm$0.005 ( 0.39$\pm$0.06 ) \\ 
00034282122 & 8116.961 & 0.117$\pm$0.010 ( 0.39$\pm$0.07 ) \\ 
00093146026 & 8118.187 & 0.102$\pm$0.010 ( 0.37$\pm$0.08 ) \\ 
00034282123 & 8123.692 & 0.149$\pm$0.014 ( 0.64$\pm$0.13 ) \\ 
00093148021 & 8124.735 & 0.081$\pm$0.008 ( 0.41$\pm$0.09 ) \\ 
00010451004 & 8126.220 & 0.100$\pm$0.006 ( 0.42$\pm$0.05 ) \\ 
00034282124 & 8130.740 & 0.109$\pm$0.015 ( 0.37$\pm$0.12 ) \\ 
00093146027 & 8132.366 & 0.122$\pm$0.011 ( 0.59$\pm$0.11 ) \\ 
00010451005 & 8135.691 & 0.091$\pm$0.005 ( 0.36$\pm$0.05 ) \\ 
00034282125 & 8138.246 & 0.151$\pm$0.033 ( 0.61$\pm$0.27 ) \\ 
00093148022 & 8140.304 & 0.115$\pm$0.013 ( 0.59$\pm$0.13 ) \\ 
00093148023 & 8142.835 & 0.119$\pm$0.009 ( 0.56$\pm$0.09 ) \\ 
00034282126 & 8144.695 & 0.115$\pm$0.009 ( 0.47$\pm$0.08 ) \\ 
00010451006 & 8145.790 & 0.097$\pm$0.005 ( 0.49$\pm$0.06 ) \\ 
00034282127 & 8151.807 & 0.104$\pm$0.012 ( 0.60$\pm$0.14 ) \\ 
00010451007 & 8155.787 & 0.094$\pm$0.005 ( 0.40$\pm$0.05 ) \\ 
00093146029 & 8159.510 & 0.145$\pm$0.020 ( 0.49$\pm$0.14 ) \\ 
00093146030 & 8164.288 & 0.039$\pm$0.030                   \\ 
00010451008 & 8166.616 & 0.094$\pm$0.005 ( 0.45$\pm$0.06 ) \\ 
00034282130 & 8172.815 & 0.094$\pm$0.009 ( 0.58$\pm$0.11 ) \\ 
00010451009 & 8176.201 & 0.093$\pm$0.005 ( 0.37$\pm$0.05 ) \\ 
00093148025 & 8184.004 & 0.108$\pm$0.008 ( 0.73$\pm$0.11 ) \\ 
00010451010 & 8185.605 & 0.094$\pm$0.007 ( 0.49$\pm$0.08 ) \\ 
00034282131 & 8194.782 & 0.115$\pm$0.009 ( 0.29$\pm$0.05 ) \\ 
00034282132 & 8202.018 & 0.106$\pm$0.010 ( 0.33$\pm$0.07 ) \\ 
00034282133 & 8207.596 & 0.115$\pm$0.014 ( 0.43$\pm$0.12 ) \\ 
00034282134 & 8210.115 & 0.073$\pm$0.013 ( 0.47$\pm$0.18 ) \\ 
00094061001 & 8211.088 & 0.086$\pm$0.010 ( 0.58$\pm$0.14 ) \\ 
00094061002 & 8213.997 & 0.087$\pm$0.009 ( 0.42$\pm$0.09 ) \\ 
00034282135 & 8216.386 & 0.103$\pm$0.008 ( 0.54$\pm$0.09 ) \\ 
00034282136 & 8222.664 & 0.091$\pm$0.008 ( 0.43$\pm$0.08 ) \\ 
00094061003 & 8224.405 & 0.097$\pm$0.007 ( 0.31$\pm$0.06 ) \\ 
00034282137 & 8229.606 & 0.093$\pm$0.011 ( 0.36$\pm$0.09 ) \\ 
00034282138 & 8237.006 & 0.099$\pm$0.009 ( 0.28$\pm$0.06 ) \\ 
00094061004 & 8239.076 & 0.116$\pm$0.008 ( 0.44$\pm$0.06 ) \\ 
00034282139 & 8242.957 & 0.117$\pm$0.011 ( 0.56$\pm$0.11 ) \\ 
00034282140 & 8252.105 & 0.096$\pm$0.009 ( 0.31$\pm$0.07 ) \\ 
00094061005 & 8253.156 & 0.099$\pm$0.008 ( 0.58$\pm$0.09 ) \\ 
00034282141 & 8257.979 & 0.094$\pm$0.009 ( 0.43$\pm$0.09 ) \\ 
00094061006 & 8266.739 & 0.097$\pm$0.007 ( 0.29$\pm$0.05 ) \\ 
00094061007 & 8281.082 & 0.109$\pm$0.008 ( 0.31$\pm$0.05 ) \\ 
00094061008 & 8294.897 & 0.086$\pm$0.007 ( 0.53$\pm$0.09 ) \\ 
00094061009 & 8308.586 & 0.091$\pm$0.007 ( 0.34$\pm$0.06 ) \\ 
00094061010 & 8322.825 & 0.095$\pm$0.007 ( 0.43$\pm$0.07 ) \\ 
00094061011 & 8337.391 & 0.114$\pm$0.009 ( 0.50$\pm$0.09 ) \\ 
00094061012 & 8351.021 & 0.095$\pm$0.007 ( 0.34$\pm$0.06 ) \\ 
00088806001 & 8356.197 & 0.078$\pm$0.007 ( 0.53$\pm$0.11 ) \\ 
00088807001 & 8358.463 & 0.107$\pm$0.008 ( 0.54$\pm$0.09 ) \\ 
00094061013 & 8365.409 & 0.110$\pm$0.008 ( 0.44$\pm$0.07 ) \\ 
00094061014 & 8379.344 & 0.091$\pm$0.007 ( 0.48$\pm$0.08 ) \\ 
00094061015 & 8393.012 & 0.091$\pm$0.007 ( 0.44$\pm$0.07 ) \\ 
00094061016 & 8407.137 & 0.092$\pm$0.007 ( 0.43$\pm$0.07 ) \\ 
00094061017 & 8421.351 & 0.108$\pm$0.008 ( 0.46$\pm$0.07 ) \\ 
00094061018 & 8435.385 & 0.098$\pm$0.007 ( 0.50$\pm$0.08 ) \\ 
00094061019 & 8448.570 & 0.095$\pm$0.007 ( 0.57$\pm$0.09 ) \\ 
00094061020 & 8463.219 & 0.086$\pm$0.007 ( 0.68$\pm$0.11 ) \\ 
00094061021 & 8476.193 & 0.113$\pm$0.013 ( 0.44$\pm$0.11 ) \\ 
00094061022 & 8490.898 & 0.114$\pm$0.012 ( 0.55$\pm$0.13 ) \\ 
00094061023 & 8505.246 & 0.140$\pm$0.024 ( 0.25$\pm$0.11 ) \\ 
00094061024 & 8518.762 & 0.095$\pm$0.014 ( 0.35$\pm$0.12 ) \\ 
00094061025 & 8522.377 & 0.081$\pm$0.012 ( 0.50$\pm$0.16 ) \\ 
\hline
\end{tabular}
\end{minipage}
\\
\tablefoot{HJD correspond to dates at mid-exposure. The hardness ratio (between brackets) is defined as $HR=H/S$. The energy bands are the same as for \xmm\ data (Table \ref{journalxmm}).}
\end{table*}

\begin{table*}
  \caption{Journal of the new RV measurements of the eclipsing binary \label{journaloptical}}
  \begin{tabular}{c c r r r r r r r r}
\hline
Date & & \multicolumn{2}{c}{He\,{\sc i} $\lambda$\,4471} & \multicolumn{2}{c}{He\,{\sc ii} $\lambda$\,4542} & He\,{\sc ii} $\lambda$\,4686 & \multicolumn{2}{c}{He\,{\sc ii} $\lambda$\,5412} & O\,{\sc iii} $\lambda$\,5592 \\
(HJD-2\,450\,000) & & \multicolumn{1}{c}{A} & \multicolumn{1}{c}{B} & \multicolumn{1}{c}{A} & \multicolumn{1}{c}{B} & & \multicolumn{1}{c}{A} & \multicolumn{1}{c}{B} & \multicolumn{1}{c}{A}\\
\hline
6455.402 & A & $-219.3$ &    74.7  & $-172.8$ &    79.0  &   260.2  & ... & ... & ... \\
6456.455 & A &   ...    &   ...    &   ...    &   ...    &   135.0  & ... & ... & ... \\
6457.444 & A &    49.9  & $-267.9$ &    53.8  & $-273.1$ &  $-93.3$ & ... & ... & ... \\
6458.450 & A &    65.3  & $-495.3$ &    72.7  & $-374.9$ & $-246.6$ & ... & ... & ... \\
6459.435 & A &    54.5  & $-229.6$ &    25.5  & $-238.7$ & $-105.2$ & ... & ... & ... \\
6460.437 & A &   ...    &   ...    &   ...    &   ...    &   103.7  & ... & ... & ... \\
6811.467 & A & $-140.5$ &   108.7  & $-113.7$ &   203.1  &   296.4  & ... & ... & ... \\
6812.512 & A &   ...    &   ...    &   ...    &   ...    &   204.2  & ... & ... & ... \\
6813.501 & A &   ...    &   ...    &   ...    &   ...    &   $-9.5$ & ... & ... & ... \\
6814.500 & A &   112.4  & $-382.1$ &    23.8  & $-348.6$ & $-178.3$ & ... & ... & ... \\
6815.482 & A &    63.3  & $-349.9$ &    94.9  & $-195.2$ & $-149.0$ & ... & ... & ... \\
6816.516 & A &   ...    &   ...    &   ...    &   ...    &    63.3  & ... & ... & ... \\
7176.659 & A &   ...    &   ...    &   ...    &   ...    &  $-70.4$ & ... & ... & ... \\
7177.573 & A &   ...    &   ...    &   ...    &   ...    & $-190.2$ & ... & ... & ... \\
7178.584 & A &   ...    &   ...    &   ...    &   ...    & $-125.4$ & ... & ... & ... \\
7179.572 & A &   ...    &   ...    &   ...    &   ...    &    65.6  & ... & ... & ... \\
7180.579 & A &   ...    &   ...    &   ...    &   ...    &   240.6  & ... & ... & ... \\
7181.572 & A &   ...    &   ...    &   ...    &   ...    &   258.1  & ... & ... & ... \\
7287.713 & H &   ...    &  ...     &   ...    & ...      &   151.0  &  $-56.8$ &  ...     &  $-72.5$\\
7289.718 & H &    22.3  & $-397.5$ &   123.7  & $-383.8$ & $-223.7$ &   145.9  & $-398.3$ &   167.7 \\
7292.643 & H & $-136.5$ &    58.3  &   ...    &   ...    &   220.0  &  ...     &  ...     &   ...   \\
7299.611 & H &   ...    &  ...     &  $-98.0$ &   142.1  &   253.1  &  $-83.9$ & ...      &  $-98.5$\\ 
7332.582 & H &   ...    &  ...     & $-235.1$ &   125.9  &   277.8  & $-132.8$ & ...      &  $-60.5$\\
7337.575 & H &   ...    &  ...     &   ...    &   ...    &    24.7  &  $-48.0$ & ...      &   ...   \\
7339.571 & H &   ...    &  ...     &   ...    &   ...    &   248.1  &  $-92.3$ &   214.8  & $-126.9$\\
7514.913 & H &    80.3  & $-431.5$ &    61.3  & $-287.3$ & $-153.7$ &   114.1  & ...      &    69.4 \\ 
7547.547 & A &    21.8  & $-391.0$ &    67.0  & $-338.2$ & $-235.4$ & ... & ... & ... \\
7547.581 & A &    46.1  & $-407.4$ &    63.3  & $-313.0$ & $-216.8$ & ... & ... & ... \\
7548.522 & A &   ...    &  ...     &   ...    &   ...    &  $-38.0$ & ... & ... & ... \\
7548.554 & A &   ...    &  ...     &   ...    &   ...    &  $-32.8$ & ... & ... & ... \\
7548.586 & A &   ...    &  ...     &   ...    &   ...    &  $-22.6$ & ... & ... & ... \\ 
7549.571 & A &   ...    &  ...     &   ...    &   ...    &   168.4  & ... & ... & ... \\
7550.568 & A &   ...    &  ...     &   ...    &   ...    &   258.3  & ... & ... & ... \\
7551.518 & A &   ...    &  ...     &   ...    &   ...    &   192.9  & ... & ... & ... \\
7843.948 & H &    77.6  & $-424.7$ &    11.6  & $-364.5$ & $-218.4$ &    47.0  & $-405.3$ &    91.3 \\
7845.969 & H &   ...    &  ...     &   ...    &   ...    &    28.3  &  $-28.2$ & ...      &  $-45.0$\\
7847.967 & H &   ...    &   ...    & $-158.0$ &     64.5 &   212.3  &  $-38.4$ & ...      &   ...   \\
7848.968 & H &   ...    &   ...    &   ...    &   ...    &    52.0  &  $-46.7$ & ...      &  $-13.1$\\
7862.958 & H &    37.3  & $-261.5$ &    61.0  & $-341.9$ & $-115.1$ &   103.8  & $-377.3$ &    50.2 \\
7864.949 & H &   ...    &   ...    &   ...    &   ...    &  $-93.6$ &   ...    & ...      &   ...   \\
7866.955 & H &   ...    &   ...    &   ...    &   ...    &   199.2  &   ...    & ...      &   ...   \\
7871.907 & H &   ...    &   ...    &   ...    &   ...    &  $-33.4$ &  $-28.5$ & ...      &    15.4 \\
7873.901 & H &   ...    &   ...    &   ...    &   ...    &   231.0  &  $-53.5$ & ...      &  $-50.4$\\
7875.899 & H &  ...     &   ...    &    61.6  & $-262.6$ & $-79.81$ &    68.7  & $-263.8$ &    73.1 \\
7879.946 & H &   ...    &   ...    &   ...    &   ...    &   205.0  &  $-41.1$ & ...      &   ...   \\
8002.466 & A &    21.6  & $-423.0$ &    77.6  & $-378.4$ & $-215.1$ & ... & ... & ... \\
8002.494 & A &    26.9  & $-444.5$ &    57.7  & $-392.1$ & $-223.7$ & ... & ... & ... \\
8003.512 & A &  ...     &   ...    &   ...    &   ...    &  $-99.2$ & ... & ... & ... \\
8004.459 & A &  ...     &   ...    &   ...    &   ...    &    99.5  & ... & ... & ... \\
8004.487 & A &  ...     &   ...    &   ...    &   ...    &    85.9  & ... & ... & ... \\
8005.491 & A &  ...     &   ...    &   ...    &   ...    &   219.5  & ... & ... & ... \\
8006.521 & A &  ...     &   ...    &   ...    &   ...    &   204.8  & ... & ... & ... \\
8007.451 & A &  ...     &   ...    &   ...    &   ...    &    10.3  & ... & ... & ... \\
8197.979 & H &  ...     &   ...    &   ...    &   ...    &   211.8  &  $-30.7$ & ...      &  $-66.9$\\
8199.979 & H &    16.6  & $-375.3$ &    96.5  & $-367.1$ & $-211.7$ &   118.8  & $-420.5$ &   ...   \\
8202.971 & H &  ...     &   ...    &   ...    &   ...    &   197.4  &  $-19.6$ & ...      &   ...   \\
8205.967 & H &  ...     &   ...    &   ...    &   ...    & $-101.6$ &    68.7  & $-346.2$ &    72.1 \\
8206.976 & H &    20.6  & $-457.3$ &   113.3  & $-353.4$ & $-222.2$ &   105.1  & $-402.6$ &   ...   \\
8211.957 & H &  ...     &   ...    &   ...    &   ...    &    24.8  &  $-45.8$ & ...      &   $-5.1$\\
\hline
\end{tabular}
\end{table*}

\begin{table*}
  \addtocounter{table}{-1}
  \caption{Continued}
  \begin{tabular}{c c r r r r r r r r}
\hline
Date & & \multicolumn{2}{c}{He\,{\sc i} $\lambda$\,4471} & \multicolumn{2}{c}{He\,{\sc ii} $\lambda$\,4542} & He\,{\sc ii} $\lambda$\,4686 & \multicolumn{2}{c}{He\,{\sc ii} $\lambda$\,5412} & O\,{\sc iii} $\lambda$\,5592 \\
(HJD-2\,450\,000) & & \multicolumn{1}{c}{P} & \multicolumn{1}{c}{S} & \multicolumn{1}{c}{P} & \multicolumn{1}{c}{S} & & \multicolumn{1}{c}{P} & \multicolumn{1}{c}{S} & \multicolumn{1}{c}{P}\\
\hline
8214.958 & H &  ...     &   ...    &   ...    &   ...    &  $-49.8$ &  $-21.8$ & ...      &    46.6 \\
8216.934 & H & $-147.1$ &    92.5  &   ...    &  ...     &   241.8  & $-109.1$ & ...      &   ...   \\
8222.921 & H &  ...     &   ...    & $-169.3$ &  ...     &   206.8  &  $-48.2$ & ...      &  $-60.7$\\
8225.945 & H &  ...     &   ...    &    41.3  & $-375.4$ & $-151.5$ &    99.6  & $-370.7$ &    33.2 \\
8228.954 & H &  ...     &   ...    &   ...    &  ...     &   123.9  &   $-6.8$ & ...      &   ...   \\
8233.958 & H &    36.9  & $-411.0$ &    54.5  & $-343.0$ & $-175.4$ &    56.1  & $-346.9$ &   ...   \\ 
8353.470 & A &  ...     &   ...    &   ...    &   ...    &  $-61.3$ & ... & ... & ... \\
8356.419 & A &  ...     &   ...    &   ...    &   ...    &   163.1  & ... & ... & ... \\
8357.417 & A &   $-1.9$ & $-221.8$ &    64.3  & $-186.8$ &  $-48.7$ & ... & ... & ... \\
8562.979 & H & 69.4 & $-433.2$ & 121.4 & $-395.1$ & $-249.4$ &    75.6  & $-397.4$ &     ... \\
8565.977 & H & ...  & ...      & ... & ...    &   198.7  &  $-15.4$ &   ...    & $-43.7$ \\
8566.972 & H & ...  & ...      & 6.8 & ...    &   225.6  &  $-10.8$ &   ...    & $-33.6$ \\\
8572.956 & H & ...  & ...      & ... & ...    &   237.2  & $-113.2$ &   ...    &      ...\\
8575.961 & H & ...  & ...      & ... & ...    & $-203.3$ &  ...     &   ...    &      ...\\
8580.913 & H & ...  & ...      & ... & ...    &   114.4  &  ...     &   ...    &      ...\\
8580.965 & H & ...  & ...      & $-57.9$ & ...&   120.1  &  $-67.1$ &   ...    & $-36.4$ \\
8581.959 & H & ...  & ...      & ... & ...    & $-118.7$ &    57.8  & $-352.5$ &      ...\\
8594.940 & H &  4.1 & $-231.1$ & ... & ...    &  $-58.0$ &    63.0  & $-295.8$ &   51.4  \\
8595.932 & H & ...  & ...      & ... & ...    & $-213.4$ &  ...     &   ...    &      ...\\
8596.912 & H & 14.0 & $-406.8$ & 46.2 & $-293.8$ & $-183.8$ &    84.1  & $-360.9$ &   34.5  \\
\hline
  \end{tabular}
  \tablefoot{The A and B letters in the table header identify RVs of the primary and secondary star respectively. All RVs are given in km\,s$^{-1}$ in the heliocentric frame of reference. The A and H in the second column stand for data taken with the Aur\'elie and HEROS spectrographs respectively.}
\end{table*}

\section{Analysis of the light curve of the eclipsing binary \label{AppendixD}}
As pointed out above, we combined all the available photometric measurements to construct an empirical $V$-band light curve consisting of points which are obtained by taking the median of the data in 50 equally spaced phase bins. Unlike the original data, this empirical light curve is essentially free of intrinsic variability. We thus took advantage of this light curve to perform a new photometric solution of the eclipsing binary with the {\tt nightfall} code (version 1.86) developed by R.\ Wichmann, M.\ Kuster and P.\ Risse\footnote{The code is available at the URL: http://www.hs.uni-hamburg.de/DE/Ins/Per/Wichmann/Nightfall.html} \citep{Wichmann}. This code relies on the Roche potential to describe the shape of the stars. For the simplest cases (two stars on a circular orbit with neither stellar spots nor discs), the model is thus fully described by six parameters: the mass-ratio, the orbital inclination, the primary and secondary filling factors (defined as the ratio of the stellar polar radius to the polar radius of the associated Roche lobe), and the primary and secondary effective temperatures. We set the mass-ratio to 3.2 and the stellar effective temperatures to 36\,000\,K. Following \citet{ant16}, we adopted a square-root limb-darkening law. Reflection effects were accounted for by considering the mutual irradiation of all pairs of surface elements of the two stars \citep{Hendry}. In accordance with \citet{lin09}, we found that including a bright spot on the secondary star significantly improves the quality of the fits.

Previous photometric solutions, not accounting for the presence of third light, yielded orbital inclinations in the range from $64^{\circ}$ to $68^{\circ}$ \citep{ls78,lin09,yas14,lau15,ant16}. As shown by \citet{caz14}, accounting for a third light contribution leads to higher inclinations. Third light arises from components C and D. According to \citet{mas01}, Star D is 2.5\,mag fainter than the combination of A, B and C (assuming that the \citet{mas01} magnitude refers to the maximum light of the eclipsing binary). This implies that the third light contribution from Star D should be $l_D/(l_A+l_B+l_C+l_D) \leq 0.1$. Our best-fit parameters for the $V$-band light curve (Table\,\ref{bestfit}) hence suggest a third light contribution due to star C around $l_C/(l_A+l_b+l_C+lD) \sim 0.13$. 

In line with previous studies \citep{ls78,lin09,yas14,lau15,ant16}, we can achieve a reasonable fit of the light curve for an overcontact configuration (see Fig.\,\ref{bestfit}). Yet, whilst the fit looks reasonable, there are two major problems which remain unsolved: the conflict between the visual brightness ratio inferred from photometry and spectroscopy, and the distance problem.

The light-curve solutions that we obtained yield a visual brightness ratio of $3.6 \pm 0.4$ between the primary and secondary star. This value is significantly larger than the brightness ratio near 1.4 inferred from the strengths of the primary and secondary spectral lines \citep{rau99}.

\citet{hal74} quoted $m_V = 9.05$ and $B - V = 1.72$ outside eclipses. This is in good agreement with the zero points of the photometric data in our analysis which suggest $m_V = 9.10 \pm 0.05$ outside eclipse. We thus adopt $m_V = 9.10 \pm 0.05$ and $B - V = 1.72 \pm 0.07$ outside eclipse, along with $(B - V)_0 = -0.27$ \citep{MP} and $R_V = 3.0$ \citep{Mas91}, resulting in $A_V = 5.97 \pm 0.21$. With the {\it Gaia}-DR2 distance modulus of $10.88^{+0.20}_{-0.18}$ \citep{bai18}, we thus estimate an absolute magnitude of $M_{V, DR2} = -7.75^{+0.28}_{-0.29}$. Adopting the bolometric corrections from \citet{MP}, we can convert the parameters of the best-fitting {\tt nightfall} models into an absolute magnitude of $M_{V, {\rm model}} = -7.33 \pm 0.10$. This value is significantly fainter than the {\it Gaia}-DR2 absolute magnitude, indicating that we are missing about one-third of the total flux of the system. 
\begin{figure}[htb]
  \begin{center}
  \resizebox{8cm}{!}{\includegraphics{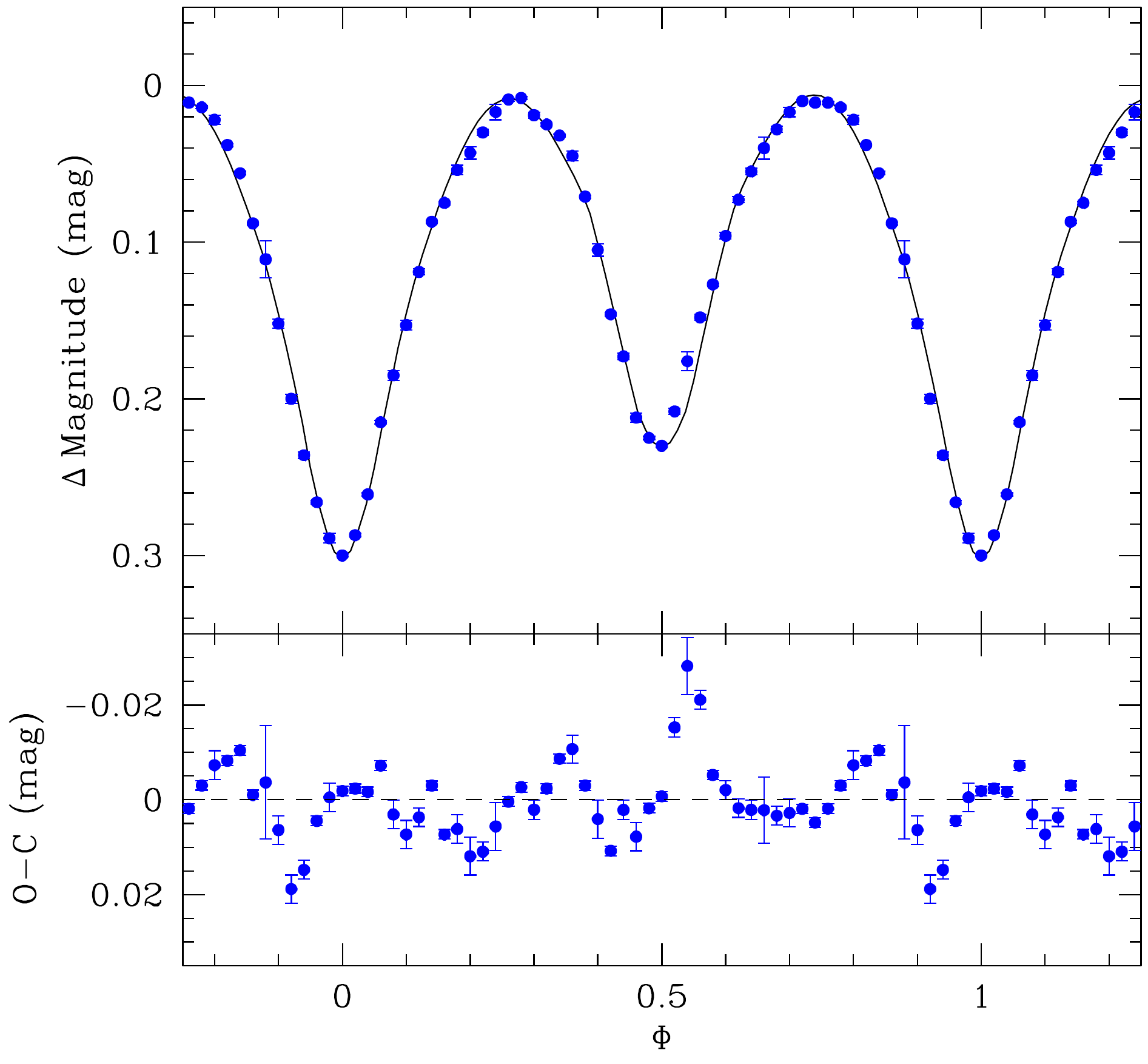}}
\end{center}  
\caption{Best-fit {\tt nightfall} solution of the $V$-band light curve of \ci. The model parameters are given in Table\,\ref{paramphoto}. In the model, the secondary features a bright spot on the side facing the primary. \label{bestfit}}
\end{figure}
\begin{table}
  \caption{Parameters of the best fit with {\tt nightfall} to the $V$-band light curve of \ci.\label{paramphoto}}
  \begin{center}
  \begin{tabular}{l c}
  \hline\hline
  Parameter & Value \\
  \hline
  $q$ & 3.2 (fixed) \\
  $fill_{\rm A} = fill_{\rm B}$ & $1.04 \pm 0.03$ \\
  $i_{\rm A+B}$\,($^{\circ}$) & $69.4_{-1.9}^{+0.6}$ \\
  $T_{\rm A}$\,(K) & 36\,000 (fixed) \\
  $T_{\rm B}$\,(K) & $25\,000 \pm 500$ \\
  $(l_{\rm C}+l_{\rm D})/(l_{\rm A}+l_{\rm B}+l_{\rm C}+l_{\rm D})$ & $0.228_{-0.074}^{+0.050}$ \\
  Spot latitude ($^{\circ}$) & 0 (fixed) \\
  Spot longitude ($^{\circ}$) & $1.7 \pm 0.5$ \\
  Spot radius ($^{\circ}$) & $36.8 \pm 1.0$ \\
  Dim factor & $1.9 \pm 0.1$ \\
  \hline
  \end{tabular}
  \end{center}
  \tablefoot{The uncertainties on $T_{\rm B}$, the spot radius and dim factor are only indicative as these parameters present a high degree of correlation.}
  \end{table}

A possible solution to the distance problem could be a higher temperature and thus higher luminosity of the primary star. Yet, we emphasize that the strength of the primary's He\,{\sc i} $\lambda$\,4471 line relative to that of the He\,{\sc ii} $\lambda$\,4542 line is fully compatible with an O6.5-7 spectral type for the primary, although this conclusion could be somewhat affected by blends with spectral features of Star C. We further note that the spectra display no N\,{\sc v} $\lambda\lambda$\,4604, 4620 absorptions, indicating a spectral type later than O4 \citep{Walborn} and thus a temperature $< 40\,000$\,K. Therefore, the primary temperature is unlikely to significantly exceed 36\,000\,K. The interstellar absorption towards Cyg\,OB2 is very large. Whilst our values of $E(B-V)$ and $A_V$ are in line with other determinations \citep{Lei82,Mas91,Han03}, \citet{Tor91} derived a higher value of $A_V = 6.40$. Adopting this value of the visual extinction would worsen the distance problem as it would bring the discrepancy between the two absolute magnitude estimates to 0.85\,mag instead of 0.42\,mag.

The distance problem would disappear if the secondary were as bright as the primary in the $V$ band, which is actually what the spectroscopic data suggest. However, this is clearly at odds with the light-curve analysis. The unequal eclipse depths indicate a primary star that should be hotter than the secondary. Therefore, to have similar optical brightnesses, the secondary should be at least as big as the primary. This is clearly not possible for a contact or overcontact configuration with $q = m_{\rm A}/m_{\rm B} = 3.1$.

As pointed out above, the light-curve model assumes that the shape of the stars is given by the Roche potential. \citet{ant16} suggested that the presence of an accretion disc around one of the components of the eclipsing binary could affect the light curve. Yet, Doppler tomography of the H$\alpha$ and He\,{\sc ii} $\lambda$\,4686 emission lines revealed no evidence of a disc-like structure around any of the stars \citep{lin09}. Therefore, there is no observational justification for including such a disc in the photometric solution. Alternatively, the presence of a strong wind around one of the components could also affect the light curve via the formation of atmospheric eclipses. As shown by \citet{ant13}, optically thick stellar winds can lead to non-equal eclipse depths for stars that have otherwise identical properties. There are some indications in favour of such a scenario. We can first cite the large difference of the systemic velocities between the primary and secondary star (see Table\,\ref{systemic} and Fig.\,\ref{RVsPcourte}), as well as the shift in systemic velocity between different lines and compared to the systemic velocities of other early-type binaries in Cyg\,OB2 \citep{Kob14}. These results indicate that both stars have a strong stellar wind, and that the secondary wind is probably denser than that of the primary. Further support comes from the appearance of variable P-Cygni type profiles at certain orbital phases. Atmospheric eclipses can indeed result in variable absorption troughs of P-Cygni profiles \citep{Auer94}. At first sight, because of the wavelength-dependence of the opacity, we would expect a stellar wind to produce light curves with differences in morphology as a function of wavelength. This is best probed with narrow-band filters that mainly encompass continuum versus filters that specifically focus on wind emission lines. Such a set of filters was used by \citet{lin09}. Whereas the He\,{\sc ii} $\lambda$\,4686 line-bearing filter showed no strong deviation from the continuum filters, significant differences were found for the He\,{\sc i} $\lambda$\,5876 line-bearing filter. Whilst the former result suggests that the opacity of the He\,{\sc ii} $\lambda$\,4686 line in the wind does not exceed that of the continuum, the latter situation was explained by \citet{lin09} as a consequence of the variable He\,{\sc i} $\lambda$\,5876 P-Cygni absorption trough. The most likely picture is thus that of a secondary wind that is optically thick in the continuum \citep[due to free electron scattering;][]{ant13}, as is the case for Wolf-Rayet stars. Under such circumstances, the ``pseudo-photosphere'' of the secondary might actually extend beyond the size of the Roche equipotential filled by the hydrostatic core of the secondary star.

From the above considerations, we see that both issues of the photometric solution could possibly be solved via the inclusion of the wind absorption into the model. Including wind parameters (mass-loss rate, wind terminal velocity and velocity law) introduces however a number of degeneracies \citep{ant13}. Designing an algorithm to solve the light curve accounting for the effects of the stellar wind and overcoming the limitations of these degeneracies is a long and strenuous task unrelated to the goal of this paper and that we defer to future work.

\section{Revised orbital solution of the eclipsing binary \label{AppendixC}}
We used our newly determined RVs (Table \ref{journaloptical}) along with the quadratic ephemerides (Eq.\,\ref{ephem2}) to revise the orbital solution of the inner binary system. For the primary star, the He\,{\sc ii} $\lambda$\,5412 and O\,{\sc iii} $\lambda$\,5592 lines consistently yield a semi-amplitude of the RV curve of $K_{\rm A} \sim 88$\,km\,s$^{-1}$ (see Fig.\,\ref{RVsPcourte}). This value is about seven percent larger than the semi-amplitude inferred for the same lines by \citet{rau99}, although the results overlap within the error bars. The RVs of the He\,{\sc i} $\lambda$\,4471 and He\,{\sc ii} $\lambda$\,4542 lines of the primary yield larger values of $K_{\rm A}$ (between 110 and 115\,km\,s$^{-1}$). The dispersion of the data points around the best-fit RV curve is larger for those two lines. These differences in $K$ and their dispersion certainly reflect the difficulties due to the lines displaying P-Cygni type profiles at some orbital phases. For the secondary, the cleanest results are obtained from the RVs of the He\,{\sc ii} $\lambda$\,5412 line. This line yields $K_{\rm B} \sim 325$\,km\,s$^{-1}$. Unfortunately, because of the change in line morphology around phase 0.25, our new secondary RV measurements are extremely scarce at this phase and this situation could bias our determination of $K_{\rm B}$. Indeed, the He\,{\sc i} $\lambda$\,4471 and He\,{\sc ii} $\lambda$\,4542 lines yield $K_{\rm B}$ values of 261 and 255\,km\,s$^{-1}$, respectively. In view of these uncertainties, we thus adopted $K_{\rm A} = (87.7 \pm 6.1)$\,km\,s$^{-1}$ from the He\,{\sc ii} $\lambda$\,5412 and O\,{\sc iii} $\lambda$\,5592 RVs, and $K_{\rm B} = (280 \pm 31)$\,km\,s$^{-1}$, i.e.\ the average of the $K_{\rm B}$ values of the three spectral lines for which the second component can be measured. The mass-ratio hence becomes $3.2 \pm 0.4$, whilst the minimum masses of the primary and secondary stars are $m_{\rm A}\,\sin^3{i_{\rm A+B}} = (25.8 \pm 8.6)$\,M$_{\odot}$ and $m_{\rm B}\,\sin^3{i_{\rm A+B}} = (8.1 \pm 1.9)$\,M$_{\odot}$, respectively. The corresponding projected orbital separation is $a_{\rm A+B}\,\sin{i_{\rm A+B}} = (47.8 \pm 4.0)$\,R$_{\odot}$.

We note that the different lines yield different apparent systemic velocities (see Table\,\ref{systemic}). This situation probably reflects the presence of optically thick stellar winds, especially for the secondary star.
\begin{table}
  \caption{Apparent systemic velocities of the absorption lines of \ci \label{systemic}}
  \begin{center}
    \begin{tabular}{c c c}
      \hline\hline
      Line & Primary (Star A) & Secondary (Star B) \\
      \hline
      He\,{\sc i} $\lambda$\,4471  & $-49.5 \pm 9.4$ & $-160 \pm 20$\\
      He\,{\sc ii} $\lambda$\,4542 & $-34.7 \pm 8.2$ & $-113 \pm 23$\\
      He\,{\sc ii} $\lambda$\,5412 & $7.1 \pm 6.4$ & $-100 \pm 18$ \\
      O\,{\sc iii} $\lambda$\,5592 & $6.0 \pm 6.6$ &  \\
      \hline
    \end{tabular}
  \end{center}
\end{table}
      
The revised orbital solution is shown in Fig.\,\ref{RVsPcourte} and summarized in Table\,\ref{orbitalparam}. Although the errors on the revised orbital solution are larger than for the solution of \citet{rau99}, the new solution is to be preferred. In fact, the errors now account for the uncertainties related to the choice of the lines in the orbital solution, whilst this was not the case in the solution of \citet{rau99}.

\begin{figure}[htb]
  \begin{center}
    \resizebox{8cm}{!}{\includegraphics{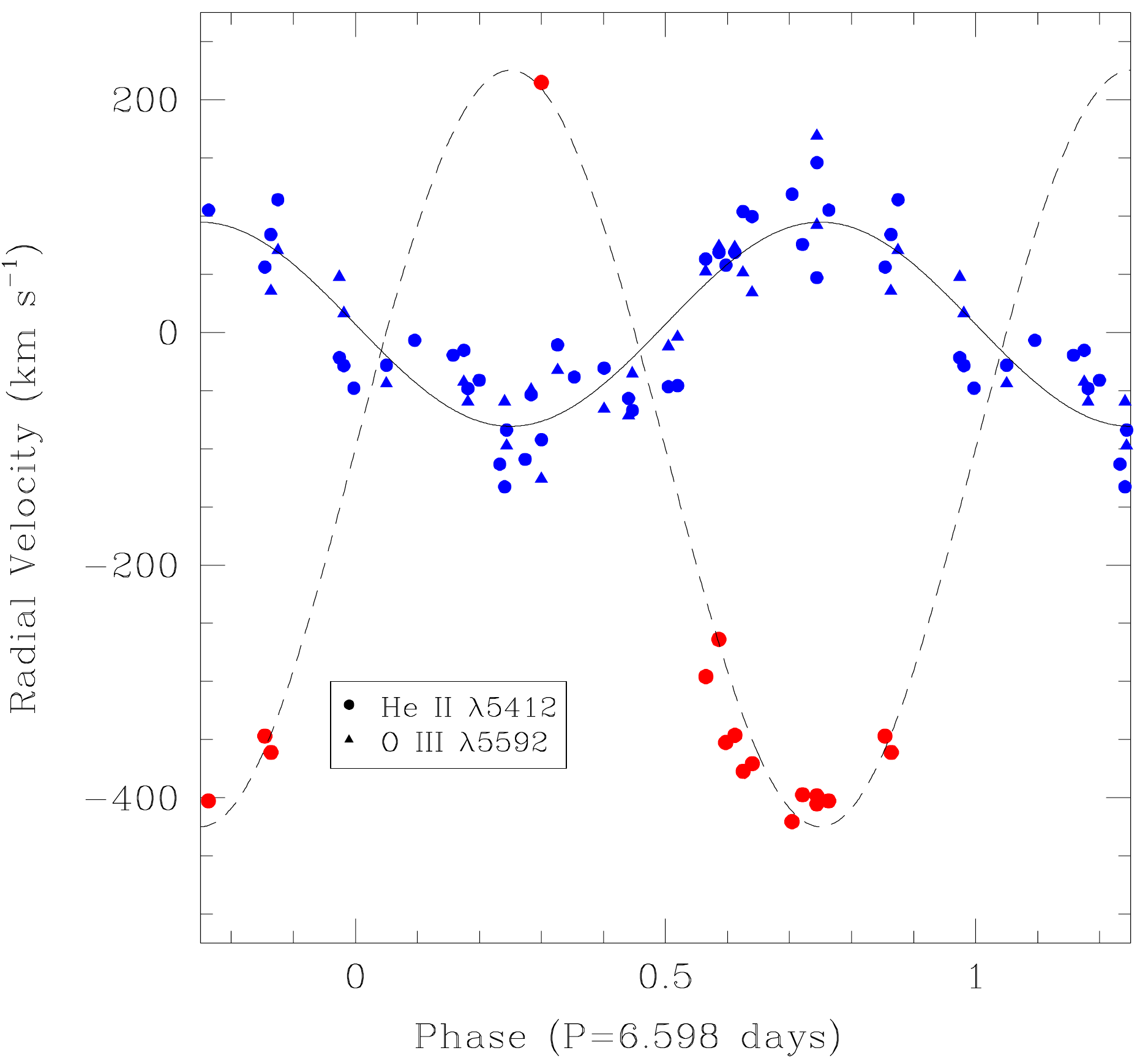}}
    \resizebox{8cm}{!}{\includegraphics{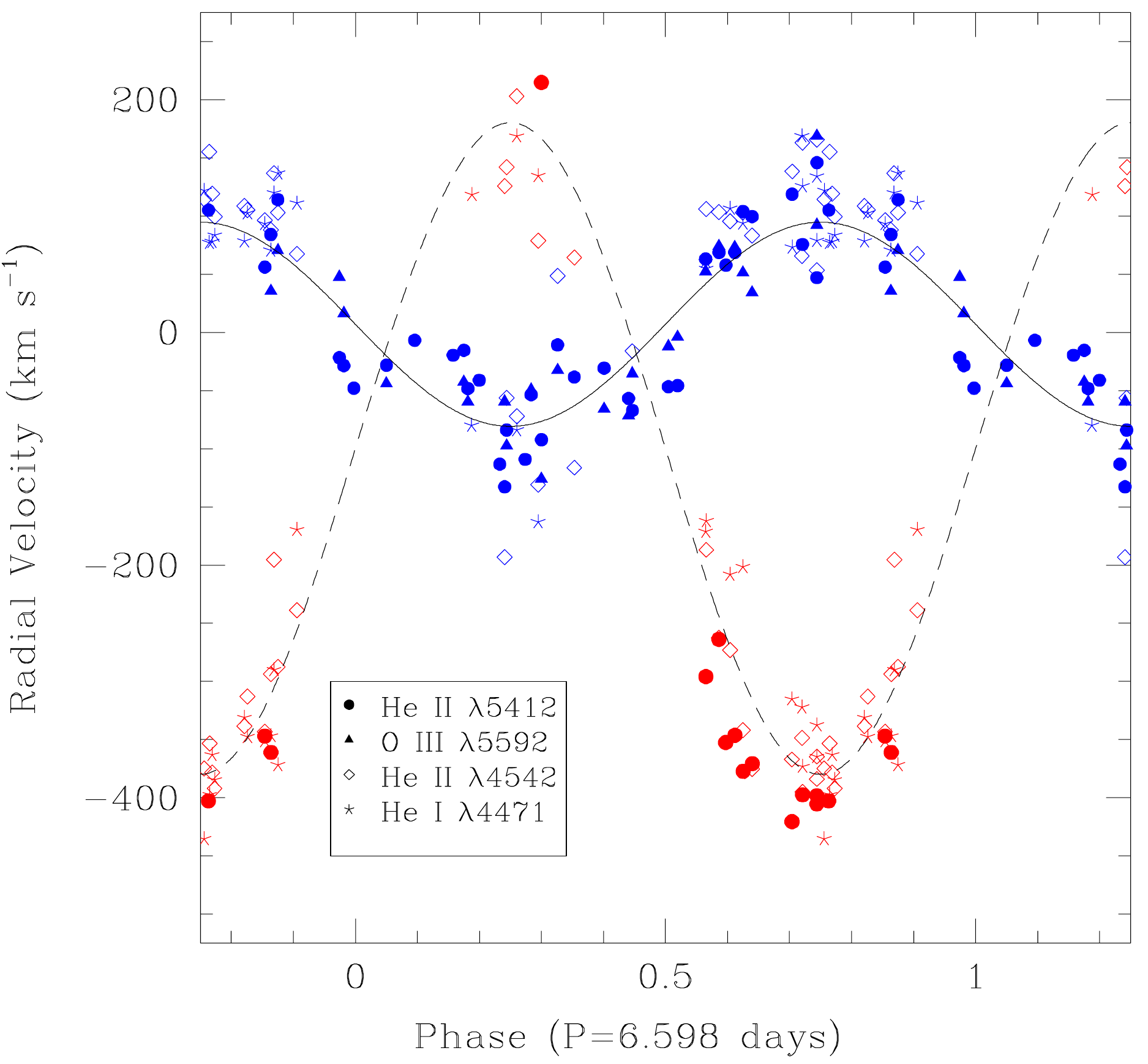}}
\end{center}  
  \caption{Radial velocity curves of the eclipsing binary. The orbital phases were computed with the quadratic ephemerides (Eq.\,\ref{ephem2}). The top panel illustrates the results obtained from the He\,{\sc ii} $\lambda$\,5412 and O\,{\sc iii} $\lambda$\,5592 RVs. Blue symbols stand for the primary star, whilst red symbols indicate the secondary. The primary RVs of O\,{\sc iii} $\lambda$\,5592 were shifted to match the systemic velocity of the He\,{\sc ii} $\lambda$\,5412 line (see Table\,\ref{systemic}). The solid and dashed curves correspond to the best-fit orbital solution based on this line. The bottom panel illustrates the full set of RV measurements along with the preferred RV curves (see Table\,\ref{orbitalparam}). For each star, the RVs of the various lines were shifted to match the systemic velocities of the He\,{\sc ii} $\lambda$\,5412 line. \label{RVsPcourte}}
\end{figure}

\section{Expression of the free-free optical depth towards a point-like source \label{tau}}
\citet{wil90} provided expressions for the free-free optical depth $\tau_{4.8}(t)$ as a function of the orbital parameters. These authors used expressions that include the tangent of the orbital inclination. Yet, for inclinations near $90^{\circ}$, it is advantageous to use expressions that involve the cotangent of the orbital inclination instead. In this appendix, we provide these expressions.

For an orbital inclination $i_{\rm AB+C} = 90^{\circ}$, we have
\begin{equation}
  \tau_{\nu}(t) = \frac{\tau_0}{2\,\left(\frac{r}{a_{\rm AB+C}}\,|\cos{(\phi_{\rm AB+C}+\omega_{\rm C})}|\right)^3}\,(\psi_1 - \sin{\psi_1}\,\cos{\psi_1})
,\end{equation}
where $\tau_0$ is a fitting parameter, $\omega_{\rm C} = \omega_{\rm AB} - \pi$ and $\psi_1$ is given by
\begin{eqnarray}
  \psi_1 = \frac{\pi}{2} + (\phi_{\rm AB+C} + \omega_{\rm C}) & {\rm if} & (\phi_{\rm AB+C} + \omega_{\rm C}) \in [0,\frac{\pi}{2}]\\
  \psi_1 = \frac{3\,\pi}{2} - (\phi_{\rm AB+C} + \omega_{\rm C}) & {\rm if} & (\phi_{\rm AB+C} + \omega_{\rm C}) \in [\frac{\pi}{2},\frac{3\,\pi}{2}]\\
  \psi_1 = (\phi_{\rm AB+C} + \omega_{\rm C}) - \frac{3\,\pi}{2} & {\rm if} & (\phi_{\rm AB+C} + \omega_{\rm C}) \in [\frac{3\,\pi}{2},2\,\pi]
.\end{eqnarray}

For an orbital inclination $i_{\rm AB+C} < 90^{\circ}$, we obtain
\begin{equation}
  \tau_{\nu}(t) = \frac{\tau_0}{\left(\frac{r}{a_{\rm AB+C}}\,|\cos{(\phi_{\rm AB+C}+\omega_{\rm C})}|\right)^3\,\sin{i}}\,(T_1 + T2)
,\end{equation}
where
\begin{equation}
  T_1 = \frac{C'}{2\,\theta^3\,A'^{3/2}}\,\left[\arctan{\left(\frac{2\,A'\,x+B'}{2\,\sqrt{A'}\,\theta}\right)}-\arctan{\left(\frac{B'}{2\,\sqrt{A'}\,\theta}\right)}\right]
  \label{T1}
,\end{equation}
\begin{eqnarray}
  T_2 & = & \frac{(B'^2 - 2\,A'\,C')\,x + B'\,C'}{A'\,(4\,A'\,C'-B'^2)\,(A'\,x^2+B'\,x+C')} \nonumber \\
  & & - \frac{B'\,C'}{A'\,C'\,(4\,A'\,C'-B'^2)}
,\end{eqnarray}
with
\begin{equation}
  x  = \tan{\psi_1}
,\end{equation}
\begin{equation}
  A' = 1 + \cot^2{i_{\rm AB+C}}\,\tan^2{(\phi_{\rm AB+C}+\omega_{\rm C})}
,\end{equation}
\begin{equation}
  B' = 2\,|\tan{(\phi_{\rm AB+C}+\omega_{\rm C})}|\,\cot^2{i_{\rm AB+C}}
,\end{equation}
\begin{equation}
  C' = 1 + \cot^2{i_{\rm AB+C}}
,\end{equation}
\begin{equation}
  \theta^2 = C' - \frac{B'^2}{4\,A'} = \frac{1 + \cot^2{i_{\rm AB+C}}\,(1+\tan^2{(\phi_{\rm AB+C}+\omega_{\rm C})})}{1 + \cot^2{i_{\rm AB+C}}\,\tan^2{(\phi_{\rm AB+C}+\omega_{\rm C})}}
,\end{equation}
and $\psi_1$ has the same definition as above. Attention must be paid in the expression of $T_1$ (equation\,\ref{T1}) to substitute $\arctan{\left(\frac{2\,A'\,x+B'}{2\,\sqrt{A'}\,\theta}\right)}$ by $\pi + \arctan{\left(\frac{2\,A'\,x+B'}{2\,\sqrt{A'}\,\theta}\right)}$ when the arctangent becomes negative. 
\end{document}